\documentclass[english]{iopart}

\usepackage{graphicx}
\usepackage{url}
\usepackage{iopams} 

\expandafter\let\csname equation*\endcsname\relax
\expandafter\let\csname endequation*\endcsname\relax

\usepackage{amssymb}
\usepackage[latin1]{inputenc}
\usepackage{amsmath}
\usepackage{epsfig}
\newcounter{fig}
\begin{document}

\title[Diagonals of rational functions]
{\Large Diagonals of rational functions, pullbacked $\, _2F_1$ hypergeometric functions and modular forms (unabrigded version)}

\vskip .3cm 

\author{Y. Abdelaziz$^\dag$  S. Boukraa$^\pounds$, C. Koutschan$^\P$,
J-M. Maillard$^\dag$}

\address{$^\dag$ LPTMC, UMR 7600 CNRS, 
Universit\'e Pierre et Marie Curie, Sorbonne Universit\'e, 
Tour 23, 5\`eme \'etage, case 121, 
 4 Place Jussieu, 75252 Paris Cedex 05, France} 

\address{$^\pounds$  \ LPTHIRM and IAESB,
 Universit\'e de Blida, Algeria}

\address{$^\P$  Johann Radon Institute for Computational and Applied Mathematics, RICAM, Altenberger Strasse 69, 
A-4040 Linz,  Austria}

\vskip .2cm 

\begin{abstract}

We recall that diagonals of rational functions naturally occur 
in lattice statistical mechanics and enumerative combinatorics. We find 
that a seven-parameter rational function of 
three variables  with a numerator equal to one (reciprocal 
of a polynomial of degree two at most) can be expressed as a pullbacked 
$\, _2F_1$ hypergeometric function. This result can be seen as the simplest 
non-trivial family of diagonals of rational functions. We focus on some subcases 
such that the diagonals of the corresponding rational functions can be written 
as a pullbacked $\, _2F_1$ hypergeometric function with two possible 
rational functions pullbacks algebraically related by modular equations, 
thus showing explicitely that the diagonal is a modular form. We then generalise 
this result to eight, nine and ten parameters families adding some selected 
cubic terms at the denominator of the rational function defining 
the diagonal. We finally show that each of these 
previous rational functions yields an infinite number of  rational functions
whose diagonals are also pullbacked $\, _2F_1$ hypergeometric functions 
and modular forms. 

\end{abstract}

\vskip .1cm

% Sorbonne Universit\'e, CNRS, Laboratoire de Physique Th\'eorique de la Mati\`ere Condens\'ee, LPTMC, F-75005 Paris, France. 
% {\bf 12th May 2018}  

\vskip .4cm


\noindent {\bf AMS Classification scheme numbers}: 34M55, 
47E05, 81Qxx, 32G34, 34Lxx, 34Mxx, 14Kxx 

\vskip .2cm


\vskip .1cm

\section{Introduction}
\label{Introduction}

It was shown in~\cite{Short,Big} that different physical related quantities, 
like the $n$-fold integrals $\, \chi^{(n)}$, corresponding 
to the $\, n$-particle contributions of the magnetic susceptibility 
of the Ising model~\cite{chi3,High,Khi6,bo-ha-ma-ze-07b},  
or the {\em lattice Green functions}~\cite{GoodGuttmann,GlasserGuttmann,DiffAlgGreen,LGF,LGF2}, are 
{\em diagonals of rational functions}~\cite{Purdue,Lipshitz,Christol84,Christol85,Christol369,Christol111}.

While showing that the $\, n$-fold integrals $\, \chi^{(n)}$ of the susceptibility of the Ising model
are diagonals  of rational functions requires some effort, seeing that the lattice 
Green functions are diagonals of rational functions nearly follows from 
their definition. For example, the lattice Green functions 
(LGF) of the $d$-dimensional face-centred cubic (fcc) lattice are given~\cite{LGF,LGF2} by:
\begin{eqnarray}
\label{LGFd}
\hspace{-0.98in}&&    
{\frac{1}{\pi^d}}\, \int_0^\pi \cdots \,
 \int_0^\pi \, {\frac{\mathrm{d}k_1\,  \cdots \,\,  \mathrm{d}k_d}{1\,-x \cdot \, \lambda_d}},
 \, \, \,  \, \, \, \, \,  \,  \,  \hbox{with:} \, \,  \,  \,  \, \, \, \, \, \, 
\lambda_d =\, { d \choose 2}^{-1} \, \, \sum_{i=1}^d \, \sum_{j=i+1}^d 
\cos(k_i) \, \cos(k_j).
\end{eqnarray}
The LGF can easily be seen to be a diagonal of a rational function: introducing the 
complex variables $\, z_j \, = \, \, e^{i\, k_j}$, $\, j \, = \, \, 1, \cdots, \, d$,
the LGF (\ref{LGFd}) can be seen as a $\, d$-fold generalization of Cauchy's contour 
integral~\cite{Short}:
\begin{equation}
  \textsf{Diag} ({\cal F})  \, \,\,  = \,  \, \, \,\,
  \frac{1}{2\pi i}\, \oint_\gamma {\cal F}(z_1,z/z_1) \, \frac{dz_1}{z_1}.
\end{equation}

Furthermore, the linear differential operators annihilating the physical quantities
 mentioned earlier $\, \chi^{(n)}$, are reducible operators. Being reducible 
they are ``breakable'' into smaller factors~\cite{High,Khi6} 
that happen to be elliptic functions, or generalizations thereof: {\em modular forms}, 
Calabi-Yau operators~\cite{CalabiYauIsing,IsingCalabi}... 
 Yet there exists a class of diagonals 
of rational functions in {\em three} variables\footnote[9]{Diagonals of rational functions 
of two variables are just algebraic functions, so one must consider 
{\em at least  three} variables to obtain special functions.} 
whose diagonals are pullbacked $\, _2F_1$ hypergeometric functions,
and in fact {\em modular forms}~\cite{DiagSelected}. These sets of diagonals of rational 
functions in {\em three} variables in~\cite{DiagSelected} 
were obtained by imposing the coefficients of the polynomial $P(x,y,z)$ appearing 
in the rational function $ \, 1/P(x,y,z) \, $ defining the diagonal to be 
$\, 0$ or $\, 1$\footnote[5]{Or $0$ or $\pm 1$ 
in the four variable case also examined in~\cite{DiagSelected}.}.

\vskip .1cm

While these constraints made room for exhaustivity, they were quite arbitrary, which raises 
the question of randomness of the sample : is the emergence of modular forms~\cite{perimeter}, 
with the constraints imposed in~\cite{DiagSelected}, an artefact of the sample? 

Our aim in this paper is to show that {\em modular forms} emerge for a much larger set  
of rational functions of three variables, than the one previously introduced 
in ~\cite{DiagSelected}, firstly because 
we obtain a whole family of rational functions whose diagonals give modular forms 
by adjoining  parameters, and 
secondly through considerations of symmetry.

\vskip .1cm

In particular, we will find that the {\em seven-parameter} rational function 
of three variables, with a numerator equal to one and a denominator 
being a polynomial of degree two at most, given by:
\begin{eqnarray}
\label{Ratfoncintro}
\hspace{-0.98in}&& \, \,  \quad 
R(x, \, y, \, z)  \, \, \, = \, \,  \quad 
 {{1} \over {
a \, \, \,+ \, b_1 \, x \,\,  + \, b_2 \, y \, \,  + \, b_3 \, z \,\,\, 
 + \, c_1 \, y\, z \,\,  + \, c_2 \, x \, z \, \,  + \, c_3 \, x\, y }},
\end{eqnarray}
can be expressed as a particular pullbacked 
$\, _2F_1$ hypergeometric function\footnote[1]{The selected  $\, _2F_1([1/12,5/12],[1],\, {\cal P})$ 
hypergeometric function is closely related
to modular forms~\cite{Youssef,Maier1}. This can be seen as a consequence of the identity 
with the Eisenstein series $\, E_4$ and $\, E_6$ and this very $\, _2F_1([1/12,5/12],[1],\, {\cal P})$ 
hypergeometric function (see Theorem 3 page 226 in~\cite{Stiller} 
and page 216 of~\cite{Shen}): 
$\,E_4(\tau) \, = \, \,  _2F_1([1/12,5/12],[1],\, 1728/j(\tau))^4$ 
(see also equation (88) in~\cite{Youssef} for $\, E_6$).} 
\begin{eqnarray}
\label{2F15HypformAintro}
\hspace{-0.7in}&&\quad \quad \quad \quad 
{{1} \over { P_2(x)^{1/4}}} \cdot \, 
 _2F_1\Bigl([{{1} \over {12}}, \, {{5} \over {12}}], \, [1],
 \, \, 1 \, - \, {{P_4(x)^2 } \over {P_2(x)^3}}\Bigr),
\end{eqnarray}
where $\, P_2(x)$ and $\, P_4(x)$ are two polynomials of degree 
two and four respectively. We then focus on subcases where 
the diagonals of the corresponding rational functions can be written 
as a pullbacked $\, _2F_1$ hypergeometric function with two 
rational function pullbacks that are algebraically related by 
{\em modular equations}\footnote[1]{Thus providing a nice illustration 
of the fact that the diagonal is a modular form~\cite{Maier1}.}. 

This seven-parameter family will then be generalized into an eight, nine and finally ten 
parameters family of rational functions that are reciprocal of a polynomial 
of three variables of degree at most three. We will finally show that each of the 
previous results yields an {\em infinite number of new exact pullbacked 
$\, _2F_1$ hypergeometric function results}, through symmetry 
considerations on monomial transformations and some 
function-dependent rescaling transformations.

\vskip .1cm 

\section{Diagonals of rational functions of three variables depending on  seven parameters }
\label{R}

\subsection{Recalls on diagonals of rational functions }
\label{recallsR}

Let us recall the definition of the diagonal of a rational function in $\, n$ variables
$\,{\cal R}(x_1, \ldots, x_n)\, = \,\, {\cal P}(x_1, \ldots, x_n)/{\cal Q}(x_1, \ldots, x_n)$, 
 where $ {\cal P}$ and $ {\cal Q}$ are polynomials of $\, x_1, \, \cdots, \, x_n\,$ 
with {\em integer coefficients} such that $\, {\cal Q}(0, \ldots, 0) \neq 0$. The 
diagonal of ${\cal R}$ is defined through its multi-Taylor expansion (for small $\, x_i$'s) 
\begin{eqnarray}
\label{defdiag}
\hspace{-0.90in}&&\quad \quad \, \, \,
{\cal R}\Bigl(x_1, \, x_2, \, \ldots, \, x_n \Bigr)
\, \, \,\, = \, \,\, \,\sum_{m_1 \, = \, 0}^{\infty}
 \, \cdots \, \sum_{m_n\, = \, 0}^{\infty} 
 \,R_{m_1,  \, \ldots, \, m_n}
\cdot  \, x_1^{m_1} \,\,  \cdots \,\, x_n^{m_n}, 
\end{eqnarray}
as the series in  {\em one variable} $\, x$:
\begin{eqnarray}
\label{defdiag2}
\hspace{-0.7in}&&\quad \quad 
Diag\Bigl({\cal R}\Bigl(x_1, \, x_2, \, \ldots, \, x_n \Bigr)\Bigr)
\, \, \, = \, \,  \quad \sum_{m \, = \, 0}^{\infty}
 \,R_{m, \, m, \, \ldots, \, m} \cdot \, x^{m}.
\end{eqnarray}
Diagonals of rational functions of two variables are algebraic functions~\cite{Fu,Denef}. Interesting 
cases of diagonals of rational functions thus require to consider rational functions  
of at least {\em three} variables. 

 \vskip .2cm 

\subsection{A seven parameters family  of rational functions of three variables}
\label{recallsRseven}

 \vskip .1cm 

We obtained the diagonal of the rational function in  {\em three} 
variables depending on {\em seven} parameters:
\begin{eqnarray}
\label{Ratfonc}
\hspace{-0.98in}&& \, \, \quad  
R(x, \, y, \, z)  \, \, \, = \, \,  \quad 
 {{1} \over {
a \, \, \,+ \, b_1 \, x \,\,  + \, b_2 \, y \, \,  + \, b_3 \, z \,\,\, 
 + \, c_1 \, y\, z \,\,  + \, c_2 \, x \, z \, \,  + \, c_3 \, x\, y }}. 
\end{eqnarray}

This result was obtained by:
\begin{itemize}
  \item{Running the {\em HolonomicFunctions}~\cite{Koutschan} package in mathematica 
for arbitrary parameters $ \, a,\, b_{1},\, \cdots,\,  c_1, \cdots\, $ and obtaining 
a large-sized second order linear differential operator $ \, L_2$.}
  \item{Running the maple command ``{\em hypergeometricsols}''~\cite{Hoeijprogram} 
for different sets of values of the parameters 
on the operator $ \, L_2$, and guessing\footnote[5]{The program ``{\em hypergeometricsols}''~\cite{Hoeijprogram} 
does not run for arbitrary parameters, hence our recourse to guessing.} the Gauss hypergeometric 
function $\, _2F_1$ with general parameters solution of $\, L_2$.}
\end{itemize}

\subsection{The diagonal of the seven parameters family  of rational functions: the general form}
\label{diagRseven}

We find the following experimental results: all these diagonals are expressed 
in terms of {\em only one pullbacked hypergeometric function}. This is worth noticing 
since, in general, when an order-two linear differential operator has 
pullbacked $\, _2F_1$  hypergeometric function solutions, the ``hypergeometricsols'' command gives the two
solutions as {\em sums of two $\, _2F_1$  hypergeometric functions}. Here, quite remarkably, the result
is ``encapsulated'' in just one pullbacked hypergeometric function.
Furthermore we find that  all these diagonals are expressed as pullbacked 
hypergeometric functions of the form
\begin{eqnarray}
\label{2F17Hyp}
\hspace{-0.7in}&&\quad \quad  \quad \quad 
 {{1} \over { P_4(x)^{1/6}}}
 \cdot \, 
_2F_1\Bigl([{{1} \over {12}}, \, {{7} \over {12}}], \, [1], 
\, {{1728 \cdot x^3 \cdot \, \, P_5(x)} \over {P_4(x)^2}}\Bigr),
\end{eqnarray}
where the two polynomials $\, P_4(x)$ and $\, P_5(x)$,
 in the  $\, \, 1728\,x^3\,  P_5(x)/P_4(x)^2$ pullback, are  polynomials 
of degree four and five in $\, x$ respectively. The pullback in (\ref{2F17Hyp}), given by 
$\, \, 1728\,x^3\,  P_5(x)/P_4(x)^2$, has the form $\,\,  1 \, - \tilde{Q}\, $ where  
$\, \, \tilde{Q}\, $  is given by the simpler expression
\begin{eqnarray}
\label{tildeQ}
\hspace{-0.7in}&&\quad \quad  \quad  \quad \quad \quad \quad  \quad  \quad  \quad 
\tilde{Q} \, \, \, = \, \, \, \, 
{{P_2(x)^3 } \over {P_4(x)^2}},
\end{eqnarray}
where $\, P_2(x)$ is a polynomial  of degree two in $\, x$. 
Recalling the identity 
\begin{eqnarray}
\label{2F1identitybis}
\hspace{-0.7in}&&\quad \,\,\,\,
_2F_1\Bigl([{{1} \over {12}}, \, {{7} \over {12}}], \, [1], \, x\Bigl)
\, \,\, \, = \, \,\,\,   (1\, -x)^{-1/12} \cdot \,
 _2F_1\Bigl([{{1} \over {12}}, \, {{5} \over {12}}], \, [1], \,{{-x} \over {1 \, -x}}\Bigr), 
\end{eqnarray}
the previous pullbacked hypergeometric function (\ref{2F17Hyp}) can be rewritten as
\begin{eqnarray}
\label{2F15Hyp}
\hspace{-0.7in}&&\quad \quad \, \,  \, \,  \quad \quad 
 {{1} \over { P_2(x)^{1/4}}} \cdot \, 
 _2F_1\Bigl([{{1} \over {12}}, \, {{5} \over {12}}], \, [1],
 \, -\, {{1728 \cdot x^3 \cdot \, P_5(x)} \over {P_2(x)^3}}\Bigr),
\end{eqnarray}
where  $\, P_5(x)$ is the same polynomial of degree five as the one
in the pullback in expression (\ref{2F17Hyp}). This new pullback 
also has the form $\, 1 \, -Q \,$ 
with $\, Q$ given by\footnote[2]{Note that $\, Q$, given 
by (\ref{Q512}), is the reciprocal of $\, \tilde{Q}$ given
 in (\ref{tildeQ}): $\, \,\, Q \, = \, \, 1/\tilde{Q}$.}:
\begin{eqnarray}
\label{Q512}
\hspace{-0.9in}&&  \,\,\,\,\, \, \, 
-\, {{1728 \cdot x^3 \cdot \, P_5(x)} \over {P_2(x)^3}}
 \, \,\, = \, \, \, 1 \, \, -Q 
\quad  \quad \quad \hbox{where:} \quad \quad  \quad \quad 
Q \, \, = \, \, \, 
{{P_4(x)^2 } \over {P_2(x)^3}}. 
\end{eqnarray}
Finding the exact result for arbitrary values of the seven parameters now
 boils down to a guessing problem.

\vskip .1cm

\subsection{Exact expression of the diagonal for arbitrary parameters $\, a$, $\, b_1$, ..., $\, c_1$, ...}
\label{exactexpress}

Now that the structure of the result is understood ``experimentally'' 
we obtain the result {\em for arbitrary parameters} 
$\, a$, $\, b_1$, $\, b_2$, $\, b_3$,  $\, c_1$, $\, c_2$, $\, c_3$.

Assuming that the
diagonal of the rational function (\ref{Ratfonc}) has
 the form explicited in the previous subsection
\begin{eqnarray}
\label{2F15HypformA}
\hspace{-0.7in}&&\quad \quad \quad \quad 
{{1} \over { P_2(x)^{1/4}}} \cdot \, 
 _2F_1\Bigl([{{1} \over {12}}, \, {{5} \over {12}}], \, [1],
 \, \, 1 \, - \, {{P_4(x)^2 } \over {P_2(x)^3}}\Bigr),
\end{eqnarray}
where $\, P_2(x)$ and $\, P_4(x)$ are two polynomials of degree 
two and four respectively:
\begin{eqnarray}
\label{P4P2}
\hspace{-0.7in}&&\quad \quad \quad \quad 
P_4(x) \, \, = \, \, \,
 A_4 \, x^4 \, + \,  A_3 \, x^3 \, + \,  A_2 \, x^2 \, + \, A_1 \, x \, + \, A_0, 
\\
\hspace{-0.7in}&&\quad \quad \quad \quad 
P_2(x) \, \, = \, \, \, B_2 \, x^2 \, + \, B_1 \, x \, + \, B_0,
\end{eqnarray}
one can write the order-two linear differential operator having 
this eight-parameter solution (\ref{2F15HypformA}), 
and identify this second order operator depending on eight arbitrary parameters, with 
the second order linear differential operator 
obtained using the {\em HolonomicFunctions}~\cite{Koutschan} program for arbitrary 
parameters. Using the results obtained for specific values of the parameters, 
one easily guesses that $\,\, A_0 \, = \, \, a^6  \, \, $ 
and $\, \,B_0 \, = \, \, a^4$. One finally gets:
\begin{eqnarray}
\label{P2}
\hspace{-0.96in}&& 
P_2(x) \, \, = \,  \, 
\nonumber \\
\hspace{-0.96in}&& 8  \cdot \, \Bigl(3\, \, a\, c_1\, c_2\, c_3 
\, \,  +2  \cdot \,  (b_1^2\, c_1^2 +\, b_2^2\, c_2^2 \,
 +\, b_3^2\, c_3^2 -\, b_1\, b_2\, c_1\, c_2 -\, b_1\, b_3\, c_1\, c_3
 -\, b_2\, b_3\, c_2\, c_3)\Bigr)  \cdot \, x^2 
\nonumber \\
\hspace{-0.96in}&& \quad \quad  \quad  \quad 
\,  -8 \cdot \, a \cdot \, 
\Bigl(a \cdot \, (b_1\, c_1 \, +\, b_2\, c_2 \, +\, b_3\, c_3) \, -3\, b_1\, b_2\, b_3\Bigr) \cdot \, x 
\,  \,\,\,  +a^4, 
\end{eqnarray}
and 
\begin{eqnarray}
\label{P4}
\hspace{-0.96in}&&\quad 
P_4(x) \, \,\, = \, \,\, \,
  216 \cdot \, c_1^2\, c_2^2\, c_3^2 \cdot \, x^4
\, \,\, \, \,
-16 \cdot \, \Bigl(9 \cdot \, a \, c_1\, c_2\, c_3  \cdot \, ( b_1\, c_1 +b_2\, c_2 \, + b_3\, c_3)
\nonumber \\
\hspace{-0.96in}&& \quad  \quad  \quad \,\,
 -6 \, \cdot \, ( b_1^2\, b_2\, c_1^2\, c_2  \,+ b_1\, b_2^2\, c_1\, c_2^2 \, +b_1^2\, b_3\, c_1^2\, c_3 \, 
 +b_1\, b_3^2\, c_1\, c_3^2 \,+ b_2^2\, b_3\, c_2^2\, c_3 \, +\, b_2\, b_3^2\, c_2\, c_3^2)
 \nonumber \\
\hspace{-0.96in}&& \quad  \quad  \quad  \quad  \quad  \,\,
+4 \cdot \, (b_1^3\, c_1^3   +\, b_2^3\, c_2^3 \, +\, b_3^3\, c_3^3) \, \, 
-3\, b_1\, b_2\, b_3\, c_1\, c_2\, c_3 
\Bigr) \cdot \, x^3
\nonumber \\
\hspace{-0.96in}&& \quad  \quad \quad 
+12 \cdot \, \Bigl( 3\, a^3\, c_1\, c_2\, c_3 \, 
+4 \cdot \, a^2 \cdot \, (b_1^2\, c_1^2 \, +b_2^2\, c_2^2 \, +b_3^2\, c_3^2) 
\nonumber \\
\hspace{-0.96in}&& \quad  \quad  \quad  \quad  \quad  \,\,
+2 \cdot \, a^2 \cdot \, (b_1\, b_2\, c_1\, c_2 \, +b_1\, b_3\, c_1\, c_3 \, +b_2\, b_3\, c_2\, c_3)
\nonumber \\
\hspace{-0.96in}&& \quad  \quad  \quad  \quad  \quad  \quad  \,\,
-12 \cdot \, a \cdot \, b_1\, b_2\, b_3 \cdot \, (b_1\, c_1+b_2\, c_2+b_3\, c_3) \, \, \, 
+18 \cdot \, b_1^2\, b_2^2\, b_3^2 \Bigr) \cdot \, x^2
\nonumber \\
\hspace{-0.96in}&& \quad  \quad \quad 
-12 \cdot \, a^3 \cdot \, \Bigl(a \cdot \, (b_1\, c_1\, +\, b_2\, c_2 \, +\, b_3\, c_3)
\, -3 \, b_1\, b_2\, b_3 \Bigr)  \cdot \, x \, \,  \,   \, \, +a^6.
\end{eqnarray}
The polynomial $\, P_5(x)$ in (\ref{Q512}), given by 
$\,P_5(x) \, = \, \,(P_4(x)^2\, -P_2(x)^3)/1728/x^3$, is a slightly larger polynomial
of the form 
\begin{eqnarray}
\label{P5}
\hspace{-0.96in}&& \quad  \quad \quad 
P_5(x)  \, = \, \, \,\,  27 \cdot \, c_1^4\, c_2^4\, c_3^4 \cdot \, x^5 
\,\, \,\, + \, \, \cdots  \, \,  \,\, +q_1 \cdot \, x   \,  \,+ \, q_0, 
\quad  \quad  \quad \quad \quad  \quad \hbox{where:} 
\nonumber \\ 
\hspace{-0.96in}&& \quad \quad \quad  
q_0   \, = \,\, \,
- \, b_1 \, b_2 \, b_3 \, a^3 \cdot  \, 
(a\, c_1 \, -b_2\, b_3) \cdot \, (a\, c_2 \, -b_1\, b_3) \cdot \, (a\, c_3 \, -b_1\, b_2).
\end{eqnarray}
The coefficient $\, q_1$ in $\, x$ reads for instance:
\begin{eqnarray}
\label{q4}
\hspace{-0.96in}&& 
q_1 \, \, = \, \, \, 
c_1\, c_2\, c_3\, (b_1\, b_2\, c_1\, c_2+b_1\, b_3\, c_1\, c_3+b_2\, b_3\, c_2\, c_3) \cdot \, a^5
\nonumber \\
\hspace{-0.96in}&&   \, 
-\Bigl(b_1^2\, b_2^2\, c_1^2\, c_2^2 +b_1^2\, b_3^2\, c_1^2\, c_3^2 +b_2^2\, b_3^2\, c_2^2\, c_3^2 \, 
-8\, b_1\, b_2\, b_3\, c_1\, c_2\, c_3 \cdot \, 
(b_1 \, c_1  +\, b_2 \, c_2 + b_3\, c_3) \Bigr) \cdot \, a^4
\nonumber \\
\hspace{-0.96in}&&  \, 
-b_1\, b_2\, b_3 \cdot \, \Bigl(57\, b_1\, b_2\, b_3\, c_1\, c_2\, c_3 \, \,
\nonumber \\
\hspace{-0.96in}&&  \,  
+ \, 8 \cdot  (b_1^2\, b_2\, c_1^2\, c_2+\, b_1^2\, b_3\, c_1^2\, c_3
+\, b_1\, b_2^2\, c_1\, c_2^2 \, +\, b_1\, b_3^2\, c_1\, c_3^2
+\, b_2^2\, b_3\, c_2^2\, c_3 +\, b_2\, b_3^2\, c_2\, c_3^2) \Bigr) \cdot \, a^3
\nonumber \\
\hspace{-0.96in}&&   \,\quad \quad \,  
+8 \, b_1^2\, b_2^2\, b_3^2  \cdot \,
  (b_1^2\, c_1^2+\, b_2^2\, c_2^2+\, b_3^2\, c_3^2) \, \cdot \, a^2
\nonumber \\
\hspace{-0.96in}&&   \,\quad \quad \, 
+ 46 \cdot \, b_1^2\, b_2^2\, b_3^2 
 \cdot \,
 (b_1\, b_2\, c_1\, c_2 +\, b_1\, b_3\, c_1\, c_3 +\, b_2\, b_3\, c_2\, c_3) \cdot \, a^2
\nonumber \\
\hspace{-0.96in}&& \, \quad \quad \quad 
-36 \cdot \, b_1^3\, b_2^3\, b_3^3 \cdot \,
 (b_1\, c_1+b_2\, c_2+b_3\, c_3) \cdot \, a 
 \, \, \, +27\, b_1^4\, b_2^4\, b_3^4. 
\end{eqnarray}
Having ``guessed'' the exact result, one can easily verify directly that this exact 
pullbacked hypergeometric result is truly the solution of the large second order linear 
differential operator obtained using the 
``{\em HolonomicFunctions}'' program~\cite{Koutschan}.

\vskip .1cm 

\subsection{Selected  subcases of these results}
\label{trivia}

When $\,\, P_2(x)^3 \, -\, P_3(x)^2 \, \, =$
$\, \,  \,-1728 \, x^3 \cdot \, P_5(x) \, \, = \, \, \, 0$,  
the pullback in (\ref{2F15HypformA}) (with (\ref{P2}), (\ref{P4}))
vanishes, and the previous exact result (\ref{2F15HypformA}),  
for the diagonal of the rational function 
(\ref{Ratfonc}), degenerates into a simple algebraic function 
(see (\ref{2F15Hyp}) and (\ref{2F15HypformA})): 
\begin{eqnarray}
\label{triv}
\hspace{-0.96in}&& \quad \quad \quad \quad
 \quad  \quad \quad \quad \quad  \quad \quad  \, 
{{1} \over { P_2(x)^{1/4}}} \, \, \, \,  = \, \, \, \,  \, {{1} \over { P_4(x)^{1/6}}}.  
\end{eqnarray}
The condition 
$\, P_2^3 \, -\, P_3^2 \, = \, \, -1728 \, x^3 \cdot \, P_5(x) \, = \, \, 0 \,\,  $ 
corresponds, for instance,  to $\,\,  c_3 \, = \, 0, \, \,b_1 \, = \, 0$, 
with the rational function 
\begin{eqnarray}
\label{Ratfonctriv1}
\hspace{-0.98in}&& \, \,  \quad \quad \quad \quad \quad \quad \quad \quad \quad 
 {{1} \over { a \, \, + b_2 \, y   +  b_3 \, z \,\, 
 +  c_1 \, y\, z  \,  + c_2 \, x \, z}}, 
\end{eqnarray}
or to  $\,\, c_3 \, = \, 0, \, \, c_1 = \, b_2 \, b_3/a, \, \, c_2 = \, b_1 \, b_3/a$, 
with the rational function:
\begin{eqnarray}
\label{Ratfonctriv2}
\hspace{-0.98in}&& \, \,  \quad \quad \quad  \quad 
 \, \,\,   {{1} \over {
a^2 \, \,+ \, a \, b_1 \, x \,\,  + \,a \,  b_2 \, y  + \,a \,  b_3 \, z \,\,\, 
 +  b_2 \, b_3 \, y\, z  + \,  b_1 \, b_3 \, x \, z }}. 
\end{eqnarray}
One easily verifies that the diagonals of the corresponding  rational functions
read respectively:
\begin{eqnarray}
\label{triv1}
\hspace{-0.96in}&& \quad \quad \quad \quad   \quad  \quad  \, 
{{1} \over { \sqrt{ a^2 \,  -4 \, b_2 \, c_2 \cdot \, x} }}, 
\quad  \quad   \quad   \quad  
{{\sqrt{a}} \over { \sqrt{ a^3 \,  +4 \, b_1 \,b_2 \, b_3 \cdot \, x} }}.
\end{eqnarray}

\vskip .1cm 

\subsection{Simple symmetries of this seven-parameter result}
\label{Simplesym}

The different pullbacks 
\begin{eqnarray}
\label{miscell}
\hspace{-0.96in}&& \quad \quad \, \,  \, \, 
{\cal P}_1 \, = \, \, -\, {{1728 \cdot x^3 \cdot \, P_5(x)} \over {P_2(x)^3}},
 \quad  \, \, 
{{1728 \cdot x^3 \cdot \, \, P_5(x)} \over {P_4(x)^2}}, \quad \quad  \, \, 
 1 \, - \, {{P_4(x)^2 } \over {P_2(x)^3}},  
\end{eqnarray}
must be compatible with some obvious symmetries. They verify the relations
\begin{eqnarray}
\label{symmP1first}
\hspace{-0.7in}&&\quad \quad \, \quad \,
{\cal P}_1(\lambda \cdot \, a, \, \, \lambda \cdot \, b_1, \, \,  \lambda \cdot \,b_2,\, 
 \,  \lambda \cdot \,b_3, \, \, \lambda \cdot \,c_1, \,\, 
  \lambda \cdot \,c_2, \,\,   \lambda \cdot \,c_3, \,\,  \, x)
\nonumber \\
\hspace{-0.7in}&&\quad \quad \quad \quad \quad \quad 
 \, \, \, \, \, = \, \, \, \,
{\cal P}_1(a, \,\, b_1, \,\,  b_2, \,\,  b_3, \,\, c_1, \, \, c_2, \,\,  c_3, \, \,\, \, x).
\end{eqnarray}
and 
\begin{eqnarray}
\label{symmP2first}
\hspace{-0.98in}&& \, \,\, \,\, \,\, \,\,
{\cal P}_1\Bigl(a, \, \, \,   \,
\lambda_1 \cdot \, b_1, \,  \,  \lambda_2 \cdot \,b_2, \,  \,  \lambda_3 \cdot \,b_3,
\,  \,\, \, \lambda_2 \, \lambda_3  \cdot \, c_1, \,\,    \lambda_1 \, \lambda_3  \cdot \,c_2, 
\, \,   \lambda_1 \, \lambda_2  \cdot \,c_3
, \,\,   \, \, {{x } \over { \lambda_1 \,  \lambda_2 \,  \lambda_3 }}\Bigr)
\nonumber \\
\hspace{-0.98in}&&\quad \quad \quad \quad \quad \quad 
 \, \, = \, \, \, \,\,
{\cal P}_1(a, \, \, b_1, \, \,  b_2, \,\,   b_3, \, \,\,  c_1, \,  c_2, \,  c_3, \,  \,\, x),
\end{eqnarray}
where $\, \lambda$, $\, \lambda_1$, $\, \lambda_2$ and   $\, \lambda_3  \, \, $ 
are arbitrary complex numbers.
A demonstration of these symmetry-invariance relations (\ref{symmP1first}) 
and (\ref{symmP2first}) is sketched in \ref{Simplesymapp}. 

\vskip .1cm

\subsection{A symmetric subcase $\, \tau \, \rightarrow \, 3 \, \tau$: $\, \, _2F_1([1/3,2/3],[1],{\cal P})$}
\label{threesymm}

\subsubsection{A few recalls on Maier's paper \\}
\label{recallMaier}

We know from Maier~\cite{Maier1} that the {\em modular equation} 
associated with\footnote[1]{$\tau$ denotes the ratio of the two periods 
of the elliptic functions that naturally emerge in the problem~\cite{Youssef}.}
 $\,\, \tau \, \rightarrow \, 3 \, \tau \,$ 
corresponds to the elimination of the $\, z$ variable between 
the two rational pullbacks:
\begin{eqnarray}
\label{j3} 
 \hspace{-0.95in}&& \quad \,  \,   \,  \,   
{\cal P}_1(z) \, \, = \, \, \,
{{ 12^3 \cdot \, z^3} \over { (z \, +27)\cdot \, (z \, +243)^3}}, 
 \quad \quad        \,   \,   
{\cal P}_2(z) \, \, = \, \, \, 
 {{ 12^3 \cdot \, z} \over { (z \, +27)\cdot \, (z \, +3)^3}}.
\end{eqnarray}
Following Maier~\cite{Maier1} one can 
also write the identities\footnote[5]{One has hypergeometric identities on 
$\, _2F_1([1/3,2/3],[1],{\cal P})$, however 
they are not associated with the involutive transformation 
$\, z \, \rightarrow \, 729/z$ as one could expect from 
 the fact that the two Hauptmoduls in  (\ref{identity1z2}) 
are exchanged by this involution: see \ref{comment}.}:
\begin{eqnarray}
\label{identity1z}
\hspace{-0.95in}&&    \quad  \quad  \quad   \, \, 
 \Bigl( 9 \cdot \,\Bigl( {{ z\, +27 } \over { z\, +243 }}  \Bigr) \Bigr)^{1/4} \cdot \,
  _2F_1\Bigl([{{1} \over {12}}, \, {{5} \over {12}}],
 \, {{1728 \, z^3} \over { (z \, +27) \cdot \, (  z\, +243)^3 }}\Bigr)
\nonumber \\ 
\label{identity1z2}
\hspace{-0.95in}&&    \,  \,  \,  \, \quad  \quad  \quad  \quad 
\,\,\,\,  = \, \, \, \,
 \Bigl( {{1} \over {9}} \cdot \,\Bigl( {{ z\, +27 } \over { z\, +3 }}  \Bigr) \Bigr)^{1/4} \cdot \,
  _2F_1\Bigl([{{1} \over {12}}, \, {{5} \over {12}}],
 \, {{1728 \, z} \over { (z \, +27) \cdot \, (  z\, +3)^3 }}\Bigr) 
\\ 
\hspace{-0.95in}&&   \,  \,  \,   \, \quad  \quad  \quad  \quad 
\,\,\,\,  = \, \, \, \,
 _2F_1\Bigl([{{1} \over {3}}, \, {{2} \over {3}}], \, [1], \,{{z} \over {z\, +27}} \Bigr).
\end{eqnarray}
Having a hypergeometric function identity (\ref{identity1z2})   
with {\em two} rational pullbacks (\ref{j3}) related by 
a {\em modular equation} provides a good heuristic way to see
that we have a {\em modular form}~\cite{Youssef,Maier1}\footnote[2]{Something 
that is obvious here since we are dealing with a $_2F_1([1/12,5/12],[1],x) \,$ 
hyperegeometric function which is known to be related  modular 
functions~\cite{Youssef,Maier1} due to its relation with the Eisenstein series 
$\, E_4$, but is less clear for other hypergeometric functions.}.

\subsubsection{The symmetric subcase \\}
\label{threesymmsub}

Let us now consider the symmetric subcase
$\, b_1 = \, b_2  = \,  b_3  = \, b \, $  and 
 $\, c_1= \, c_2 = \,  c_3  = \, c$.
If we take that limit in our previous general expression (\ref{2F15HypformA}),  
we obtain the solution of the 
order-two linear differential operator annihilating 
the diagonal\footnote[9]{Called the ``telescoper''~\cite{Telescopers,Telescopers2}.} in the form 
\begin{eqnarray}
\label{2F15Hypform}
\hspace{-0.7in}&&\quad  \quad \,  
{{1} \over { P_2(x)^{1/4}}} \cdot \, 
 _2F_1\Bigl([{{1} \over {12}}, \, {{5} \over {12}}], \, [1],
 \, \, 1 \, - \, {{P_4(x)^2 } \over {P_2(x)^3}}\Bigr)
\nonumber \\
\hspace{-0.7in}&&\quad \quad  \quad \quad \quad 
\, \, \, = \, \,\, {{1} \over { P_2(x)^{1/4}}} \cdot \, 
 _2F_1\Bigl([{{1} \over {12}}, \, {{5} \over {12}}], \, [1],
 \, -\, {{1728 \cdot \, x^3 \cdot \,  P_5(x)} \over {P_2(x)^3}}\Bigr),
\end{eqnarray}
with
\begin{eqnarray}
\label{P2P4sym}
\hspace{-0.9in}&& \quad  \quad  \quad  \quad 
P_2(x) \,\,  = \, \, \, \, \, 
a \cdot \, (24 \cdot \, c^3 \cdot \, x^2 \, -24 
\cdot \, b \cdot \,(a\, c \, -b^2) \cdot \,x \, \, +a^3),
\\
\label{P2P4symP4}
\hspace{-0.9in}&& \quad  \quad  \quad  \quad 
P_4(x) \,\,  = \, \, \, \, 
216 \cdot \, c^6 \cdot \, x^4 \, \, 
-432 \cdot \,b \,c^3 \cdot \,(a\,c\,\,  -b^2)\cdot \, x^3 \, 
\nonumber  \\
\hspace{-0.9in}&&\quad \quad \quad  \quad \quad  \quad  \quad \, \,  \,\,  \,
 +36\cdot \, (a^3\, c^3 \, +6 \cdot \, a^2\, b^2 \, c^2
 \, -12 \cdot \,a \, b^4 \, c \, +6 \cdot \, b^6) \cdot \, x^2
\nonumber  \\
\hspace{-0.9in}&&\quad \quad \quad  \quad \quad \quad  \quad  \quad  \quad  \, \,\, \, \, 
\, -36 \cdot \,a^3 \, b \cdot \,(a\, c \, \, -b^2) \cdot \, x \, \, \, +a^6.
\end{eqnarray}
and: 
\begin{eqnarray}
\label{P5sym}
\hspace{-0.9in}&& \, \, \,  \, 
P_5(x) \, \,  = \, \, \,\, 
 (27\, c^3\, x^2 \, -27 \, b \cdot \, (a\, c \, -b^2) \cdot \, x \, \, +a^3)
 \cdot \, (c^3\, x \, -b \cdot \, (a\, c \, -b^2))^3. 
\end{eqnarray}
In this symmetric case, one can write the pullback in (\ref{2F15Hypform}) as follows:  
\begin{eqnarray}
\label{Hauptmodul}
\hspace{-0.9in}&& \, \quad  \quad  \quad  \quad   \quad   \quad 
-\, {{1728 \cdot \, x^3 \cdot \,  P_5(x)} \over {P_2(x)^3}} 
\,  \, \, = \, \, \, \, 
{{ 12^3 \cdot \, z^3} \over { (z \, +27)\cdot \, (z \, +243)^3}}, 
\end{eqnarray}
where $\, z$ reads:
\begin{eqnarray}
\label{Hauptmodulz}
\hspace{-0.9in}&& \, \quad  \quad  \quad  \quad  \quad  \quad \, 
z \, \, = \, \, \, 
 - \,  {{ 9^3 \cdot \, x  \cdot \, (c^3 \cdot \, x \, \, -b \cdot \, (a\,c \, -b^2)) 
} \over { 
27 \cdot \, c^3 \cdot \, x^2 \, -27 \cdot \,b \cdot \,(a\,c \, -b^2)\cdot \, x \, \, +a^3}}. 
\end{eqnarray}

Injecting the expression (\ref{Hauptmodulz}) for $\, z$ in $\, {\cal P}_2(z)$ 
given by (\ref{j3}), one gets another pullback
\begin{eqnarray}
\label{HauptmodulP2} 
\hspace{-0.95in}&& \quad \quad \quad \quad \quad \quad \quad \quad  \quad  \, 
{\cal P}_2(z) \, \, = \, \, \,  \, 
-\, 1728   \cdot \, x \cdot  \, {{\tilde{P}_5 } \over {\tilde{P}_2(x)^3}},  
\end{eqnarray}
with
\begin{eqnarray}
\label{P5sym2}
\hspace{-0.98in}&& \,  \,  \, \, \,  \,    \,  \, 
\tilde{P}_5(x) \, \,  = \, \, \, \, 
 (27\, c^3\, x^2 \, -27 \, b \cdot \, (a\, c \, -b^2) \cdot \, x \, \, +a^3)^3
 \cdot \, (c^3\, x \, -b \cdot \, (a\, c \, -b^2)). 
\end{eqnarray}
and:
\begin{eqnarray}
\label{P2xz}
\hspace{-0.9in}&& \quad  \quad \quad  \, 
\tilde{P}_2(x) \, = \, \, \, \, 
a \cdot \, 
(-216 \cdot \, c^3 \cdot \, x^2 \, 
+216 \cdot \, b \cdot \,(a\, c \, -b^2) \cdot \,x \, \, +a^3),
\end{eqnarray}
In this case the diagonal of the rational function can be written as a 
single hypergeometric function with two different pullbacks
\begin{eqnarray}
\label{Many}
\hspace{-0.7in}&& \quad   \quad  \quad
{{1} \over { P_2(x)^{1/4}}} \cdot \, 
 _2F_1\Bigl([{{1} \over {12}}, \, {{5} \over {12}}], \, [1],
 \, -\, {{1728 \cdot x^3 \cdot \, P_5(x)} \over {P_2(x)^3}}\Bigr)
\nonumber \\
\hspace{-0.7in}&& \quad  \quad \quad    \quad \quad   \, \, 
\, = \, \,\,\,
{{1} \over { \tilde{P}_2(x)^{1/4}}} \cdot \, 
 _2F_1\Bigl([{{1} \over {12}}, \, {{5} \over {12}}], \, [1], 
\, -\, {{1728 \cdot \, x \cdot \, \tilde{P}_5(x)} \over {\tilde{P}_2(x)^3}}\Bigr), 
\end{eqnarray}
with the relation between the two 
pullbacks given by the {\em modular equation} associated with~\cite{Youssef,Maier1} 
$\, \tau \, \rightarrow \, \, 3 \, \tau$:
\begin{eqnarray}
\label{modcurve3}
\hspace{-0.95in}&& \quad \,    \,   \,   \,  \,  \,   
 2^{27} \cdot \, 5^9 \cdot \, {Y}^{3} {Z}^{3}\cdot \, (Y +Z)
\, \, \,    + 2^{18} \cdot \, 5^6  \cdot \,{Y}^{2} {Z}^{2} \cdot \, 
(27\,{Y}^{2} -45946 \,Y Z +27\,{Z}^{2}) \, 
\nonumber \\ 
\hspace{-0.95in}&& \quad \quad  \quad   \,    \,  \,   
+ 2^{9} \cdot \, 5^3 \cdot \, 3^5 \cdot \, \,
Y Z \cdot \, (Y +Z)  \cdot \, ({Y}^{2} +241433\,Y Z +{Z}^{2}) 
\nonumber \\ 
\hspace{-0.95in}&& \quad \quad  \quad   \,     \,  \,   
 +729 \cdot  \,({Y}^{4}\, + \,{Z}^{4}) \,\, \, 
 -779997924 \cdot \, (Y {Z}^{3} \, +{Y}^{3} Z) \,\, 
 + 31949606 \cdot 3^{10} \cdot \, {Y}^{2} {Z}^{2} 
\nonumber \\ 
\hspace{-0.95in}&& \quad \quad  \quad   \,   \,  \,   
+2^9 \cdot \, 3^{11} \cdot \, 31 \cdot \,Y \, Z \cdot \, (Y +Z)
\, \, \, - 2^{12} \cdot \, 3^{12} \cdot \,Y Z\,\,\,\,\, = \,\,\,\,\,0. 
\nonumber   
\end{eqnarray}

\vskip .1cm

\subsubsection{Alternative expression for the symmetric subcase \\}
\label{alternative}

Alternatively, we can obtain the exact expression of the diagonal using 
directly the ``{\em HolonomicFunctions}'' program~\cite{Koutschan}  
for arbitrary parameters $\, a$, $\, b$
and $\, c$ to get an order-two linear differential operator annihilating that diagonal. 
Then, using ``{\em hypergeometricsols}''\footnote[2]{We use M. van Hoeij 
``hypergeometricsols'' program~\cite{Hoeijprogram} for many values 
of $\, a$, $  \, b$ and $\, c$, and then perform some guessing.} 
we obtain the solution of this  second order linear differential 
operator in the form 
\begin{eqnarray}
\label{canalso2}
\hspace{-0.95in}&&  \quad   \quad   \quad    \quad    \quad    \quad  
 {{1} \over {a}} \cdot \,
  _2F_1\Bigl([{{1} \over {3}}, \, {{2} \over {3}}], \, [1], \,\, 
 -\, {{27} \over {a^3}} \cdot \, x \cdot \, (c^3\, x \, -b \cdot \, (a\, c \, -b^2))\Bigr), 
\end{eqnarray}
which looks, at first sight, different from (\ref{2F15Hypform})
with (\ref{P2P4sym}) and (\ref{P2P4symP4}).
This last expression (\ref{canalso2}) is 
 compatible with the form (\ref{2F15Hypform})
as a consequence of the identity:
\begin{eqnarray}
\label{identity0}
\hspace{-0.95in}&&    \,\, \,   \,  \, 
\Big( {{9 \, -8 \, x} \over {9}} \Bigr)^{1/4} \cdot \, 
 _2F_1\Bigl([{{1} \over {3}}, \, {{2} \over {3}}], \, [1], \,x \Bigr) 
\,\,\, = \, \, \, \,
  _2F_1\Bigl([{{1} \over {12}}, \, {{5} \over {12}}],
 \, {{64 \, x^3 \cdot \, (1\, -x)} \over { (9 \, -8 \, x)^3 }}\Bigr).
\end{eqnarray}
The reduction of the (generic) $\, _2F_1([1/12, 5/12],[1],{\cal P})$  
hypergeometric function to a $\, _2F_1([1/3,2/3],[1],{\cal P})$ form
corresponds to a selected $\, \tau \, \rightarrow  \, 3 \, \tau$ 
modular equation situation (\ref{j3}) well described in~\cite{Maier1}.

These results can also be expressed
in terms of $\, _2F_1([1/3,1/3],[1], \, {\cal P})$  
pullbacked hypergeometric functions~\cite{Maier1} 
using the identities 
\begin{eqnarray}
\label{identity13}
\hspace{-0.96in}&& \,\,   \quad 
_2F_1\Bigl([{{1} \over {3}}, \, {{1} \over {3}} ], \, [1], \, x) 
\, \, \, = \, \,  \,\, 
 (1\, -x)^{-1/3} \cdot \, _2F_1\Bigl([{{1} \over {3}}, \, {{2} \over {3}} ], \, [1],
 \, \, - \, {{x} \over {1\, -x}})
\\
\hspace{-0.95in}&&    \quad   \quad  \quad 
 = \, \, \, 
  \Bigl( (1\, -9\,x)^3 \cdot \, (1 \, -\,x) \Bigr)^{-1/12} \cdot \, 
 _2F_1\Bigl([{{1} \over {12}}, \, {{5} \over {12}} ], \, [1], \, \, 
   -\, {{64 \,\, x } \over {(1 \, -9\, x)^3 \cdot \, (1\,-\,x) }}  \Bigr), 
\nonumber 
\end{eqnarray}
or: 
\begin{eqnarray}
\label{identity13}
\hspace{-0.96in}&& \,\,  \,  \quad 
_2F_1\Bigl([{{1} \over {3}}, \, {{1} \over {3}} ], \, [1], \,
 - \, {{x} \over {27}}) \, \, = \, \,  \,
 \Bigl(1\, +{{x} \over {27}} \Bigr)^{-1/3} \cdot
 \, _2F_1\Bigl([{{1} \over {3}}, \, {{2} \over {3}} ], \, [1],
 \, \,  \, {{x} \over {x \, +27}}\Bigr)
\\
\hspace{-0.95in}&&    \quad  \quad   \quad   \quad 
\, \,\, \,  = \, \, \, 
  \Bigl( {{(x+3)^3 \cdot \, (x+27) } \over {729 }} \Bigr)^{-1/12} \cdot \, 
 _2F_1\Bigl([{{1} \over {12}}, \, {{5} \over {12}} ], \, [1], 
\, {{1728 \, x } \over {(x+3)^3 \cdot \, (x+27) }}  \Bigr). 
\nonumber 
\end{eqnarray}

\vskip .1cm

\subsection{A non-symmetric subcase $\, \tau \, \rightarrow \, 4 \, \tau$: $\, \, _2F_1([1/2,1/2],[1],{\cal P})$.}
\label{threeasymm}

Let us consider the non-symmetric subcase 
$\, b_1 = \, b_2  = \,  b_3  = \, b \, $  and 
 $\, c_1 = \,  c_2  = \, 0$,  $\, c_3 = \,  b^2/a$.
The pullback in (\ref{2F15Hypform}) reads:
\begin{eqnarray}
\label{4tau}
\hspace{-0.95in}&&  \quad  \,      \quad   \quad  \, 
{\cal P}_1   \, \, = \, \, \, \,
-\, {{1728 \cdot x^3 \cdot \, P_5(x)} \over {P_2(x)^3}}
\, \, \,\, = \, \, \, \,\, 
{{ 1728 \cdot \, a^3 \, b^{12} \cdot \, x^4 \cdot \, (16\, b^3\, x\, + \, a^3)
} \over { (16\, b^6 \, x^2 \, + \, 16\, a^3\, b^3\, x \, + a^6)^3}}.
\end{eqnarray}
This pullback can be seen as the first of the two Hauptmoduls 
\begin{eqnarray}
\label{4tauHaupt}
\hspace{-0.95in}&& \quad   \quad  \,   \, \, \, \,  
{\cal P}_1   \, \, = \, \, \, \, \,
 {{1728 \cdot z^4 \cdot \, (z\, +16) } \over { (z^2\, + \, 256 \, z \, + 4096)^3}}, 
\quad \quad \, \, \,  
{\cal P}_2   \, \, = \, \, \, \, \, 
{{1728 \cdot z \cdot \, (z\, +16) } \over { (z^2\, + \, 16 \, z \, + 16)^3}},
\end{eqnarray}
provided $\, z$ is given by\footnote[2]{These two expressions are related 
by the involution $\,\, z \,\, \leftrightarrow   \, \,   -\, 16 \, z/(z \, +16)$. }:
\begin{eqnarray}
\label{4tauZ}
\hspace{-0.95in}&& 
\quad \quad \quad \quad \quad \quad 
 z \, \, = \, \, \, {{256\, b^3 \, x } \over {a^3}}
 \quad  \quad  \, \, 
\hbox{ or:}  \, \, \, \,  \quad \quad  \,  \, \,  
z \, \, = \, \, \, {{ -256 \, b^3 \cdot \, x} \over {a^3 \, + \, 16 \, b^3 \, x}}.
\end{eqnarray}
These exact expressions (\ref{4tauZ}) of $\, z$ in terms of $\,x$ give 
exact rational expressions of the second Hauptmodul $\, {\cal P}_2  \, $ 
in terms of $\, x$:
\begin{eqnarray}
\label{4tausecond}
\hspace{-0.95in}&&  \quad  \quad    \quad  \quad    \quad  
 {\cal P}_2^{(1)}  \, \, = \, \, \, 
{{ 1728 \cdot  \,  a^{12}\, b^3 \cdot \,  x \cdot \, (a^3 \, +16 \, b^3\, x)^4 } \over {
( 4096 \, b^6 \, x^2\, +256 \, a^3 \, b^3\,  x  \, + a^6 )^3 }}
 \quad   \quad   \quad   \quad \quad     
\hbox{ or:} 
\\
\label{4tausecondbis}
 \hspace{-0.95in}&& \quad   \quad  \quad  \quad   \quad   
   {\cal P}_2^{(2)}  \, \, = \, \, \, 
{{ -1728 \cdot  \,  a^3\, b^3 \cdot  \,  x \cdot \, (a^3 \, +16 \, b^3\, x)^4 } \over {
(256 \, b^6 \, x^2\, - \, 224 \, a^3 \, b^3 \, x  \, + a^6 )^3 }}. 
\end{eqnarray}
These two pullbacks (\ref{4tau}),  (\ref{4tausecond})  and (\ref{4tausecondbis})
(or $\,{\cal P}_1$ and $\,{\cal P}_2$ in (\ref{4tauHaupt}))
are related by a  {\em modular equation}
 corresponding\footnote[1]{See page 20 in~\cite{Youssef}.} to 
$\, \tau \, \rightarrow \, \, 4 \, \tau$.

This subcase thus corresponds to the diagonal of the rational function 
being expressed in terms of a {\em modular form} associated 
to an identity on a hypergeometric function:
\begin{eqnarray}
\label{identi} 
\hspace{-0.95in}&&    \quad \quad  \quad  \,  \,   \,       \, 
(16\, b^6 \, x^2 \, + \, 16\, a^3\, b^3\, x \, + a^6)^{-1/4}
 \cdot \, _2F_1\Bigl([{{1} \over {12}}, \, {{5} \over {12}}], \, [1], \, {\cal P}_1 \Bigr) 
\nonumber \\
 \hspace{-0.95in}&&  \quad \quad \quad \quad \quad  \quad   \, 
\, \, = \, \, \, \, 
( 4096 \, b^6 \, x^2\, +256 \, a^3 \, b^3\,  x  \, + a^6 )^{-1/4}
 \cdot \, _2F_1\Bigl([{{1} \over {12}}, \, {{5} \over {12}}], \, [1], \, {\cal P}_2^{(1)} \Bigr)
\nonumber \\
 \hspace{-0.95in}&&  \quad \quad \quad \quad \quad \quad    \, 
\, \, = \, \, \, \, 
(256 \, b^6 \, x^2\, - \, 224 \, a^3 \, x  \, + a^6)^{-1/4}
 \cdot \, _2F_1\Bigl([{{1} \over {12}}, \, {{5} \over {12}}], \, [1], \, {\cal P}_2^{(2)} \Bigr)
\nonumber \\
 \hspace{-0.95in}&&  \quad \quad \quad \quad \quad  \quad  
 \, \,\, = \, \, \, \, 
_2F_1\Bigl([ {{1} \over {2}}, \, {{1} \over {2}}], \, [1],
 \,  \, - \, {{16\cdot \, b^3} \over {a^3}} \cdot \, x \Bigr). 
\end{eqnarray}
The last equality is a consequence of the identity: 
\begin{eqnarray}
\label{identity1}
\hspace{-0.95in}&&    \quad   \,  \,  \,
_2F_1\Bigl([ {{1} \over {2}}, \, {{1} \over {2}}],
 \, [1], \, - \, {{x} \over {16}} \Bigr)
 \,\, \, 
\\ 
\hspace{-0.95in}&&  \quad \quad \quad   \,   \, \, \,  \,  \, = \, \, \,  \,  \,
 2 \cdot \, (x^2 \, +16\,x\, +16)^{-1/4}  \cdot \, 
_2F_1\Bigl([{{1} \over {12}}, \, {{5} \over {12}}], \, [1], 
\,  {{ 1728 \cdot \, x  \cdot \, (x+16) } \over {(x^2 \, +16\,x\, +16)^3} } \Bigr).
 \nonumber
\end{eqnarray}

Similarly, the elimination of $\, x$ between the pullback $\, X \, = \, {\cal P}_1$ 
(given by (\ref{4tau})) and  $\, Y \, = \, {\cal P}_2^{(1)}$
gives the {\em same modular equation} (representing
 $\, \tau \, \rightarrow \, \, 4 \, \tau$) 
than the elimination of $\, x$ between the pullback $\, X \, = \, {\cal P}_1$ 
(given by (\ref{4tau})) and  $\, Y \, = \, {\cal P}_2^{(2)}$, namely:
\begin{eqnarray}
\label{mod4}
\hspace{-0.95in}&& 
\,  825^9 \cdot \, X^6\, Y^6\,  \, \,   \, 
 -  389 \cdot \,  11^6  \cdot \,5^{16} 
\cdot \, 3^{10} \cdot \, 2^6  \cdot \, X^5 \, Y^5 \cdot \, (X+Y)
\nonumber \\
 \hspace{-0.95in}&&   \,  \,  \, 
 + 11^3  \cdot \,5^{12} \cdot \, 3^7 \cdot \,2^{4} \, \cdot  \, X^4\, Y^4\,  \cdot \, 
\Bigl(26148290096 \cdot \, (X^2+Y^2)\, -15599685235 \cdot \, X\, Y\Bigr)
\nonumber \\
 \hspace{-0.95in}&&  \quad   \, 
-105955481959 \cdot \, 5^{10} \cdot \, 3^7 \cdot \,2^{15} 
\cdot \, X^3 \, Y^3\, \cdot \,  (X+Y)\,  \cdot \, 
 (X^2+Y^2)
\nonumber \\
 \hspace{-0.95in}&&   \quad  \, 
+503027637092599 \cdot \, 5^{10} \cdot \, 3^7 \cdot \,2^{6}
 \cdot \, X^4 \, Y^4 \, \cdot \,  (X+Y)\, 
\nonumber \\
 \hspace{-0.95in}&&  \, 
+ 5^6 \cdot \, 3^4 \cdot \,2^{16}  \cdot \, X^2\, Y^2\,  \cdot \, 
\Bigl(1634268131  \cdot \, (X^4+Y^4) \, \,+1788502080642816 \cdot \, X^2\, Y^2 
\nonumber \\
 \hspace{-0.95in}&&   \,  \quad  \quad  \quad \quad \quad \quad \quad  \quad 
 +848096080668355 \cdot \, (X^3\, Y \,+ X\, Y^3) \Bigr)
\nonumber 
\end{eqnarray}
\begin{eqnarray}
 \hspace{-0.95in}&&  \, 
- \,  5^4 \cdot \, 3^4 \cdot \,2^{22} \cdot \, \, X \, Y \cdot \, (X+Y) \cdot \, 
\Bigl(389  \cdot \, (X^4\, +Y^4) \, \, \, +41863592956503 \cdot \, X^2\, Y^2
\nonumber \\
 \hspace{-0.95in}&&   \, \quad  \quad  \quad \quad \quad  \quad  \quad  \quad \quad 
 \,  -54605727143 \cdot \, (X^3\, Y\, +X\, Y^3) \Bigr)
\nonumber \\
 \hspace{-0.95in}&&   \, 
+ 2^{24}  \cdot \, 
 \Bigl(X^6 +Y^6 \,  \, \,  +561444609 \cdot \, (X^5\, Y \, +X\, Y^5)
\nonumber \\
 \hspace{-0.95in}&&   \, \quad \quad \quad \quad \, \,
 +1425220456750080 \cdot \, (X^4\, Y^2\, +X^2\, Y^4) \,\, \,
+2729942049541120 \cdot \, X^3\, Y^3\Bigr)
\nonumber \\
 \hspace{-0.95in}&&   \, 
-\, 5 \cdot \, 3^7 \cdot \,2^{34} \cdot \,  \, X\, Y\,  
\cdot \, (X+Y)\, \cdot\, (391\, X^2 \, -12495392 \, X\, Y \,  +391\, Y^2)
\\
 \hspace{-0.95in}&&  
+ \, 31 \cdot \, 3^7 \cdot \,2^{40} \cdot \,  \, X\, Y \cdot \, (X +2\, Y)\, \cdot \,(2\, X+Y)
\, \, \, \, \, 
- 3^9 \cdot \,2^{42} \cdot \,  \, X \, Y \cdot \, (X+Y)
\, \,\, \, = \, \,\,  \,\, 0.
\nonumber
\end{eqnarray}
The elimination of  $\, x$ between the pullback $\, X \, = \, {\cal P}_2^{(1)}$ 
(given by (\ref{4tau})) and  the pullback $\, Y \, = \, {\cal P}_2^{(2)}$ also 
gives\footnote[2]{This result can be also seen in the $\, z$ variable (see 
(\ref{4tauZ})): see the details in \ref{IntheZ}.} 
the {\em same modular equation} (\ref{mod4}).

\vskip .1cm
\vskip .1cm

\subsection{$ _2F_1([1/4,3/4],[1],{\cal P})$ subcases: walks in the quarter plane}
\label{1434}

\vskip .1cm

The diagonal of the rational function
\begin{eqnarray}
\label{diag3quart0}
\hspace{-0.98in}&& \,\, \,\,\,\,\,\,
{{2} \over { 2 \,\,   + \, (x+y+z) \, +\,x\, z +1/2 \cdot x\, y }} 
\, \,\, \,= \, \, \,\,\,
 {{4} \over { 4 \, \,  + 2 \cdot \, (x+y+z)\,  +\, 2 \, x\, z +\, x\, y }},
\end{eqnarray}
is given by the pullbacked hypergeometric function:
\begin{eqnarray}
\label{ident3quart0a}
\hspace{-0.96in}&& \,\, \quad \quad \quad \quad \, \,
\Bigl(1 \, +{{3} \over {4}} \cdot \, x^2\Bigr)^{-1/4} \cdot \,
 _2F_1\Bigl([{{1} \over {12}}, \, {{5} \over {12}}],\, [1],
 \, {{27\, x^4\cdot \, (x^2+1) } \over { (3\,x^2 \,+4)^3}} \Bigr)
 \nonumber \\
\label{ident3quart0}
\hspace{-0.96in}&& \,\, \quad \quad \quad\quad \quad \quad \quad\quad 
\, \,  \,\,  \,\, = \, \,  \, \, 
 _2F_1\Bigl([{{1} \over {4}}, \, {{3} \over {4}}],\, [1], \, -x^2),
\end{eqnarray}
which is reminiscent of the hypergeometric series number 5 and 15 
in Figure 10 of Bostan's
 HDR~\cite{HDRBostan}. Such 
 pullbacked hypergeometric function (\ref{ident3quart0})
corresponds to the rook walk problems~\cite{Rook,76,77}.

Thus the diagonal of the rational function corresponding to the simple rescaling
$(x, \, y, \, z) \,\, \longrightarrow $
$\, \,(\pm \sqrt{-1} \, x, \, \pm \sqrt{-1} \,y, \, \pm \sqrt{-1} \,z)$ of (\ref{diag3quart0}) 
namely 
\begin{eqnarray}
\label{diag3quart}
\hspace{-0.96in}&& \,\, \quad \quad \quad  \quad \quad \quad
R_{\pm} \, \, = \, \, \, 
{{2} \over { 2 \, \,\, \pm \sqrt{-1} \cdot \, (x+y+z)\, -\,x\, z\, -1/2 \cdot x\, y }} 
\end{eqnarray}
or the diagonal of the rational function $\, (R_{+} \, + \, R_{-})/2$ reading
\begin{eqnarray}
\label{diag3quartBIS}
\hspace{-0.98in}&& \,\,  \quad \quad \quad
 \,{\frac {  4 \cdot \,  (4\, -xy \, -2\,xz)}{
{y}^{2}{x}^{2} +4\,{x}^{2}yz+4\,{x}^{2}{z}^{2}+4\,{x}^{2}-8\,xz+4\,{y}^{2}+8\,yz+4\,{z}^{2} \, +16}}, 
\end{eqnarray}
becomes (as a consequence of  identity (\ref{ident3quart0})):
\begin{eqnarray}
\label{ident3quart}
\hspace{-0.96in}&& \,\, \quad \quad \quad \quad \quad \, \, 
\Bigl(1 \, -{{3} \over {4}} \cdot \, x^2\Bigr)^{-1/4} \cdot \,
 _2F_1\Bigl([{{1} \over {12}}, \, {{5} \over {12}}],\, [1],
 \, {{27\, x^4\cdot \, (1\, -x^2) } \over { (4 \, -3\,x^2)^3}} \Bigr) 
\nonumber \\ 
\hspace{-0.98in}&& \quad \quad \quad  \quad  \quad \quad \quad \quad \quad \quad 
\, \, = \, \, \, \, 
  _2F_1\Bigl([{{1} \over {4}}, \, {{3} \over {4}}],\, [1], \, x^2\Bigr).
\end{eqnarray}

\vskip .2cm 

{\bf Remark:}  $\, _2F_1([1/4,3/4],[1],{\cal P})$ hypergeometric functions can be 
also seen as modular forms corresponding to identities with {\em two} pullbacks 
related by a modular equation. For example the following identity:
\begin{eqnarray}
\label{newident}
\hspace{-0.98in}&&  \quad \quad \quad   \quad  \quad  \, \, 
_2F_1\Bigl([{{1} \over {4}}, \, {{3} \over {4}}], \, [1],
 \,   {{ x^2} \over {(2\, -x)^2 }} \Bigr)
\nonumber \\ 
\hspace{-0.98in}&&    \quad \quad \quad \quad \quad \quad \quad \quad 
\,  \, = \,  \, \, 
 \Bigl( {{2\, -x} \over { 2 \cdot \, (1 \,  + \, x)}} \Bigr)^{1/2}      \cdot \, 
_2F_1\Bigl([{{1} \over {4}}, \, {{3} \over {4}}], \, [1], 
\,    {\frac { 4\, {x}}{ (1 \, + \, {x})^{2}}} \Bigr),
\end{eqnarray}
where the two rational pullbacks \begin{eqnarray}
\label{SchwaSymmrhoAB}
\hspace{-0.98in}&& \,\, \, \, \quad  \quad \quad \quad \quad \quad  \quad \quad 
A \, \, = \, \, \,   {\frac { 4\, {x}}{ (1 \, + \, {x})^{2}}},
  \quad  \quad \quad 
B \, \, = \, \, \,   {{ x^2} \over {(2\, -x)^2 }}, 
\end{eqnarray}
are related by the {\em asymmetric}\footnote[5]{At first sight one expects the 
two pullbacks (\ref{SchwaSymmrhoAB}) in a relation like (\ref{Mod2}) to be 
on the same footing, the {\em modular equation}  between these two  pullbacks
being {\em symmetric}: see for instance~\cite{Youssef}. This paradox is explained 
in detail in \ref{identitiesmodular}} modular equation:
 \begin{eqnarray}
\label{Mod2}
\hspace{-0.98in}&&  \,  \, \, \, 
81\cdot \, A^2\, B^2 \,  -18 \, A\, B \cdot \, (8\,B\,+A) \,  \, 
+(A^2 \, +80 \cdot \,A\,B\, +64\, B^2)
\,  \, \, -64 \, B\,\, \,  = \,  \,\,\,0. 
\end{eqnarray}
The modular equation (\ref{Mod2}) gives the following expansion for $\, B$ seen 
as an {\em algebraic series}\footnote[3]{We discard the 
other root expansion 
$\,    B \, = \, \,   1 \, +A \, +{\frac{5}{4}}{A}^{2} \, +{\frac{25}{16}}{A}^{3}
\, +{\frac{31}{16}}{A}^{4} \,  \, + \cdots \, \,\, $ since $\, B(0) \, \ne \, 0$. } in $\, A$:
 \begin{eqnarray}
\label{Mod2ser}
\hspace{-0.98in}&& \quad   \, \,   \,  \, 
B \, \, = \, \, \,  \, {\frac{1}{64}}{A}^{2} \, \,  \, 
+{\frac{5}{256}}{A}^{3} \, \,  \, 
+{\frac{83}{4096}}{A}^{4}\, \,  +{\frac{163}{8192}}{A}^{5} \, \,  \, 
+{\frac{5013}{262144}}{A}^{6} \,\,  \, 
\, + \, \,\,  \cdots
\end{eqnarray}
More details are given in \ref{identitiesmodular}.
 
\vskip .1cm

\subsection{The generic case: modular forms, with just one rational pullback }
\label{modulargeneric}

\vskip .1cm

The previous pullbacks in the pullbacked $\, _2F_1$
hypergeometric functions can be seen (and should be seen)
as {\em Hauptmoduls}~\cite{Maier1}. It is only in certain  
cases like in sections (\ref{threesymm}) or (\ref{threeasymm}) 
that we encounter the situation underlined by  Maier~\cite{Maier1}
of a representation of a modular form as a pullbacked hypergeometric function 
with {\em two} possible {\em rational pullbacks}, related by a 
 modular equation of {\em genus zero}. These 
selected situations are recalled in \ref{modularAPP}. 

Simple examples of modular equations of genus zero with {\em rational pullbacks}  
include reductions of the generic $\, _2F_1([1/12,5/12],[1], \, {\cal P})$ 
hypergeometric function to selected hypergeometric functions like
 $\, _2F_1([1/2,1/2],[1], \, {\cal P})$,  $\, _2F_1([1/3,2/3],[1], \,{\cal P})$, 
$\, _2F_1([1/4,3/4],[1], \, {\cal P})$, 
and also~\cite{Shen} $\, _2F_1([1/6,5/6],[1], \,{\cal P})$
(see  for instance~\cite{Garvan}).  

However, in the generic situation corresponding to (\ref{2F15HypformA})
 we have a single hypergeometric function with two different pullbacks $\, A$ and $\, B$ 
\begin{eqnarray}
\label{modula}
\hspace{-0.95in}&& \quad \quad  \quad  \quad  \quad 
 _2F_1\Bigl([{{1} \over {12}}, \, {{5} \over {12}}], \, [1], \, A  \Bigr)
\, = \, \, \, \,\, 
G \cdot \,
 _2F_1\Bigl([{{1} \over {12}}, \, {{5} \over {12}}], \, [1], \,  B  \Bigr), 
\end{eqnarray}
with $\, G$  an algebraic function of $\, x$, and where  $\, A$ and $\, B$ 
{\em are related by an algebraic modular equation}, but one of the two pullbacks
say $\, A$ is a rational function given by (\ref{Q512}) 
where $\, P_2(x)$ and $\, P_4(x)$ are given by (\ref{P2}), (\ref{P4}). The two pullbacks 
$\, A$ and $\, B$  are also 
related by a Schwarzian equation 
that can be written in a symmetric way in $\, A$ and $\, B$:
\begin{eqnarray}
\label{SchwaSymmeq1}
\hspace{-0.98in}&& \,\, \, \,   \quad   \quad \quad   \quad \quad  
 {{1} \over {72}}\,{\frac {32 \,{B}^{2}\,-41\,B\,+36}{{B}^{2} \cdot \, (B \, -1)^{2}}}  \,
\cdot \, \Bigl({{d B } \over {dx}}\Bigr)^2 \,
\, + \, \, \,  \{B, \, x\} 
 \nonumber \\
\hspace{-0.98in}&& \,\,   \quad \quad   \quad \quad   \quad \quad \quad \quad
\,\, \, \,= \, \, \,\,\, 
 {{1} \over {72}}\,{\frac {32 \,{A}^{2} \, -41 \,A \, +36}{{A}^{2} \cdot \,  (A \, -1)^{2}}}  \,
\cdot \, \Bigl({{d A } \over {dx}}\Bigr)^2 \,
\, + \, \, \,  \{A, \, x\}.
\end{eqnarray}
One can rewrite the exact 
expression (\ref{2F15HypformA}) in the form 
\begin{eqnarray}
\label{2F15HypformArewrite}
\hspace{-0.7in}&&\quad \quad  \quad \quad  \, 
{{1} \over { P_2(x)^{1/4}}} \cdot \, 
 _2F_1\Bigl([{{1} \over {12}}, \, {{5} \over {12}}], \, [1],
 \, \, 1 \, - \, {{P_4(x)^2 } \over {P_2(x)^3}}\Bigr)
\nonumber \\
\hspace{-0.7in}&&\quad \quad \quad \quad \quad \quad\quad \quad
\, \, = \, \, \, \,  {\cal B} \cdot \, 
 _2F_1\Bigl([{{1} \over {12}}, \, {{5} \over {12}}], \, [1],
 \, \, B\Bigr),
\end{eqnarray}
where $\, {\cal B}$ is an algebraic function, 
and $\, B$ is another pullback related to the 
{\em rational} pullback $\, A \, = \, \, 1 \, -\, P_4(x)^2/P_2(x)^3$
by a modular equation. The pullback $\, B$ is an {\em algebraic function}.
In the generic case, {\em only one 
of the two pullbacks} (\ref{2F15HypformArewrite})
{\em can be expressed as a rational function}: see  \ref{modularAPP}
for more details.

\vskip .1cm
\vskip .1cm 

\section{Eight, nine and ten-parameters generalizations}
\label{conclus}
 
Adding  randomly terms in the denominator of (\ref{Ratfonc}) yields diagonals 
annihilated by minimal linear differential operators 
of order higher than two: this is what happens when
quadratic terms like  $\, x^2$, $\, y^2$ or $\, z^2$ are added for example. 
This leads to irreducible telescopers~\cite{Telescopers,Telescopers2} 
(i.e. minimal order linear differential operators 
annihilating the diagonals) of {\em higher orders} than the previous order two, 
or to telescopers~\cite{Telescopers} of quite high orders that are not 
irreducible, but factor into 
many irreducible factors, one of them being of order larger than two. 

With the idea of keeping  the linear differential operators annihilating
the diagonal of order two, we were able to generalize the seven-parameter 
family (\ref{Ratfonc}) by carefully choosing the terms added to the 
quadratic terms in (\ref{Ratfonc}) and still keep the linear differential 
operator annihilating the diagonal of order two.

 \vskip .1cm

\subsection{Eight-parameter rational functions giving pullbacked $\, _2F_1$ hypergeometric functions for their diagonals }
\label{step}

Adding the cubic term $\, x^2 \, y$ to the denominator of (\ref{Ratfonc})
yields the rational function: 
\begin{eqnarray}
\label{Ratfoncplus}
\hspace{-0.98in}&& 
R(x, \, y, \, z)  \, \, \, = \, \,  \, \,  \,
 {{1} \over {
a \, \, \,+ \, b_1 \, x \, + \, b_2 \, y \,  + \, b_3 \, z \,\,
 + \, c_1 \, y\, z \, + \, c_2 \, x \, z \,  + \, c_3 \, x\, y
 \, \, + \, \, d \, x^2 \, y }}. 
\end{eqnarray}
After obtaining the diagonal of (\ref{Ratfoncplus}) for several sets of 
values of the parameters, one can make the educated guess that the
diagonal of the rational function (\ref{Ratfoncplus}) has the form
\begin{eqnarray}
\label{2F15HypformAplus}
\hspace{-0.7in}&&\quad \quad \quad \quad 
{{1} \over { P_3(x)^{1/4}}} \cdot \, 
 _2F_1\Bigl([{{1} \over {12}}, \, {{5} \over {12}}], \, [1],
 \, \, 1 \, - \, {{P_4(x)^2 } \over {P_3(x)^3}}\Bigr),
\end{eqnarray}
where $\, P_3(x)$ and $\, P_4(x)$ are two polynomials of degree 
three and four respectively:
\begin{eqnarray}
\label{P4P3}
\hspace{-0.7in}&&\quad \quad \quad  \, \,
P_4(x) \, \, = \, \, \,
 {\cal A}_4 \, \, x^4 \, + \,  {\cal A}_3 \,  x^3 \, + \,  {\cal A}_2 \, x^2 \, +  {\cal A}_1 \, x \, +  {\cal A}_0, 
\\
\hspace{-0.7in}&&\quad \quad \quad  \, \,
P_3(x) \, \, = \, \, \,  {\cal B}_3 \,  x^3 \, +  {\cal B}_2 \,  x^2 \, +  {\cal B}_1 \, x \, +  {\cal B}_0, 
\end{eqnarray}
and where the coefficients $\, {\cal A}_i$ and $\, {\cal B}_j$ are at most 
quadratic expressions in the parameter $\, d \, \, $ appearing in 
the denominator of (\ref{Ratfoncplus}). 
The pullback in (\ref{2F15HypformAplus}) has the form
\begin{eqnarray}
\label{P4P3oftheform}
\hspace{-0.98in}&& \quad \quad \quad \quad \quad \quad  \quad  \quad  \quad 
1 \, - \, {{P_4(x)^2 } \over {P_3(x)^3}} 
 \,  \,  \,  \, = \,   \,  \,  \,  \,  
 {{ 1728 \, x^3 \, P_6} \over { P_3(x)^3}}, 
\end{eqnarray}
where
\begin{eqnarray}
\label{namely9}
\hspace{-0.98in}&& \quad   \quad  \quad  \quad  \, \, 
P_4 \, = \, \,\, \, \,  p_4  \, \, \, \,
 +216 \cdot \, b_3^2 \, c_1^2 \cdot \, d^2 \cdot\, x^4 \, \, \,\, +d \cdot \, u_1 \cdot \, x^4 \, \, 
  \nonumber \\ 
\hspace{-0.98in}&&   \quad \quad  \quad   \quad    \quad \quad  \quad    \quad   
+ a \, d \cdot \, u_2 \cdot \, x^3  \,  \,  \,  \, \,
 -144 \cdot \, a \, b_1 \, b_3 \, c_1^2 \, d \cdot\, x^3
\nonumber \\ 
\hspace{-0.98in}&& \quad  \quad   \quad  \quad \quad \quad \quad \quad  \quad 
  \,\, \, 
-144 \cdot \, b_2 \, b_3^2 \, d \cdot \, (b_1 \, c_1 \, +4\, b_2 \, c_2 \, -2\, b_3 \,  c_3) \cdot \, x^3 
 \nonumber \\ 
\hspace{-0.98in}&& \quad  \quad   \quad  \quad \quad \quad \quad \quad  \quad 
 \, \,   +36 \, a^2 \, \cdot \, ( a \, b_3 \,c_1 \, \,-2  \, b_2 \, b_3^2) \cdot \, d \cdot\, x^2, 
 \nonumber \\ 
\hspace{-0.98in}&&\quad \quad \quad \quad  \, \,
P_3 \, = \, \, \,\,  \,\, 
  p_2  \, \, \, \, -48 \cdot \, c_1^2\, c_2 \cdot \, d \cdot \, x^3
 \, \, \,  +24 \, b_3  \cdot \, (a \, \, c_1 \, -2 \, b_2 \, b_3) \cdot  \, d \cdot \, x^2, 
\end{eqnarray}
with
\begin{eqnarray}
\label{namelywhere}
\hspace{-0.98in}&&\quad \quad  \quad \quad  \quad  \quad  \quad 
u_1  \, = \, \,\, 144 \cdot \, ( 2 \,\, b_1 \,  c_1^3 \, c_2  \, \,  \,   -4 \,\, b_2 \,  c_1^2 \, c_2^2
  \, \,   \, -\,\, b_3\,  c_1^2 \, c_2\, c_3),
 \nonumber \\ 
\hspace{-0.98in}&&
 \quad \quad  \quad \quad  \quad  \quad  \quad 
u_2  \, = \, \,\,\, 72 \cdot \, (10\,\, b_2 \, b_3 \,  c_1\, c_2  \, \, -\,  a\,  c_1^2\, c_2 \, 
\, \,  \,  - 2\,\,  b_3^2 \,  c_1\, c_3), 
\end{eqnarray}
and where the polynomials  $\, p_2$ and $\, p_4$  denote the  polynomials  $\, P_2(x)$ and  $\, P_4(x)$
given by (\ref{P2}) and  (\ref{P4}) in section (\ref{R}): 
 $\, p_2$ and $\, p_4$   correspond to the $\, d \, = \, 0 \, $ limit. 

\vskip .1cm 

\subsection{Nine-parameter rational functions giving pullbacked $\, _2F_1$ hypergeometric functions for their diagonals}
\label{twostep}

Adding now another cubic term $ \, y \, z^2$ to the denominator of (\ref{Ratfoncplus}) 
\begin{eqnarray}
\label{Ratfoncplusplus}
\hspace{-0.98in}&&
\quad  \quad 
 {{1} \over {
a \,\, \,+ \, b_1 \, x \, + \, b_2 \, y \,  + \, b_3 \, z \,\,\,
 + \, c_1 \, y\, z \, + \, c_2 \, x \, z \,  + \, c_3 \, x\, y 
\, \,\, + \, \, d \, x^2 \, y \, \, \, + \, \, e \, y \, z^2 }}, 
\end{eqnarray}
also yields linear differential  operator annihilating 
the diagonal of (\ref{Ratfoncplusplus}) of order two.
After computing the second order linear differential  operator annihilating 
the diagonal of (\ref{Ratfoncplusplus}) for several values
of the parameters with the ``{\em HolonomicFunctions}'' 
program~\cite{Koutschan}, and, in a second step, obtaining
their pullbacked hypergeometric solutions using the 
maple command ``{\em hypergeometricsols}''~\cite{Hoeijprogram}, 
we find that the diagonal of the rational function (\ref{Ratfoncplusplus}) 
has the form
\begin{eqnarray}
\label{2F15HypformAplusplus}
\hspace{-0.7in}&&\quad \quad \quad \quad \quad 
{{1} \over { P_4(x)^{1/4}}} \cdot \, 
 _2F_1\Bigl([{{1} \over {12}}, \, {{5} \over {12}}], \, [1],
 \, \, 1 \, - \, {{P_6(x)^2 } \over {P_4(x)^3}}\Bigr),
\end{eqnarray}
where $\, P_4(x)$ and $\, P_6(x)$ are two polynomials of degree 
four and six respectively:
\begin{eqnarray}
\label{P6P4two}
\hspace{-0.98in}&& \quad  \quad   \,   \, \, \, \, 
P_4(x) \, \, = \, \,\, \,  \, \, 
  p_2 \, \, \, \,  \,   + \, \,  16 \cdot  \,d^2 \cdot  \,e^2 \cdot  \, x^4 \, \,  
\nonumber \\
\hspace{-0.98in}&& \quad \quad  \quad \quad  \, \, \,  \, 
 -16 \cdot \, \Bigl(3 \cdot \, c_2 \cdot\, (  c_1^2 \cdot \,d \, +\, c_3^2 \cdot \, e)
 \, \,   \,  +  (b_1 \,c_1  \, +b_3 \, c_3 \, -14 \, b_2 \, c_2) \cdot \, d \, e \Bigr) \cdot  \, x^3 
\nonumber \\
\hspace{-0.98in}&& \quad \quad  \quad  \quad   \, \, \,  \, 
 \,  \,  \, +8 \cdot  \,(3 \, a \, b_3 \, c_1  \, d \, +3 \, a \,  b_1\, c_3 \, e \,  \, 
      - \,  a^2 \, d \, e \,  \,  -6 \, b_2 \, b_3^2 \, d \, -6  \, b_2 \, b_1^2 \, e) \cdot  \, x^2,  
\end{eqnarray}
and 
\pagebreak 
\begin{eqnarray}
\label{P6P4two}
\hspace{-0.98in}&&  \,  \quad   \quad \quad 
P_6(x) \, \, = \,\,  \,\, \,\,\,  \, 
p_4 \, \,\, \,\,  -12 \cdot \, a^4 \,  d \, e \,  \cdot \, x^2  
\nonumber \\
\hspace{-0.98in}&&  \quad \quad  \quad  \quad  \quad \, \,\,\,\, \,\, 
+   36 \cdot \, a^2 \,  
\Bigl(  \, b_3 \cdot \  (a \, c_1 \, -2 \, b_2 \, b_3) \cdot  \, d \,
  + \, b_1 \cdot \, (a \, c_3 \, -2 \, b_1 \, b_2) \cdot  \, e   \Bigr) \cdot \, x^2 
\nonumber \\
\hspace{-0.98in}&& \quad  \quad   \quad \quad  \quad  \quad \, \,\,
 \,  \, \,  -72 \cdot \, a\, c_1 \cdot  \,
 (a \, c_1\, c_2 \,-10 \,b_2 \, b_3 \, c_2 \, +2\, b_3^2 \, c_3) \cdot \, d   \cdot \,x^3
 \nonumber \\
\hspace{-0.98in}&& \quad  \quad \quad  \quad  \quad  \quad   \quad  \quad \, \,\,
\, \, \, -72 \cdot \, a\, c_3 \cdot  \,
 (a \, c_2 \, c_3 \,-10 \, b_1 \, b_2 \, c_2 \, +2\,  b_1^2 \,c_1) \cdot \, e  \cdot \,x^3
\nonumber \\
\hspace{-0.98in}&& \quad \quad \quad  \quad \quad  \quad  \quad  \quad   \, \,\,
 \,  \,  -144 \cdot \,  b_2\, b_3^2 \cdot \,
 (b_1 \, c_1 \, +4\, b_2 \, c_2 \, -2\, b_3 \, c_3)  \cdot \, d  \cdot \,x^3
 \nonumber \\
\hspace{-0.98in}&& \quad \quad \quad  \quad \quad   \quad  \quad  \quad  \, \,\,
\,  \,   -144 \cdot \,  b_2\, b_1^2 \cdot \, 
 (b_3 \, c_3 \, +4\, b_2 \,c_2 \, -2\, b_1 \, c_1)  \cdot \, e   \cdot \,x^3
\nonumber \\
\hspace{-0.98in}&&   \quad  \quad \quad  \quad  \quad  \, \,\,
  \,  -144  \cdot \, a \, b_1 \, b_3 \cdot \, (c_1^2  \cdot \, d  \, +  \,c_3^2  \cdot \,e) \cdot \,x^3
 \nonumber \\
\hspace{-0.98in}&&  \quad \quad \quad  \quad  \quad  \quad  \quad  \quad \, \,\,
 \, +24 \cdot a \, \, (a\, b_3\, c_3  \, +\,a\, b_1\, c_1 \, 
-20 \, a\, b_2 \, c_2  \, + \, 30 \, b_1\, b_2 \, b_3)  \cdot \, d \cdot \, e \cdot \, x^3 
\nonumber \\
\hspace{-0.98in}&&  \quad  \quad  \quad  \quad \quad   \, \,\,
\, \,  +216  \cdot  \, 
( b_3^2 \, c_1^2 \cdot \, d^2  \, \, + \, b_1^2 \, c_3^2 \cdot \, e^2) \cdot \, x^4 
 \nonumber \\
\hspace{-0.98in}&&   \quad \quad \quad \quad  \quad  \quad  \quad \quad\, \,\,
-  \, 144  \cdot \,  c_1^2  \,\, c_2 \cdot \,
 (b_3 \, c_3 \, +4\, b_2 \, c_2 \, -2\, b_1 \,  c_1) \cdot \, d  \cdot \, x^4 
 \nonumber \\
\hspace{-0.98in}&&   \quad \quad\quad \quad  \quad  \quad  \quad  \quad \, \,\,
-  \, 144  \cdot \, \, c_3^2  \,\, c_2 \cdot \,
 (b_1 \, c_1 \, +4\,b_2 \,  c_2 \, -2\, b_3 \,  c_3) \cdot \, e   \cdot \, x^4 
\nonumber \\
\hspace{-0.98in}&& \quad  \quad  \quad  \quad  \, \,  \, 
 + 48 \cdot \, a^2 \,  d^2 \cdot \, e^2  \cdot  \,x^4 \, 
\, \,\, +96  \cdot \, 
(b_1^2 \, c_1^2  \, +b_3^2 \, c_3^2 \,  +22 \,b_2^2 \, c_2^2)  \cdot  \, d  \cdot  \,e  \cdot  \,x^4
        \nonumber \\
\hspace{-0.98in}&&   \quad  \quad  \quad  \quad  \quad   \,  \, \, 
 -144 \cdot \, \Bigl( (a\, b_3 \, c_1 \, +4 \, b_2\, b_3^2) \cdot \, d \,\,
          + \, (a\, b_1 \, c_3 \, +4 \, b_2 \, b_1^2)  \cdot \, e \Bigr) 
    \cdot  \, d  \cdot  \,e  \cdot  \,x^4
  \nonumber \\
\hspace{-0.98in}&& \quad  \quad \quad   \quad 
      \,  +48 \cdot  \, (b_1 \, b_3 \, c_1 \,c_3 \,  + 15\, a\, c_1\, c_2\, c_3 
\, -20\, b_1 \, b_2 \,  c_1\, c_2  \, -20\, b_2 \, b_3 \, c_2\, c_3)  
    \cdot  \, d  \cdot  \,e  \cdot  \,x^4
\nonumber \\
\hspace{-0.98in}&&  \quad \quad \quad \quad  \quad    \quad  \quad  \,  \, \,  
 +96 \cdot \, (b_1\, c_1 \, +22\, b_2\, c_2\, +b_3\, c_3) \cdot \, d^2  \cdot \, e^2  \cdot \, x^5 
\nonumber \\
\hspace{-0.98in}&&  \quad \quad \quad \quad \quad   \quad  \quad  \quad  \quad  \,  \, \,  
   -576 \, c_2 \,\cdot  \, (c_3^2  \cdot \,  e  \, \, 
 \,  + \,  c_1^2  \cdot \, d )  \cdot \, d \, e \, \cdot \, x^5      
\nonumber \\
\hspace{-0.98in}&&   \quad \quad  \quad \quad 
\quad  \quad    \quad \quad \quad  \quad  \quad   \quad\quad  \, 
  -64 \cdot \, d^3  \cdot \, e^3  \cdot \, x^6,  
\end{eqnarray}
where the polynomials  $\, p_2$ and $\, p_4$  are the  polynomials  $\, P_2(x)$ and  $\, P_4(x)$
of degree two and four in $\, x$ given by (\ref{P2}) and  (\ref{P4}) in section (\ref{R}): 
 $\, p_2$ and $\, p_4$   correspond to the $\, d \, = \, e \, = \, 0 \, $ limit

Note that the $\, d \,\leftrightarrow \,\, e$ symmetry corresponds to keeping $\, c_2$ fixed, 
but changing $\, c_1\, \leftrightarrow \,\, c_3$
(or equivalently  $\, y$ fixed,  $\, x \,\leftrightarrow \,\, z$). 

\vskip .1cm 
\vskip .1cm 

{\bf Remark 1:} The nine-parameter family (\ref{Ratfoncplusplus}) singles out $\, x$ and $\, y$, 
but of course, similar families that single out $\, x$ and $\, z$, or  single out $\, y$ and $\, z$
exist, with similar results (that can be obtained permuting the three variables 
$ \, x$, $\, y$ and $\, z$).

\vskip .1cm 
\vskip .1cm 

{\bf Remark 2:} Note that the simple symmetries arguments displayed in section (\ref{Simplesym}) 
for the seven-parameter family straightforwardly generalize for this 
nine-parameter family. 
The pullback $\, {\cal H}$ in (\ref{2F15HypformAplusplus})  verifies (as it should)  
\begin{eqnarray}
\label{symmP2gen}
\hspace{-0.98in}&& \, \,\quad \quad \quad   \quad   
{\cal H}\Bigl(a, \, \, \,   \,
\lambda_1 \cdot \, b_1, \,  \,  \lambda_2 \cdot \,b_2, \,  \,  \lambda_3 \cdot \,b_3,
\,  \,\, \, \lambda_2 \, \lambda_3  \cdot \, c_1, \,\,    \lambda_1 \, \lambda_3  \cdot \,c_2, 
\, \,   \lambda_1 \, \lambda_2  \cdot \,c_3, 
\nonumber \\
\hspace{-0.98in}&&\quad \quad \quad \quad \quad \quad   \quad  \quad \quad 
\quad \quad \quad  \quad \, \, \, \lambda_1^2 \, \lambda_2  \cdot \,d, 
 \, \,\,\,  \, \lambda_3^2 \, \lambda_2  \cdot \,e 
, \,\,   \, \,\,\,  {{x } \over { \lambda_1 \,  \lambda_2 \,  \lambda_3 }}\Bigr)
\nonumber \\
\hspace{-0.98in}&&\quad \quad \quad \quad \quad \quad \quad   
 \, \, = \, \, \, \,\,
{\cal H}(a, \, \, b_1, \, \,  b_2, \,\,   b_3,
 \, \,\,  c_1, \,  c_2, \,  c_3, \,\, \, d, \,\, \, e, \,   \,\, x),
\end{eqnarray}
and:
\begin{eqnarray}
\label{symmP2genbis}
\hspace{-0.98in}&& \, \,\quad  \quad \quad  \quad   
{\cal H}\Bigl( \lambda \cdot \, a, \, \,\lambda \cdot \,  b_1, \, \,  \lambda \cdot \, b_2, 
\,\,  \lambda \cdot \,  b_3, \, \,\,  \lambda \cdot \, c_1, \, \lambda \cdot \,  c_2,
 \, \lambda \cdot \,  c_3, \,\, \, \lambda \cdot \,  d,  \,\, \, \lambda \cdot \,  e, \,   \,\, x)
\nonumber \\
\hspace{-0.98in}&&\quad \quad \quad \quad \quad   \quad \quad \quad 
 \, \, = \, \, \, \,\,
{\cal H}(a, \, \, b_1, \, \,  b_2, \,\,   b_3, 
\, \,\,  c_1, \,  c_2, \,  c_3, \,\,\,  d, \,\,\,  e,  \,   \,\, x).
\end{eqnarray}

\vskip .1cm 

\subsection{Ten-parameter rational functions giving pullbacked $\, _2F_1$ hypergeometric functions for their diagonals}
\label{Threesteps}

\vskip .1cm 

Adding the {\em three cubic terms}\footnote[2]{An equivalent family of ten-parameter 
rational functions amounts to adding 
$\, \,x \, y^2$,  $\,\,y \, z^2$ and $\, \,z \, x^2$.} 
$\,\, x^2 \, y$,  $\,\,y^2 \, z$ and $\,\, z^2 \, x$ to the denominator of (\ref{Ratfonc})
we get the rational function: 
\begin{eqnarray}
\label{Ratfoncplusplusplus}
\hspace{-0.98in}&&
 R(x, \, y, \, z)  \, \, \, = \, \,  
 \\
\hspace{-0.98in}&&
 {{1} \over {
a \, \, \,  + \, b_1 \, x \,  + \, b_2 \, y \,   + \, b_3 \, z \,\,\, 
 + \, c_1 \, y\, z \, + \, c_2 \, x \, z \, + \, c_3 \, x\, y
 \, \, \, + \, \, d_1 \, x^2 \, y \, \,   + \, \, d_2 \, y^2 \, z \,   + \, \, d_3 \, z^2 \, x  }}. 
\nonumber 
\end{eqnarray}
While (\ref{Ratfoncplusplusplus}) is {\em not} a generalization  of  (\ref{Ratfoncplusplus}), 
it is a generalization  of (\ref{Ratfoncplus}) .

After computing the second order linear differential operator annihilating 
the diagonal of (\ref{Ratfoncplusplusplus}) for several values
of the parameters with the ``{\em HolonomicFunctions}'' 
program~\cite{Koutschan}, and, in a second step, 
their pullbacked hypergeometric solutions using ``{\em hypergeometricsols}''~\cite{Hoeijprogram},
we find that the diagonal of the rational function (\ref{Ratfoncplusplusplus}) 
has the experimentally observed form:
\begin{eqnarray}
\label{2F15HypformAplusplusplus}
\hspace{-0.7in}&&\quad \quad \quad \quad \quad 
{{1} \over { P_3(x)^{1/4}}} \cdot \, 
 _2F_1\Bigl([{{1} \over {12}}, \, {{5} \over {12}}], \, [1],
 \, \, 1 \, - \, {{P_6(x)^2 } \over {P_3(x)^3}}\Bigr).
\end{eqnarray}
Furthermore, the pullback in (\ref{2F15HypformAplusplusplus}) is seen to be of the form: 
\begin{eqnarray}
\label{oftheform}
\hspace{-0.98in}&& \quad  \quad \, \quad  \quad \quad \quad \quad \quad \quad 
 1 \, - \, {{P_6(x)^2 } \over {P_3(x)^3}}  \, \, \, \, = \, \, \, \, \,
{{1728 \, x^3 \cdot \, P_9} \over { P_3(x)^3 }}.  
\end{eqnarray}
The polynomial $\, P_3(x)$ reads 
\begin{eqnarray}
\label{oftheformP3}
\hspace{-0.99in}&&  
P_3(x) \,   = \, \,    p_2 \, \, \,  
-24 \cdot \, \Bigl(9 \cdot \, a \cdot \, d_1\, d_2\, d_3  \, \,
   -6 \cdot \, 
(b_1\, c_3 \cdot \, d_2\, d_3 \, +\, b_2\, c_1 \cdot \, d_1\, d_3 \, +\, b_3\, c_2 \cdot \, d_1\, d_2) 
\nonumber \\
\hspace{-0.99in}&&  \quad  \quad  \quad  \quad  \quad  \quad 
\, \, 
     +2 \cdot \,
 (c_1^2\, c_2\, d_1 +\, c_1\, c_3^2\, d_3 \, +\, c_2^2\, c_3 \, d_2) \Bigr) \cdot \, x^3
 \\
\hspace{-0.99in}&&  
 +24 \cdot \, \Bigl(a \cdot \, (b_1\, c_2\, d_2 \, +\, b_2\, c_3\, d_3 \, +\, b_3\, c_1\, d_1) 
\, \, -2 \, \cdot \,
 (b_1^2\, b_3\, d_2 \, +\, b_1\, b_2^2\, d_3 \, +\, b_2\, b_3^2\, d_1) \Bigr) \cdot \, x^2, 
\nonumber 
\end{eqnarray}
where $\, p_2$ is the  polynomial  $\, P_2(x)$ 
of degree two in $\, x$ given by (\ref{P2})  in section (\ref{R}): 
 $\, p_2$   corresponds to the
 $\, d_1 \, = \, d_2 \, = \, d_3 \, = \, \, 0 \, \, $ limit. 
The expression of the polynomial $\, P_6(x)$ is more involved. It reads:
\begin{eqnarray}
\label{oftheformP6}
\hspace{-0.99in}&&  \quad  \quad  \quad  \quad  \quad  \quad  \quad  \quad  \quad  \quad 
P_6(x) \,  \,  = \, \, \, \,    p_4 \, \, \,  + \, \, \Delta_6(x), 
\end{eqnarray}
where $\, p_4$ is the  polynomial  $\, P_4(x)$ 
of degree four in $\, x$ given by (\ref{P4})  in section (\ref{R}).  
The expression of polynomial $\, \Delta_6(x)$ of degree six in $\, x$ 
is quite large and is given in \ref{DeltaP6}. 

\vskip .1cm 
\vskip .1cm 

{\bf Remark 1:} A set of results and subcases (sections (\ref{justtwosteps1}) and  (\ref{justtwosteps2})),
can be used to ``guess'' the general exact expressions of the polynomials $\, P_3(x)$ and $\, P_6(x)$ 
in (\ref{2F15HypformAplusplusplus}) for the ten-parameters family (\ref{Ratfoncplusplusplus}).  
From the subcase $\, d_3 \, = \, 0$ of section (\ref{justtwosteps}) below, it is easy 
to see that one can deduce similar exact results for $\, d_1 \, = 0$ or  $\, d_2 \, = 0$: 
it just amounts to performing some cyclic transformation 
$\, x \, \rightarrow \, y \,  \rightarrow \, z \,  \rightarrow \,x$
which corresponds to  transformation 
$\, b_1 \, \rightarrow \, b_2 \, \rightarrow \, b_3 \, \rightarrow \, b_1$,
 $\, c_1 \, \rightarrow \, c_2 \, \rightarrow \, c_3 \, \rightarrow \, c_1$, 
$\, d_1 \, \rightarrow \, d_2 \, \rightarrow \, d_3 \, \rightarrow \, d_1$.
One can see  $\, P_3$ 
and $\, P_6(x)$ as   $\, p_2$ and $\, p_4$ given by (\ref{P2}) and  (\ref{P4}) plus some 
corrections given, in \ref{polynomials}, 
by (\ref{P3screw}) and (\ref{P5screw}) for $\, d_3 \, = \, 0$, and 
similar corrections\footnote[1]{Taking care of the double counting !}  
for $\, d_1 \, = \, 0$ and  $\, d_2 \, = \, 0$, plus corrections 
of the form $\, d_1 \, d_2\, d_3 \times \, $ something. These last terms are the 
most difficult to get. We already know some of these terms from (\ref{simpleThreeP3P6}) 
and (\ref{simpleThreeP3P6bis}) in section (\ref{justtwosteps1}) below. 
Furthermore, the symmetry constraints (\ref{symmP2genbislast}) and (\ref{symmP2genlast}) below,
as well as other constraints corresponding to the symmetric subcase 
of section (\ref{justtwosteps2}),
give additional constraints on the kind of allowed final correction terms.

\vskip .1cm 

{\bf Remark 2:} Note, again, that the simple symmetries arguments 
displayed in section (\ref{Simplesym}) 
for the seven-parameter family straightforwardly generalize for this 
ten-parameter family. 
The $\, {\cal H}$ pullback (\ref{oftheform}) in (\ref{2F15HypformAplusplusplus}) 
 verifies (as it should):  
\begin{eqnarray}
\label{symmP2genlast}
\hspace{-0.98in}&& \, \,\quad \quad \quad   \quad   
{\cal H}\Bigl(a, \, \, \,   \,
\lambda_1 \cdot \, b_1, \,  \,  \lambda_2 \cdot \,b_2, \,  \,  \lambda_3 \cdot \,b_3,
\,  \,\, \, \lambda_2 \, \lambda_3  \cdot \, c_1, \,\,    \lambda_1 \, \lambda_3  \cdot \,c_2, 
\, \,   \lambda_1 \, \lambda_2  \cdot \,c_3, 
\nonumber \\
\hspace{-0.98in}&&\quad \quad \quad \quad \quad \quad   \quad  \quad \quad 
\quad \quad \quad  \quad \, \, \, \lambda_1^2 \, \lambda_2  \cdot \,d_1,  
\, \,\,\,  \, \lambda_2^2 \, \lambda_3  \cdot \, d_2, 
 \,\,  \, \lambda_3^2 \, \lambda_1  \cdot \, d_3,  
 \,\,   \, \,\,\,  {{x } \over { \lambda_1 \,  \lambda_2 \,  \lambda_3 }}\Bigr)
\nonumber \\
\hspace{-0.98in}&&\quad \quad \quad \quad \quad \quad \quad   
 \, \, = \, \, \, \,\,
{\cal H}(a, \, \, b_1, \, \,  b_2, \,\,   b_3, \, \,\,  c_1, \,  c_2, 
\,  c_3, \,\, \, d_1, \,\,\,  d_2,  \,  \,\,  d_3,  \,   \,\, x),
\end{eqnarray}
and:
\begin{eqnarray}
\label{symmP2genbislast}
\hspace{-0.98in}&& \, \, \quad   \quad   
{\cal H}\Bigl( \lambda \cdot \, a, \, \,\lambda \cdot \,  b_1, \, \,  \lambda \cdot \, b_2, 
\,\,  \lambda \cdot \,  b_3, \, \,\,  \lambda \cdot \, c_1, \, \lambda \cdot \,  c_2,
 \, \lambda \cdot \,  c_3, 
\,\, \, \lambda \cdot \,  d_1,  \,\, \, \lambda \cdot \,  d_2, \,\, \, \lambda \cdot \,  d_3, \,   \,\, x)
\nonumber \\
\hspace{-0.98in}&&\quad \quad \quad \quad \quad   \quad \quad \quad 
 \, \, = \, \, \, \,\,
{\cal H}(a, \, \, b_1, \, \,  b_2, \,\,   b_3,
 \, \,\,  c_1, \,  c_2, \,  c_3, \,\,\,  d_1, \,\,\,  d_2,  \,  \,\,  d_3,  \,   \,\, x).
\end{eqnarray}

\vskip .1cm 

{\bf Remark 3:} Do note that adding arbitrary sets of cubic terms 
yields telescopers~\cite{Telescopers} of order larger than two: 
{\em the corresponding diagonals 
are no longer pullbacked $\, _2F_1$ hypergeometric functions}. 

\vskip .1cm 
\vskip .1cm 
\vskip .1cm 

Let us just now  focus on simpler subcases whose results are easier 
to obtain than in the general case  (\ref{Ratfoncplusplusplus}).

\vskip .1cm 

\subsubsection{Noticeable subcases of (\ref{Ratfoncplusplusplus}): a nine-parameter rational function\\} 
\label{justtwosteps}

\vskip .1cm 

Instead of adding  three cubic terms, let us  add two cubic terms. This 
amounts to restricting the rational function (\ref{Ratfoncplusplusplus}) 
to the $\, d_3 \, = \, 0$ subcase
\begin{eqnarray}
\label{Ratfoncplusplusplusjusttwo}
\hspace{-0.98in}&& \,\, \,\, \,\, \,\, \,\,\,\,
 {{1} \over {
a \, \, +  b_1 \, x \,   +  b_2 \, y \,   +  b_3 \, z \,\,  
 +  c_1 \, y\, z \,\,  +  c_2 \, x \, z \, \,  +  c_3 \, x\, y
 \, \, +  d_1 \, x^2 \, y \, \, \,  +   d_2 \, y^2 \, z }}, 
\end{eqnarray}
which {\em cannot be reduced to the nine parameter family} (\ref{Ratfoncplusplus})
even if it looks similar.  The
diagonal of the  rational function (\ref{Ratfoncplusplusplusjusttwo})
has the {\em experimentally observed} form
\begin{eqnarray}
\label{2F15HypformAplusplusplusbis}
\hspace{-0.7in}&&\quad \quad \quad \quad \quad 
{{1} \over { P_3(x)^{1/4}}} \cdot \, 
 _2F_1\Bigl([{{1} \over {12}}, \, {{5} \over {12}}], \, [1],
 \, \, 1 \, - \, {{P_5(x)^2 } \over {P_3(x)^3}}\Bigr),
\end{eqnarray}
where $\, P_3(x)$ and  $\, P_5(x)$ are two polynomials of degree
respectively three and five in $\, x$. 
Furthermore the pullback in (\ref{2F15HypformAplusplusplusbis}) 
has the form: 
\begin{eqnarray}
\label{oftheformbis}
\hspace{-0.98in}&& \quad  \quad \, \quad  \quad \quad \quad \quad \quad \quad 
 1 \, - \, {{P_5(x)^2 } \over {P_3(x)^3}} 
 \, \, \, \, = \, \, \, \, \,
{{1728 \, x^3 \cdot \, P_7} \over { P_3(x)^3 }}.  
\end{eqnarray}
The two polynomials  $\, P_3(x)$ and  $\, P_5(x)$ are given in \ref{polynomials}. 

\vskip .1cm 

\subsubsection{Cubic terms  subcase of (\ref{Ratfoncplusplusplus})\\}
\label{justtwosteps1}

\vskip .1cm 

A  simple subcase of (\ref{Ratfoncplusplusplus})
 corresponds to $\,\, b_1 \, = \, b_2 \, = b_3 \, = \, \, c_1 \, = \, c_2 \, = c_3 \, = \, 0$,
namely to the rational function:
\begin{eqnarray}
\label{anothersimple}
\hspace{-0.98in}&& \quad   \quad   \quad   \quad   \quad   \quad   \quad  
 R(x, \, y, \, z)  \, \, \, = \, \,  \,   \,  
 {{1} \over { a \,  \, \, 
+ d_1 \cdot \, x^2 \, y \, \, \,  +  d_2 \cdot\, y^2 \, z \,\,   +  d_3 \cdot \, z^2 \, x  }}, 
\nonumber 
\end{eqnarray}
whose diagonal reads
\begin{eqnarray}
\label{simpleThree}
\hspace{-0.98in}&& \quad  \, \, \quad   \quad  
 _2F_1\Bigl([{{1} \over {3}}, \, {{2} \over {3}}], \, [1], 
 \, \,  \,  \,  -27 \cdot \, {{d_1 \, d_2 \, d_3} \over {a^3}} \cdot \, x^3 \Bigr)
\\
\hspace{-0.98in}&& \quad  \, \, \quad \quad  \quad   \quad \,  \, \,  = \, \, \, \, 
\Bigl( 1 \, -216 \,  \cdot \, {{d_1 \, d_2 \, d_3} \over {a^3}} \cdot \, x^3 \Bigr)^{-1/4} 
\cdot \, 
 _2F_1\Bigl([{{1} \over {12}}, \, {{5} \over {12}}], \, [1], 
 \, \, 1 \, - \, {{P_6(x)^2} \over {P_3(x)^3}}\Bigr), 
\nonumber 
\end{eqnarray}
with:
\begin{eqnarray}
\label{simpleThreeP3P6}
\hspace{-0.98in}&& \quad  \, \, \, \,  \, \,  \quad  \quad   
P_3(x) \, \, = \, \,\, 
 -216 \cdot \, a \,\, d_1\, \, d_2\, \, d_3 \cdot \, x^3 \, \,\, +a^4,
\\
\label{simpleThreeP3P6bis}
\hspace{-0.98in}&& \quad  \, \,  \, \, \, \,  \quad  \quad 
P_6(x) \, \, = \, \,\, 
-\, 5832 \cdot  \, d_1^2 \, \, d_2^2\,\,  d_3^2 \cdot \, x^6
\,\, \, \,  +540 \cdot \, a^3\, \, d_1 \, d_2\, d_3 \cdot \, x^3\,\, \,  +a^6.
\end{eqnarray}
Relation (\ref{simpleThree}) actually corresponds to the hypergeometric identities:
\begin{eqnarray}
\label{hyperiden}
\hspace{-0.99in}&&  \quad    \, \,  \, \, 
 _2F_1\Bigl([{{1} \over {3}}, \, {{2} \over {3}}], \, [1],  \, -27 \, X \Bigr) 
 \\
\hspace{-0.99in}&& \quad \quad \quad  \, \, \,   = \, \, 
\Bigl( 1 \, -216  \, X  \Bigr)^{-1/4} \cdot \, 
 _2F_1\Bigl([{{1} \over {12}}, \, {{5} \over {12}}], \, [1],
 \,  - \, {{ 1728 \, X \cdot \, (1 \, + \, 27 \, X)^3  } \over {(1 \, -216\, X )^3 }} 
\Bigr)
\nonumber \\
\hspace{-0.99in}&& \quad \quad  \, \, \,  \,   \, \,  \, \,    \, \, = \, \,\,  
\Bigl( 1 \, -216 \,  \cdot \, X  \Bigr)^{-1/4} \cdot \, 
 _2F_1\Bigl([{{1} \over {12}}, \, {{5} \over {12}}], \, [1],
 \, \, 1 \, - \, {{ ( 1 \, +540\, X \, -5832\, X^2)^2 } \over {(1 \, -216\, X )^3 }} \Bigr).
\nonumber 
\end{eqnarray}

\vskip .1cm

\subsubsection{A symmetric subcase of (\ref{Ratfoncplusplusplus})\\}
\label{justtwosteps2}

Let us also consider another simple very symmetric subcase of (\ref{Ratfoncplusplusplus}). For 
 $\, b_1 \, = \, b_2 \, = \, b_3 \, = \, b$,   $\, c_1 \, = \, c_2 \, = \, c_3 \, = \, c$, 
 $\, d_1 \, = \, d_2 \, = \, d_3 \, = \, d$, the diagonal reads\footnote[2]{Note that trying 
to mix the two previous subcases imposing  
$\, b_1  = \, b_2  = \, b_3 \, = \, b$,   $\, \, c_1  = \, c_2  = \, c_3 = \, c \, $
with $\, d_1$, $\, d_2$ , $\, d_3$ no longer equal, do not yield 
a $\, _2F_1([1/3,2/3],[1],{\cal P}) \, \, $ hypergeometric function. }
\begin{eqnarray}
\label{simpleThreesym}
\hspace{-0.98in}&& \quad  \quad \quad \quad \quad \quad \quad  \, \, 
{{1} \over { a \, - 6 \, \, d  \cdot \,  x}} \cdot \, 
 _2F_1\Bigl([{{1} \over {3}}, \, {{2} \over {3}}], \, [1],  \, \,  \,  {\cal P}\Bigr),
\end{eqnarray}
where the pullback $\, {\cal P}$ reads:
\begin{eqnarray}
\label{simpleThreepull}
\hspace{-0.98in}&& \quad   
{\cal P} \, \, = \, \, \, 
- \, {{ 27 \, x \cdot \, \Bigl( a^2 \,  d \, - a\, b\, c \, +b^3
\, \, \, + (c^3 \, - 3 \, b \, c \,  d \, -3 \,a \,  d^2) \cdot \, x 
\,    \, + 9 \, \, d^3  \cdot \,  x^2 
\Bigr)  } \over { (a \,\,   - 6  \,\, d \, \cdot \,  x)^3}}.
\end{eqnarray}
At first sight the hypergeometric result (\ref{simpleThreesym}) 
with the pullback (\ref{simpleThreepull}) does not seem to 
be in agreement with the  hypergeometric result (\ref{simpleThree})
of section (\ref{justtwosteps1}). In fact these two results are in agreement
as a consequence of the hypergeometric identity:
\begin{eqnarray}
\label{hypergeomidentity}
\hspace{-0.98in}&& \quad   \, \,     \, \, 
 {{1} \over { 1 \, -6 \, X}}
  \cdot \, \,  _2F_1\Bigl([{{1} \over {3}}, \, {{2} \over {3}}], \, [1], \, \,  \,  \, 
 - {{ 27 \cdot \, X \cdot (1 \, -3 \, X \, + \, 9 \, X^2)} \over {(1 \, -6 \, X)^3 }} \Bigr)
\nonumber \\
\hspace{-0.98in}&& \quad  \quad \, \,   \, \,  \quad  \, \, \, \, \, \, = \, \, \, 
_2F_1\Bigl([{{1} \over {3}}, \, {{2} \over {3}}], \, [1], 
 \, \,  \,  \,  -27 \cdot \, X^3 \Bigr)
\quad \quad \, \,  \, \, \, \hbox{with:} \quad  \quad \, \, \, 
X \,\,  = \, \, \, {{d \, \cdot \, x} \over {a}}.
\end{eqnarray}
This hypergeometric result (\ref{simpleThree}) can also be rewritten 
in the form (\ref{2F15HypformAplusplusplus}) where 
the two polynomials $\, P_3(x)$ and $\, P_6(x)$ read respectively: 
\begin{eqnarray}
\label{simpleThreepullP3P6}
\hspace{-0.98in}&& \quad   
P_3(x) \, \, = \, \,  \, 
- \, 72 \cdot \, d \cdot \, (3\,a{d}^{2} \, -6\,bcd \, +2\,{c}^{3}) \cdot \, {x}^{3}
\, \, +24   \cdot \, ( 3\,abc \, d \, +a{c}^{3} \, -6\,{b}^{3} \, d) \cdot \,  {x}^{2} \, \, 
\nonumber \\ 
\hspace{-0.98in}&& \quad \quad \quad       \quad      \quad      \quad   
 -24 \cdot \,a \, b \cdot \, (ac \, -{b}^{2}) \cdot \,  x
    \, \, \, \,\, \, +{a}^{4}, 
\label{simpleThreepullP3P6bis}
\\
\hspace{-0.98in}&& \quad   
P_6(x) \, \, = \, \,  \, - \, 5832 \cdot \,{d}^{6} \cdot \, {x}^{6}
\, \, \, +3888 \cdot \,c \, {d}^{3} \cdot \,  (3\,b \, d \, -{c}^{2}) \cdot \,  {x}^{5}
\nonumber \\ 
\hspace{-0.98in}&& \quad \quad \quad \quad 
\, \, -216 \cdot \, (18\,abc \, {d}^{3} \,+18\,{b}^{3} \, {d}^{3}
 \, -12\,a{c}^{3}{d}^{2}  \, -9\,{b}^{2}{c}^{2} \, {d}^{2}
    \, \, +6\,b{c}^{4} \, d \, \, \, -{c}^{6}) \cdot \, {x}^{4}
\nonumber \\ 
\hspace{-0.98in}&& \quad \quad \quad \quad 
\, \, +108 \cdot \, (5\,{a}^{3} \, {d}^{3} 
\, \, -18\,{a}^{2}bc \, {d}^{2} 
  \,\,   -2\,{a}^{2}{c}^{3} \, d \, +12\,a{b}^{2}{c}^{2} \, d 
\,  \, +24\,a {b}^{3} \, {d}^{2} \, -4\,a \, b{c}^{4}
\nonumber \\
\hspace{-0.98in}&& \quad \quad \quad \quad \quad \quad \quad \quad \quad \quad 
\,  \, -12\,{b}^{4}c \, d  +4\,{b}^{3}{c}^{3}) \cdot \, {x}^{3}
\nonumber \\ 
\hspace{-0.98in}&& \quad \quad \quad \quad 
\, \,\, +36 \cdot \, ( 3\,{a}^{3}bc \, d \, -6\,{a}^{2}{b}^{3} \, d 
\, +{a}^{3}{c}^{3} +6\,{a}^{2}{b}^{2}{c}^{2} \, -12\,a{b}^{4}c+6\,{b}^{6}) \cdot \,  {x}^{2}
\nonumber \\ 
\hspace{-0.98in}&& \quad \quad \quad \quad \quad \quad \quad \quad \quad 
\, \, -36 \cdot \,{a}^{3} \, b  \cdot \, (ac \, -{b}^{2}) \cdot \,  x \,  \,  \, \, \, +{a}^{6}.
\end{eqnarray}

\vskip .1cm 
\vskip .1cm 

\section{Transformation symmetries of the diagonals of rational functions}
\label{moregener}
 
\vskip .1cm 

The previous results can be expanded through symmetry considerations. 

We are first going to see that performing 
monomial transformations on each of the previous (seven-parameter, 
eight, nine or ten-parameter) rational functions yields an 
{\em infinite number} of rational functions whose diagonals are 
pullbacked $\, _2F_1$ hypergeometric functions.
\subsection{$(x, \, y, \, z) \, \rightarrow \, \, (x^n, \, y^n, \, z^n) \, $ symmetries}
\label{xnynzn}
 
We have a first remark: once we have an exact result for a diagonal, 
we immediately get another diagonal by changing $(x, \, y, \, z)$
 into $(x^n, \, y^n, \, z^n)$
for any positive integer $\, n$ in the rational function. As a result
we obtain a new expression for the diagonal changing $\, x$ into $\, x^n$.

A simple example amounts to revisiting the fact that the diagonal of (\ref{diag3quart})
given below is the hypergeometric function (\ref{ident3quart}). Changing $(x, \, y, \, z)$
into  $(8\, x^2, \,8\, y^2, \, 8\,z^2)\, $ in (\ref{diag3quart}), 
one obtains  the pullbacked $\, _2F_1$ hypergeometric function
number 5 or 15 in Figure 10 of Bostan's HDR~\cite{HDRBostan} (see also~\cite{Rook,76,77})
\begin{eqnarray}
\label{ident3quartC}
\hspace{-0.98in}&&  \quad \quad \quad  \quad \quad \quad \quad \quad \quad \quad 
 _2F_1\Bigl([{{1} \over {4}}, \, {{3} \over {4}}],\, [1], \, 64 \, x^4),
\end{eqnarray}
can be seen as the diagonal of
\begin{eqnarray}
\label{diag3quart8}
\hspace{-0.96in}&& \,\, \quad \quad \quad \quad  \quad \,   \,   
{{1} \over { 2 \, \, \,   + 8 \, \sqrt{-1} \cdot \, (x^2+y^2+z^2) 
      \,\,    -\,64 \, x^2 \, z^2 \, \,  \,  -\, 32 \cdot x ^2\, y^2 }}, 
\end{eqnarray}
which is tantamount to saying that the transformation 
$ \, (x, \, y, \, z) \, \rightarrow \, (x^n, \, y^n, \, z^n) \, \, $
is a symmetry. 

\vskip .1cm 

\subsection{Monomial transformations on rational functions}
\label{monomial}

More generally, let us consider the monomial transformation
\begin{eqnarray}
\label{monomial}
 \hspace{-0.98in}&&  \,  \,  \, \quad \quad  \quad  \quad 
(x, \, y, \, z) \, \,\, \quad \,   \longrightarrow \,  \, \, \,\quad  \, 
M(x, \, y, \, z) \,\,\, = \, \,\,  (x_M, \, y_M, \, z_M) 
\nonumber \\
\hspace{-0.98in}&& \quad \quad \quad \quad \quad  \,  \,  \,\,  \,
\, \, \, \, = \, \, \, 
 \Bigl(x^{A_1} \cdot \, y^{A_2} \cdot \, z^{A_3}, 
\, \, \,   x^{B_1} \cdot \, y^{B_2} \cdot \, z^{B_3},
 \, \, \, x^{C_1} \cdot \, y^{C_2} \cdot \, z^{C_3} \Bigr), 
\end{eqnarray}
where the $\, A_i$'s,  $\, B_i$'s and   $\, C_i$'s are positive integers such that 
$\, A_1 \, = \, A_2 \, = \, A_3 \,$ is excluded (as well as 
 $\, B_1 \, = \, B_2 \, = \, B_3$ 
as well as  $\, C_1 \, = \, C_2 \, = \, C_3$),
and that the determinant of the $\, 3 \, \times \, 3$ matrix 
\begin{eqnarray}
\label{3x3}
\hspace{-0.95in}&& \,\quad \,\quad \, \quad \quad \quad \,\quad \,\quad   \, 
\quad \quad \quad 
\left[ \begin {array}{ccc} 
                  A_1&B_1&C_1 \\ 
\noalign{\medskip}A_2&B_2&C_2 \\ 
\noalign{\medskip}A_3&B_3&C_3 
\end {array} \right],  
\end{eqnarray}
is not equal to zero\footnote[9]{We want the rational function 
$\, \tilde{{\cal R}} \, = \,\,  \,{\cal R}(M(x, \, y, \, z))$ 
deduced from  the monomial transformation (\ref{monomial}) 
to remain a rational function of three variables
and not of two, or one, variables. }, and that:
\begin{eqnarray}
\label{defdiagBIScond}
\hspace{-0.90in}&&\quad \quad    \quad   \quad  \, \, \,\,
\, A_1\, +B_1\, +C_1 \,\, = \, \, \,\, A_2\, +B_2\, +C_2 \, \, = \, \,\,  \, A_3\, +B_3\, +C_3.
\end{eqnarray}
We will denote by $\, n$ the integer in these three equal\footnote[1]{For $\, n \, = \, 1$ 
the $\, 3 \times 3$ matrix (\ref{3x3}) is stochastic and  transformation (\ref{monomial}) 
is  a {\em birational  transformation}.} sums (\ref{defdiagBIScond}): 
$\, n \, = \, A_i\, +B_i\, +C_i$.
The condition (\ref{defdiagBIScond}) is introduced in order to impose that 
the product\footnote[5]{Recall that taking the diagonal of a rational function 
of three variables extracts, in the multi-Taylor expansion (\ref{defdiag}), only the terms 
that are $\, n$-th power of the {\em product}  $\, x\, y\, z$.} 
of $\, x_M\, y_M\, z_M$ is an integer power of the product of $\, x\, y\, z$:    
$\, \,x_M\, y_M\, z_M\, \, = \,\,\, (x\, y\, z)^n$.

If we take a  rational function  ${\cal R}(x, \, y, \, z)$ in  three variables 
and  perform a monomial transformation (\ref{monomial}) 
$\, (x, \, y, \, z) \, \rightarrow \, M(x, \, y, \, z)$, 
on the rational function $ \,{\cal R}(x, \, y, \, z)$, we 
get  another rational function  that we  denote by
$\, \tilde{{\cal R}} \, = \,\,  \,{\cal R}(M(x, \, y, \, z))$. 
Now the diagonal of $\, \tilde{{\cal R}}$ 
is the diagonal of  $ \,{\cal R}(x, \, y, \, z)$
where we have changed $\, x$ into $\, x^n$:
\begin{eqnarray}
\label{defdiag2BIS}
\hspace{-0.7in}&&\quad \quad 
\Phi(x) \, \, = \, \, \, \, Diag\Bigl({\cal R}\Bigl(x, \, y, \, z \Bigr)\Bigr),
 \quad \quad 
 Diag\Bigl(\tilde{{\cal R}}\Bigl(x, \, y, \, z \Bigr)\Bigr)
\,  \, = \, \,  \,  \Phi(x^n).
\end{eqnarray}
A demonstration of this result is sketched in \ref{monomialdiag}.

\vskip .1cm 

From the fact that the diagonal of the rational function 
\begin{eqnarray}
\label{monomialex}
\hspace{-0.95in}&&  \,  \quad \quad \quad  \quad  \quad \quad  \quad 
 {{1} \over { 1 \,\, \,  +x \, +y \, +z \,\,  + 3\cdot \, (x\, y\, + \, y\, z + \, x\, z)  }},
\end{eqnarray}
is the hypergeometric function
\begin{eqnarray}
\label{monomialex2F1}
\hspace{-0.95in}&&  \, \quad  \quad  \quad \quad  \quad    \quad  \quad \quad 
 _2F_1 \Bigl([{{1 }  \over {3 }}, \, {{ 2}  \over {3 }}], \, [1],
 \,\, \,  27 \, x \cdot \, (2\, -27\, x)  \Bigr),    
\end{eqnarray}
one deduces immediately that 
 the diagonal of the rational function (\ref{monomialex}) 
transformed by  the monomial transformation $  \, (x, \, y, \, z) \, \,  \longrightarrow \, \,   \,
 (z, \, \, \,   x^2\, y, \, \, \, y \, z) \,$
\begin{eqnarray}
\label{monomialex}
\hspace{-0.95in}&&  \,  \quad  \quad  \quad \quad  \quad   \,   
 {{1} \over { 1 \,\,  \, + y\, z \,\, +x^2\, y  
   \, \, + 3 \cdot \, ( y\, z^2 \, + \, x^2\, y \, z \, + \, x^2\, y^2 \, z)  }},
\end{eqnarray}
is the pullbacked hypergeometric function
\begin{eqnarray}
\label{monomialex2F1x2}
\hspace{-0.95in}&&  \,  \quad  \quad  \quad \quad  \quad  \quad  \quad   \quad  
 _2F_1 \Bigl([{{1 }  \over {3 }}, \, {{ 2}  \over {3 }}], \, [1], 
\, \,\,  27 \, x^2 \cdot \, (2\, -27\, x^2)  \Bigr),    
\end{eqnarray}
which is (\ref{monomialex2F1}) where $\,\, x \,\, \rightarrow \,\, x^2$. 

\vskip .1cm

To illustrate the point further, from the fact that the diagonal 
of the rational function 
\begin{eqnarray}
\label{monomialexmore}
\hspace{-0.95in}&&  \,  \quad \quad \quad  \quad  \quad \quad  \quad 
 {{1} \over {
 1 \,\, \,  +x \, +y \, +z \,\,  + \,3 \, x\, y\, + \,5 \, y\, z +\, 7 \, x\, z  }},
\end{eqnarray}
is the hypergeometric function
\begin{eqnarray}
\label{monomialex2F1nn}
\hspace{-0.98in}&&  \,  \, \,  \,  
{{1} \over { (2712\,\, x^2 \,\, -96\,\, x \,\, +1)^{1/4} }}
\\
\hspace{-0.98in}&&  \, \quad  \, \, \,  \,   \, 
\times \,  \, 
 _2F_1 \Bigl([{{1 }  \over {12 }}, \, {{ 5}  \over {12 }}], \, [1],
 \,\, \,   
 1\,  \, - {{(2381400\, x^4 \, -181440\, x^3 \, +7524\,x^2-144\, x+1)^2} \over {
 (2712\,\, x^2 \, -96\,\, x \,\, +1)^3}}  \Bigr),
\nonumber     
\end{eqnarray}
one deduces immediately that  the diagonal of the rational function 
 (\ref{monomialexmore}) transformed by the monomial transformation 
$\, (x, \, y, \, z) \, \,  \rightarrow \, \,  \, \,   \,
 (x\, z, \, \, \,   x^2\, y, \, \, \, y^2 \, z^2)$ 
\begin{eqnarray}
\label{monomialexMM}
\hspace{-0.95in}&&  \,  \quad  \quad  \quad \quad  \quad   \,   
 {{1} \over {
 1 \,\, \,  +x\, z \, +x^2\, y \, +y^2\, z^2 \,\, 
 + \,3 \, x^2\, y^3\, + \,5 \, x\, y^2\, z^3 +\, 7 \, x^3 \, y \, z  }},
\end{eqnarray}
is the  hypergeometric function
\begin{eqnarray}
\label{monomialex2F1next}
\hspace{-0.98in}&&  \,  \, \,  
{{1} \over { (2712\,\, x^6 \,\, -96\,\, x^3 \,\, +1)^{1/4} }}
\\
\hspace{-0.98in}&&  \, \quad  \, \, \,    
\times \,  \, 
 _2F_1 \Bigl([{{1 }  \over {12 }}, \, {{ 5}  \over {12 }}], \, [1],
 \,\, \, 
 1\, - {{(2381400\, x^{12} \, -181440\, x^{9} \, +7524\,x^6 \, -144\, x^3 \, +1)^2} \over {
 (2712\,\, x^6 \, -96\,\, x^3 \,\, +1)^3}}  \Bigr),
\nonumber     
\end{eqnarray}
which is nothing but (\ref{monomialex2F1nn}) where $\, x$ has been changed into $\, x^3$.

We have the same result for more involved rational functions
and more involved monomial transformations.

\vskip .1cm 
\vskip .2cm 

\subsection{More symmetries on diagonals}
\label{breath}

Other transformation symmetries of the diagonals 
include the function-dependent rescaling transformation
\begin{eqnarray}
\label{Scaling}
\hspace{-0.96in}&& \,   \quad  \quad    \quad \, 
 (x, \, y, \, z) \, \, \, \, \, \,  \longrightarrow \quad \quad
 \Bigl(F(x\,y\, z) \cdot \, x, \quad  \,F(x\,y\, z) \cdot \, y, \, \quad  F(x\,y\, z) \cdot \,z\Bigr),
\end{eqnarray}
where $\, F(x\,y\, z) \, $ is a rational function\footnote[1]{More generally one 
can imagine that  $\, F(x\,y\, z)$ is the series expansion of an algebraic function.} 
of the product of the three variables $ \, x$, $\, y$ and $\, z$. Under such a 
transformation the previous diagonal $\, \Delta(x)$ 
becomes $\, \Delta(x \cdot \,F(x)^3)$. 

For instance, changing
\begin{eqnarray}
\hspace{-0.96in}&& \,\, \,  \quad \quad \quad \, 
 (x, \, y, \, z) \, \, \, \,  \longrightarrow   \quad \quad
 \Bigl( {{x} \over { 1\, + \, 7\ x\, y \, z}}, \, \,\,  {{y} \over { 1\, + \, 7\ x\, y \, z}},
 \,  \,\,  {{z} \over { 1\, + \, 7\ x\, y \, z}} \Bigr),
\end{eqnarray}
the rational function 
\begin{eqnarray}
\label{Figure10aa}
\hspace{-0.96in}&& \,\, \quad \quad \quad \quad \quad \quad \quad \quad \quad 
{{1} \over { 1 \,\, \, -x \,\, -y \, -z \,\,\,  +\, y\, z}}, 
\end{eqnarray} 
whose diagonal is 
$\, _2F_1([1/2,\, 1/2], \, [1], \, 16 \, x)$,  
becomes the rational function
\begin{eqnarray}
\label{Figure10abecomes}
\hspace{-0.98in}&& 
{{(1\, + \, 7\ x\, y \, z )^2} \over { 
1  \,\, -x-y-z \,\,+y\,z\,  \, +14\,x\,y\,z   \, -7\,x^2\,y\,z \,-7\,x\,y^2\,z \,-7\,x\,y\,z^2 
\,  \, +49\, x^2\,y^2\,z^2 }}, 
\end{eqnarray} 
which has the following diagonal:
\begin{eqnarray}
\label{Figure10abecomesdiag}
\hspace{-0.96in}&& \,\,  \quad 
_2F_1\Bigl([{{1} \over {2}}, \, {{1} \over {2}}], \, [1],
 \, {{16  \, x } \over {(1\, + \, 7\ x)^3 }}   \Bigr)
 \, \,  = \, \, \, \, \, 
  1 \, \, \,\, +4 \,x \,\, \,-48 \,x^2\, \, \,+64 \,x^3 \,\, +3024 \,x^4
\nonumber \\
\hspace{-0.96in}&& \,\,  \quad  \quad \quad  \quad
\,\, -13524 \,x^5  \,-245196 \,x^6 
\,+1933152 \,x^7 \,+21288192 \,x^8\, \,-263440460 \,x^9 \,
\nonumber \\
\hspace{-0.96in}&& \,\,  \quad  \quad \quad  \quad \quad  \quad \quad  \quad
-1758664568\, x^{10} \, \,+34575759792\, x^{11} 
 \,\, \,\, +  \, \, \, \cdots 
\end{eqnarray}  
To illustrate the point further take
\begin{eqnarray}
\label{TTT}
\hspace{-0.98in}&& \,\,    \quad  \quad  \, 
 (x, \, y, \, z) \, \, \, \,  \longrightarrow   \,  \quad
 \Bigl( x \cdot \, F, \, \,\,  y \cdot \, F, \,  \,\,  z \cdot \, F\Bigr), 
  \quad  \quad  \quad  \quad  \quad  \,\,  \hbox{with:} 
\\
\label{FFF}
\hspace{-0.98in}&& \quad  \quad  \quad  \quad  \quad  \quad \quad  \, 
F \,\, = \, \,
 {{ 1\, + \, 2\, x\, y \, z} \over {
 1\, + \, 3\, x\, y \, z \, + \, 5 \, x^2\, y^2 \, z^2 }}
 \, \, \, = \,\,  \, \Phi(x\, y\, z), 
 \\
\label{PPP}
\hspace{-0.96in}&& \,\,    \quad \quad 
  \hbox{where:}   \quad  \quad  \quad \quad  \quad  \quad
\Phi(x) \,\,  = \, \, {{ 1\, + \, 2\,x} \over { 1\, + \, 3\, x\, + \, 5 \, x^2}}, 
\end{eqnarray}
the rational function 
\begin{eqnarray}
\label{Figure10aagg}
\hspace{-0.96in}&& \,\, \quad \quad \quad \quad \quad \quad \quad \quad \quad 
{{1} \over {
 1 \,\, \, +x \,\, +y \, +z \,\,\,  +\, y\, z +\, x\, z \, + \, x\,y }},  
\end{eqnarray} 
whose diagonal is 
$\, _2F_1([1/3,\, 2/3], \, [1], \, -27 \, x^2)$,  
becomes the rational function $\, P(x, \, y, \, z)/Q(x, \, y, \, z)$, 
where the numerator $\, P(x, \, y, \, z)$ and the denominator $\, Q(x, \, y, \, z)$, 
read respectively: 
\begin{eqnarray}
\label{Figure10aann}
\hspace{-0.98in}&&   \quad  \quad  
P(x, \, y, \, z)  \, \, = \, \,  \,
 (1\, + \, 3\, x\, y \, z \, + \, 5 \, x^2\, y^2 \, z^2)^2,  
\\
\hspace{-0.98in}&&  \quad   \quad 
Q(x, \, y, \, z)  \, \, = \, \,\,  \, 25\,{x}^{4}{y}^{4}{z}^{4} \, \,
+10 \cdot \,({x}^{4}{y}^{3}{z}^{3}+\,{x}^{3}{y}^{4}{z}^{3}+\,{x}^{3}{y}^{3}{z}^{4}) \, 
+30\,{x}^{3}{y}^{3}{z}^{3} 
\nonumber \\
\hspace{-0.98in}&&  \quad   \quad  \quad  \quad 
\,\, \,\, 
  +4 \cdot \,({x}^{3}{y}^{3}{z}^{2} \, +\,{x}^{3}{y}^{2}{z}^{3}\, +\,{x}^{2}{y}^{3}{z}^{3})
\,\, +11 \cdot \,({x}^{3}{y}^{2}{z}^{2} \, +\,{x}^{2}{y}^{3}{z}^{2} \, +\,{x}^{2}{y}^{2}{z}^{3})
\nonumber \\
\hspace{-0.98in}&&  \quad   \quad  \quad  \quad 
\, \,\, \, +19\,{x}^{2}{y}^{2}{z}^{2} \,
+4 \cdot \,({x}^{2}{y}^{2}z \, +\,{x}^{2}y{z}^{2} \, +\,x{y}^{2}{z}^{2}) \, \,
+5 \cdot \,({x}^{2}yz \, +\,x{y}^{2}z \, +\,xy{z}^{2})
\nonumber \\
\hspace{-0.98in}&&  \quad   \quad  \quad  \quad 
\,\,\,  \, +6\,xyz \, \,  \, 
+xy+xz+yz \,  \, \,+x+y+z \,\,  \, +1. 
\end{eqnarray} 
The diagonal of this last rational function is equal to:
\begin{eqnarray}
\label{Figure10aann}
\hspace{-0.98in}&&   \quad  \quad  \quad   \quad    \quad  \,\,
 _2F_1\Bigl([ {{1} \over {3}}, \,{{2} \over {3}}], \, [1], 
\, \, \,  -27 \cdot \, \Bigl(x \cdot \, \Phi(x)^3  \Bigr)^2\Bigr)
\nonumber \\
\hspace{-0.98in}&&  \quad   \quad \quad  \quad  \quad  \quad  \quad    \quad 
 \,  \,\,\, = \, \,  \, \,
 _2F_1\Bigl([ {{1} \over {3}}, \,{{2} \over {3}}], \, [1], \, \, \,
  -\, 27 \,  x^2 \cdot \, \Bigl({{  1\, + \, 2\, x} \over {1\, + \, 3\, x \, + \, 5 \, x^2 }}\Bigr)^6 \Bigr). 
\end{eqnarray} 

Let us give a final example: let us consider again the rational function (\ref{monomialexmore})
whose diagonal is (\ref{monomialex2F1nn}), and let us consider the 
same function-rescaling transformation (\ref{TTT}) with (\ref{FFF}). One finds that the 
diagonal of the rational function 
\begin{eqnarray}
\label{monomialexmorefinal}
\hspace{-0.95in}&&  \,  \,  \,  \,  \,  \,  \, 
 {{1} \over {
 1 \,\, \,  + F \cdot \, x \, +F \cdot \, y \, + F \cdot \, z 
   \,\,  \, + \,3  \cdot \, F^2 \cdot \, x\, y\, 
+ \, 5 \cdot \,F^2 \cdot \,  y\, z +\, 7  \cdot \, F^2 \cdot \,  x\, z  }},
\end{eqnarray}
is the hypergeometric function
\begin{eqnarray}
\label{monomialex2F1final}
\hspace{-0.98in}&&  \,  \, \quad \quad 
{{1} \over { (2712\,\, x^2 \, \Phi(x)^6 \,\, -96\,\, x \,  \, \Phi(x)^3 \, +1)^{1/4} }}  \,  \times \,\, 
 _2F_1 \Bigl([{{1 }  \over {12 }}, \, {{ 5}  \over {12 }}], \, [1], 1 \, - {\cal H} \Bigr),  
\end{eqnarray}
where the pullback $\,\, 1 \, -{\cal H} \, $ reads: 
\begin{eqnarray}
\label{monomialex2F1finalH}
\hspace{-0.99in}&&  \,  \, 
 1\, \, - {{(2381400\, x^4 \, \Phi(x)^{12} \, -181440\, x^3 \, \Phi(x)^9
 \, +7524\,x^2  \, \Phi(x)^6 \, -144\, x  \, \Phi(x)^3 \, +1)^2} \over {
 (2712\,\, x^2  \, \Phi(x)^6 \, -96\,\, x  \, \Phi(x)^3 \, \, +1)^3}}  \Bigr).
\nonumber     
\end{eqnarray}
The pullbacked hypergeometric function  (\ref{monomialex2F1finalH}) 
is nothing but (\ref{monomialex2F1nn}) where $\, x$ has been changed 
into $\, x \, \Phi(x)^3$.

\vskip .1cm 

A demonstration of these results is sketched in \ref{rescalingdiag}. 

\vskip .1cm 

Thus for each rational function belonging to one of the seven, eight, nine or ten parameters
families of rational functions yielding a pullbacked $\, _2F_1$ hypergeometric function 
one can deduce from the  transformations (\ref{Scaling})
an {\em infinite number} of other rational functions,  with denominators of degree much
 {\em higher} than two or three.

\vskip .1cm 
\vskip .1cm 

One can combine these two sets of transformations, the monomial transformations (\ref{monomial}) 
and the function-dependent rescaling  transformations (\ref{Scaling}), thus yielding 
from each of the (seven, eight, nine or ten parameters) rational functions 
of the paper an {\em infinite number}  of rational functions of quite high degree 
yielding  pullbacked $\, _2F_1$ hypergeometric (modular form) exact results 
for their diagonals. 

\vskip .3cm 

\section{Conclusion}
\label{conclus}
 
\vskip .1cm 
\vskip .1cm 

We found here that a seven-parameter rational function of 
three variables  with a numerator equal to one 
and a of polynomial denominator of degree two at most, can be expressed as a pullbacked 
$\, _2F_1$ hypergeometric function. We generalized that result to eight, then nine and ten 
parameters, by adding specific cubic terms.
We focused on subcases where 
the diagonals of the corresponding rational functions are
pullbacked $\, _2F_1$ hypergeometric function with two possible 
rational function pullbacks algebraically related by {\em modular equations}, 
thus obtaining the result that the diagonal is a {\em modular form}\footnote[1]{Differently
from the usual definition of modular forms in the $\, \tau$ variables.}. 

\vskip .1cm 

We have finally seen that simple monomial transformations, as well as a simple function rescaling
of the three (resp. $N$) variables, are symmetries of the diagonals of 
rational functions of  three (resp. $N$) variables. Consequently each of our previous families of
rational functions, once transformed by these symmetries yield an {\em infinite number} 
of families of rational functions of  three variables (of higher degree) whose diagonals are also
pullbacked $\, _2F_1$ hypergeometric functions and, in fact, modular forms.

Since diagonals of rational functions emerge naturally in integrable lattice 
statistical mechanics and enumerative combinatorics,
exploring the kind of exact results we obtain 
for diagonals of rational functions (modular forms, Calabi-Yau operators, 
pullbacked $\, _nF_{n-1}$ hypergeometric functions, ...) is an important 
systematic work to be performed to provide results and tools in integrable 
lattice statistical mechanics and enumerative combinatorics.

\vskip .3cm 
\vskip .3cm 

{\bf Acknowledgments.}  
J-M. M. would like to thank G. Christol for many enlightening 
discussions on diagonals of rational functions. 
S. B. would like to thank the LPTMC and the CNRS for kind support. 
We would like to thank  A. Bostan 
for useful discussions on creative telescoping.
C.K. was supported by the Austrian Science Fund (FWF): P29467-N32.
Y. A. was supported by the Austrian Science Fund (FWF): F5011-N15.
We thank the Research Institute for Symbolic Computation,
 for access to the RISC software packages.  We thank M. Quaggetto 
for technical support. 
This work has been performed 
without any support of the ANR, the ERC or the MAE, or any PES of the CNRS. 

\vskip .1cm 

\vskip .3cm 
\vskip .2cm 

\appendix

\section{Simple symmetries of the diagonal of the rational function (\ref{Ratfonc}) }
\label{Simplesymapp}

Let us recall the  pullbacks (\ref{miscell}) in section (\ref{Simplesym}),
that we denote  $\, {\cal P}_1$.

\vskip .1cm 

\subsection{Overall parameter symmetry }
\label{Simplesymoverall}

The seven parameters are defined up to an overall parameter 
(they must be seen as homogeneous variables).
Changing $(a, \, b_1, \,  b_2, \,  b_3, \, c_1, \,  c_2, \,  c_3)$ into
 $(\lambda \cdot \, a, \, \lambda \cdot \, b_1, \,  \lambda \cdot \,b_2,$
$ \,  \lambda \cdot \,b_3, \, \lambda \cdot \,c_1, \,  \lambda \cdot \,c_2, \,  \lambda \cdot \,c_3)$
the rational function $\, R$ given by (\ref{Ratfonc}) and its diagonal $\, Diag(R)$ 
are changed into $\, R/\lambda \,$ and $\, Diag(R)/\lambda$. It is thus clear that 
the previous pullbacks (\ref{miscell}), which totally ``encode'' the exact expression
of the diagonal as a pullbacked hypergeometric function, must be invariant 
under this transformation. This is actually the case:
\begin{eqnarray}
\label{symmP1}
\hspace{-0.7in}&&\quad \quad \, \quad \,
{\cal P}_1(\lambda \cdot \, a, \, \, \lambda \cdot \, b_1, \, \,  \lambda \cdot \,b_2,\, 
 \,  \lambda \cdot \,b_3, \, \, \lambda \cdot \,c_1, \,\, 
  \lambda \cdot \,c_2, \,\,   \lambda \cdot \,c_3, \,\,  \, x)
\nonumber \\
\hspace{-0.7in}&&\quad \quad \quad \quad \quad \quad 
 \, \, \, \, \, = \, \, \, \,
{\cal P}_1(a, \,\, b_1, \,\,  b_2, \,\,  b_3, \,\, c_1, \, \, c_2, \,\,  c_3, \, \,\, \, x).
\end{eqnarray}
This result corresponds to the fact that $\, P_2(x)$ (resp. $P_4(x)$) is a 
{\em homogeneous polynomial} in the seven 
parameters $ \,a, \, b_1, \cdots, c_1, \cdots$ of degree two (resp. four ). 

\subsection{Variable rescaling symmetry }
\label{Simplesymrescal}

On the other hand, the rescaling of the three variables $(x, \, y, \, z)$ in (\ref{Ratfonc}),  
$(x, \, y, \, z) \, \rightarrow \, \, $
$(\lambda_1 \cdot x, \,\, \lambda_2 \cdot y, \,\, \lambda_3 \cdot z) \, \, $
is a change of variables that is compatible with the operation 
of taking the diagonal of the rational function $\, R$.

When taking the diagonal and performing this change of variables, the monomials 
in the multi-Taylor expansion of  (\ref{Ratfonc}) transform as:
\begin{eqnarray}
\label{amnp}
\hspace{-0.7in}&&\quad \quad \quad
a_{m, \, n, \, p} \cdot \, x^m \, y^n \, z^p \, \,\quad  \longrightarrow \, \, \,\,  \quad
 a_{m, \, n, \, p} \cdot \, \lambda_1^{m}  \cdot \, \lambda_2^{n} 
 \cdot \, \lambda_3^{p}  \cdot \, x^m \, y^n \, z^p.
\end{eqnarray}
Taking the diagonal yields 
\begin{eqnarray}
\label{diagamnp}
\hspace{-0.7in}&&\quad \quad \quad \quad \, \,\,  
a_{m, \, m, \, m} \cdot \, x^m  \, \,\quad  \longrightarrow \, \, \, \,  \quad
  a_{m, \, m, \, m}   \cdot \, (\lambda_1\, \lambda_2 \, \lambda_3)^{m} \cdot \, x^m.
\end{eqnarray}
Therefore it amounts to changing 
$\, x \, \rightarrow \, \, \lambda_1\, \lambda_2 \, \lambda_3 \cdot \, x$.
With that  rescaling 
$(x, \, y, \, z) \, \rightarrow $
$\, \, (\lambda_1 \cdot x, \,\, \lambda_2 \cdot y, \,\, \lambda_3 \cdot z) \, \, $
the diagonal of the rational function remains invariant if one changes the seven
parameters as follows:
\begin{eqnarray}
\label{asfollows}
\hspace{-0.7in}&&\quad   
(a, \, \,\, b_1, \,  b_2, \,  b_3, \, \, \,c_1, \,  c_2, \,  c_3)
 \quad  \quad \longrightarrow 
\nonumber \\
\hspace{-0.7in}&& \quad  \quad  \quad \quad 
(a, \,  \,\lambda_1 \cdot \, b_1, \,  \,  \lambda_2 \cdot \,b_2,
 \,  \,  \lambda_3 \cdot \,b_3, \, \, \lambda_2 \, \lambda_3  \cdot \, c_1,
 \,  \,  \lambda_1 \, \lambda_3  \cdot \,c_2, 
\, \,   \lambda_1 \, \lambda_2  \cdot \,c_3). 
\end{eqnarray}
One deduces that the pullbacks (\ref{miscell})  verify: 
\begin{eqnarray}
\label{symmP2}
\hspace{-0.98in}&& \, \, \,\, \,\, \quad  \quad 
{\cal P}_1\Bigl(a, \, \, \,   \,
\lambda_1 \cdot \, b_1, \,  \,  \lambda_2 \cdot \,b_2, \,  \,  \lambda_3 \cdot \,b_3,
\,  \,\, \, \lambda_2 \, \lambda_3  \cdot \, c_1, \,\,    \lambda_1 \, \lambda_3  \cdot \,c_2, 
\, \,   \lambda_1 \, \lambda_2  \cdot \,c_3
, \,\,   \, \, {{x } \over { \lambda_1 \,  \lambda_2 \,  \lambda_3 }}\Bigr)
\nonumber \\
\hspace{-0.98in}&&\quad \quad \quad \quad \quad \quad 
 \, \, = \, \, \, \,\,
{\cal P}_1(a, \, \, b_1, \, \,  b_2, \,\,   b_3, \, \,\,  c_1, \,  c_2, \,  c_3, \,  \,\, x).
\end{eqnarray}

\vskip .1cm 

\section{Comment on $\, _2F_1([1/3,2/3],[1],{\cal P})$  as a modular form}
\label{comment}

From identity (\ref{identity1z2}) of section (\ref{threesymm}) 
\begin{eqnarray}
\label{identity1zapp}
\hspace{-0.95in}&&    \,  \,  \quad   \quad \, 
 _2F_1\Bigl([{{1} \over {3}}, \, {{2} \over {3}}], \, [1], \,{{z} \over {z\, +27}} \Bigr) 
 \\ 
\label{identity1z2app}
\hspace{-0.95in}&&    \, \,  \quad  \quad  \quad  \quad 
\, = \, \, \, \,
 \Bigl( 9 \cdot \,\Bigl( {{ z\, +27 } \over { z\, +243 }}  \Bigr) \Bigr)^{1/4} \cdot \,
  _2F_1\Bigl([{{1} \over {12}}, \, {{5} \over {12}}],
 \, {{1728 \, z^3} \over { (z \, +27) \cdot \, (  z\, +243)^3 }}\Bigr)
 \\ 
\label{identity1z2bisapp}
\hspace{-0.95in}&&    \, \,  \quad  \quad  \quad  \quad 
\, = \, \, \, \,
 \Bigl( {{1} \over {9}} \cdot \,\Bigl( {{ z\, +27 } \over { z\, +3 }}  \Bigr) \Bigr)^{1/4} \cdot \,
  _2F_1\Bigl([{{1} \over {12}}, \, {{5} \over {12}}],
 \, {{1728 \, z} \over { (z \, +27) \cdot \, (  z\, +3)^3 }}\Bigr),
\end{eqnarray}
it is tempting to imagine an identity relating  $\, _2F_1([1/3,2/3],[1],{\cal P})$
with two different pullbacks.

Since switching the last two Hauptmoduls in (\ref{identity1z2app}) and (\ref{identity1z2bisapp}) 
amounts to performing the involutive transformation 
$\, z \, \rightarrow \, 729/z$, it is tempting to imagine that the first $\, _2F_1$
hypergeometric function (\ref{identity1zapp}) is related to itself with 
$\, z \, \rightarrow \, 729/z$, namely that
\begin{eqnarray}
\label{729}
\hspace{-0.95in}&&    \,  \quad   \quad  \quad  \quad 
 _2F_1\Bigl([{{1} \over {3}}, \, {{2} \over {3}}], \, [1], \,{{27} \over {z\, +27}} \Bigr) 
\,\, = \, \, \, \,
 _2F_1\Bigl([{{1} \over {3}}, \, {{2} \over {3}}], \, [1], \, 1 \,-\,{{z} \over {z\, +27}} \Bigr), 
\end{eqnarray}
is related to\footnote[2]{It corresponds to a trivial pullback change 
$\, p \, = \, z/(z+27) \, \rightarrow \, 1 \, -p$.} 
\begin{eqnarray}
\label{729bis}
\hspace{-0.95in}&&    \,  \quad   \quad  \quad  \quad 
 _2F_1\Bigl([{{1} \over {3}}, \, {{2} \over {3}}], \, [1], \,{{27} \over {729/z\, +27}} \Bigr) 
\,\, = \, \, \, \,
 _2F_1\Bigl([{{1} \over {3}}, \, {{2} \over {3}}], \, [1], \,\, \,{{z} \over {z\, +27}} \Bigr). 
\end{eqnarray}
This is the case, since (\ref{729}) and (\ref{729bis}) are solutions of the same linear ODE, 
but this does not mean that one can 
deduce an identity on the different pullbacks (\ref{729}) and (\ref{729bis}): the relation between 
these two hypergeometric functions (\ref{729}) and (\ref{729bis}) corresponds to 
a connection matrix~\cite{ze-bo-ha-ma-05c}.  A direct identity 
on $\, _2F_1([1/3,2/3],[1],{\cal P})$,  does however exist:
\begin{eqnarray}
\label{doesexist}
\hspace{-0.95in}&&    \,  \quad   \quad    \quad   \quad   \quad 
 _2F_1([{{1} \over {3}}, \, {{2} \over {3}}], \, [1], \, x^3)
\nonumber \\
\hspace{-0.95in}&&    \,  \quad  \quad   \quad  \quad    \quad    \quad  \quad 
 \, \, =  \,\, \, {{1} \over { 1\, + \, 2 \, x}} \cdot \, 
_2F_1\Bigl([{{1} \over {3}}, \, {{2} \over {3}}], \, [1], 
\,  \, {{ 9 \, x \cdot \, (1\, + \, x \, +\, x^2)} \over {  (1\, + \, 2 \, x)^3 }} \Bigr).
\end{eqnarray}
Identity (\ref{doesexist}) corresponds\footnote[5]{Using the relation (\ref{identity13}).} 
to the identity on $\, _2F_1([1/3,1/3],[1],{\cal P})$:
\begin{eqnarray}
\label{doesexist2}
\hspace{-0.95in}&&    \,  \quad   \quad  \quad  
 _2F_1\Bigl([{{1} \over {3}}, \, {{1} \over {3}}], \, [1], \, -\, {{x^3}  \over { 1\, -x^3}} \Bigr)
\nonumber \\
\hspace{-0.95in}&&    \,  \quad   \quad    \quad  \quad 
 \, \, =  \,\, \, \Bigl({{1 \, +x \, +x^2} \over { (1\, - \, x)^2 }} \Bigr)^{1/3} \cdot \, 
_2F_1\Bigl([{{1} \over {3}}, \, {{1} \over {3}}], \, [1],
 \,  \, - \, {{ 9 \, x \cdot \, (1\, + \, x \, +\, x^2)} \over {  (1\, - \, x)^3 }} \Bigr).
\end{eqnarray}

 \vskip .1cm 

\section{Comments on the $\, \tau \, \rightarrow \, 4 \, \tau$ modular equation (\ref{mod4}).}
\label{IntheZ}

The fact that in section (\ref{threeasymm}), the three Hauptmoduls (\ref{4tau}), 
(\ref{4tausecond}) and (\ref{4tausecondbis})
can be introduced  for the $\, \tau \, \rightarrow \, \, 4 \, \tau$ 
modular equation (\ref{mod4}), can be revisited in the $\, z$ variable.
 (see equation (\ref{4tauHaupt}). 
Recalling $\, {\cal P}_1 $ and $\, {\cal P}_2$  given in (\ref{4tauHaupt}),
and performing the (involutive) change of variable 
$\, z \, \rightarrow \, -16\, z/(z+16)$, on  $\, {\cal P}_2$,  
we get a third Hauptmodul  $\, {\cal P}_3$ 
\begin{eqnarray}
\label{4tauHauptthird}
\hspace{-0.96in}&&  \quad   \quad   \quad     \,   \,
{\cal P}_3  \,  = \, \,
 -\, {{1728 \cdot z \cdot \, (z\, +16)^4 } \over { (z^2\, - \, 224 \, z \, + 256)^3}}
 \,  \, = \, \,   \, 
\Bigl({{1728 \cdot z} \over {(z\, +16)^3}}\Bigr) \circ \,  
\Bigl(-\, {{4096 \, z } \over {(z+16)^2 }}\Bigr),
\end{eqnarray}
to be compared\footnote[1]{As it should
 $\, z \, \rightarrow \, -16\, z/(z+16)\, $  changes 
$\, z \cdot \, (z+16)$ into $\, -4096 \, z/ (z+16)^2$.} with:
\begin{eqnarray}
\label{4tauHauptthird2}
\hspace{-0.96in}&&  \quad   \quad   \quad    \, \,
{\cal P}_2  \,  = \, \,
 \, {{1728 \cdot z \cdot \, (z\, +16) } \over { (z^2\, +16 \, z \, + 16)^3}}
 \, \,  = \, \, \,
  \Bigl({{1728 \cdot z} \over {(z\, +16)^3}}\Bigr)  \circ \, \Bigl(z \cdot \, (z+16)\Bigr).
\end{eqnarray}
One also has: 
\begin{eqnarray}
\label{4tauHauptthird3}
\hspace{-0.96in}&&  \quad   \quad   \quad    \, \,
{\cal P}_1  \,  = \, \,
 \, {{1728 \cdot z^4 \cdot \, (z\, +16) } \over { (z^2\, +256 \, z \, + 4096)^3}}
 \, \,  = \, \, \,
  \Bigl({{1728 \cdot z} \over {(z\, +16)^3 }}\Bigr)  \circ \,
 \Bigl({{ 4096 \cdot \, (z+16)} \over {z^2 }}\Bigr)
\\
\label{4tauHauptthird3B}
\hspace{-0.96in}&&  \quad   \quad  \quad   \quad  \quad     \quad   
\, \,  = \, \, \,    \Bigl({{1728 \cdot z^2} \over {(z\, +256)^3 }}\Bigr)  \circ \,
 \Bigl( {{z^2} \over {z\, +16}} \Bigr).
\end{eqnarray}
These three Hauptmoduls have to be compared with the Hauptmodul:
\begin{eqnarray}
\label{4tauHauptthird3last}
\hspace{-0.96in}&&  \quad   \quad   \quad  \quad    \, \,
{\cal P}_0  \,  = \, \,
 \, {{1728 \cdot z^2 \cdot \, (z\, +16)^2 } \over { (z^2\, +16 \, z \, + 256)^3}}
 \, \,  = \, \, \,
  \Bigl({{1728 \cdot z} \over {(z\, +16)^3 }}\Bigr)  \circ \,
 \Bigl({{ z^2} \over { z \, +16 }}\Bigr)
\\
\label{4tauHauptthird3lastB}
\hspace{-0.96in}&&  \quad   \quad   \quad  \quad  \quad   \quad    \quad   
\, \,  = \, \, \,    \Bigl({{1728 \cdot z^2} \over {(z\, +256)^3 }}\Bigr)  \circ \,
 \Bigl(z \cdot  \, (z\, +16)\Bigr).
\end{eqnarray}
Note that the elimination of $\, z$, between this last Hauptmodul $\, {\cal P}_0$
{\em and each of the three  Hauptmoduls} $\, {\cal P}_1$, $\, {\cal P}_2$ , $\, {\cal P}_3$, 
gives the $\, \tau \, \rightarrow \, 2 \, \tau$ 
modular equation (see (\ref{modularcurveapp}) below) 
instead of the  $\, \tau \, \rightarrow \, 4 \, \tau$ modular equation (\ref{mod4}).

The decomposition of the Hauptmoduls $\, {\cal P}_3$, $\, {\cal P}_2$ and  $\, {\cal P}_1$ 
given by  (\ref{4tauHauptthird}), (\ref{4tauHauptthird2}) 
 and  (\ref{4tauHauptthird3}) suggests 
to substitute 
\begin{eqnarray}
\label{substitute}
\hspace{-0.96in}&&  \quad   \quad   \quad \quad   \quad    \quad    \quad    \, \, 
X \, = \, {{1728 \tilde{X}} \over { (\tilde{X}\, +16)^3}}, 
\quad \quad \quad   \quad 
Y \, = \, {{1728 \tilde{Y}} \over { (\tilde{Y}\, +16)^3}}, 
\end{eqnarray}
in the $\, \tau \, \rightarrow \, 4 \, \tau$ modular equation (\ref{mod4}).
This change of variable transforms  the LHS of the modular equation (\ref{mod4})
into the product of  four polynomials: 
\begin{eqnarray}
\label{pol1}
\hspace{-0.96in}&&   
  {\hat p}_1(\tilde{X},\, \tilde{Y}) \, = \, \,\,\,\,
  \tilde{X}^2\, \tilde{Y}^2 \,\, \,\, -3 \cdot \, 2^{16} \cdot \, \tilde{X} \,\tilde{Y}
 \,\,\, \,-2^{24} \cdot \, (\tilde{X} +\tilde{Y}), 
\end{eqnarray}
\begin{eqnarray}
\label{pol2}
\hspace{-0.96in}&&  
  {\hat p}_{2,1}(\tilde{X},\, \tilde{Y})
 \, = \, \,  \, \, {\hat p}_{2,2}(\tilde{Y},\, \tilde{X}) 
\,  \, = \, \, \,\,
\tilde{X}^4\, \tilde{Y}^3 \,\,\, +96\, \tilde{X}^3\, \tilde{Y}^3
\, +196608\, \tilde{X}^3\, \tilde{Y}^2 \, \,
+2352\, \tilde{X}^2\, \tilde{Y}^3
\nonumber \\
 \hspace{-0.95in}&&   \, \quad \quad \,  \, 
+16777216\, \tilde{X}^3\, \tilde{Y} \, -7335936\, \tilde{X}^2\, \tilde{Y}^2 \,
 +10496\, \tilde{X}\, \tilde{Y}^3-\tilde{Y}^4 \, 
+805306368\, \tilde{X}^2\, \tilde{Y}
\nonumber \\
 \hspace{-0.95in}&&   \, \quad \quad \,  \, 
+9633792\, \tilde{X}\, \tilde{Y}^2 \, +1610612736\, \tilde{X}\, \tilde{Y} \, +68719476736\, \tilde{X}, 
\end{eqnarray}
\begin{eqnarray}
\label{pol4}
\hspace{-0.96in}&&  
  {\hat p}_4(\tilde{X},\, \tilde{Y}) \, = \, \,\tilde{X}^7\, \tilde{Y}^5 \, 
+\tilde{X}^5\, \tilde{Y}^7 +96\, \tilde{X}^7\, \tilde{Y}^4+144\, \tilde{X}^6\, \tilde{Y}^5
+144\, \tilde{X}^5\, \tilde{Y}^6 \, 
+96\, \tilde{X}^4\, \tilde{Y}^7
\nonumber \\
 \hspace{-0.95in}&&   \, \quad
 \, +2352\, \tilde{X}^7\, \tilde{Y}^3-182784\, \tilde{X}^6\, \tilde{Y}^4
+13968\, \tilde{X}^5\, \tilde{Y}^5-182784\, \tilde{X}^4\, \tilde{Y}^6 
\, +2352\, \tilde{X}^3\, \tilde{Y}^7
\nonumber \\
 \hspace{-0.95in}&&   \, \quad \, 
 +\tilde{X}^8\, \tilde{Y}+10496\, \tilde{X}^7\, \tilde{Y}^2
+7674625\, \tilde{X}^6\, \tilde{Y}^3-1300992\, \tilde{X}^5\, \tilde{Y}^4 
-1300992\, \tilde{X}^4\, \tilde{Y}^5
\nonumber \\
 \hspace{-0.95in}&&   \, \quad \,
+7674625\, \tilde{X}^3\, \tilde{Y}^6+10496\,
+7674625\, \tilde{X}^3\, \tilde{Y}^6+10496\, \tilde{X}^2\, \tilde{Y}^7 
+\tilde{X}\, \tilde{Y}^8+192\, \tilde{X}^7\, \tilde{Y}
\nonumber \\
 \hspace{-0.95in}&&   \, \quad \,-8122320\, \tilde{X}^6\, \tilde{Y}^2 
\, +1526542992\, \tilde{X}^5\, \tilde{Y}^3
+700465152\, \tilde{X}^4\, \tilde{Y}^4 \, +1526542992\, \tilde{X}^3\, \tilde{Y}^5 
\nonumber \\
 \hspace{-0.95in}&&   \, \quad
\, -8122320\, \tilde{X}^2\, \tilde{Y}^6+192\, \tilde{X}\, \tilde{Y}^7
 \,+13920\, \tilde{X}^6\, \tilde{Y}
+759331584\, \tilde{X}^5\, \tilde{Y}^2
\nonumber \\
 \hspace{-0.95in}&&   \, \quad \,+56157592368\, \tilde{X}^4\, \tilde{Y}^3
 \, +56157592368\, \tilde{X}^3\, \tilde{Y}^4
+759331584\, \tilde{X}^2\, \tilde{Y}^5 \, +13920\, \tilde{X}\, \tilde{Y}^6
\nonumber \\
 \hspace{-0.95in}&&   \, \quad \,+472576\, \tilde{X}^5\, \tilde{Y} \, 
-13144356607\, \tilde{X}^4\, \tilde{Y}^2+229377672192\, \tilde{X}^3\, \tilde{Y}^3
\nonumber \\
 \hspace{-0.95in}&&   \, \quad \, \, -13144356607\, \tilde{X}^2\, \tilde{Y}^4
+472576\, \tilde{X}\, \tilde{Y}^5+7547184\, \tilde{X}^4\, \tilde{Y}
+39849037920\, \tilde{X}^3\, \tilde{Y}^2
\nonumber \\
 \hspace{-0.95in}&&   \, \quad \, \, +39849037920\, \tilde{X}^2\, \tilde{Y}^3
+7547184\, \tilde{X}\, \tilde{Y}^4\, +49771008\, \tilde{X}^3\, \tilde{Y}
\nonumber \\
 \hspace{-0.95in}&&   \, \quad \,-13195144656\, \tilde{X}^2\, \tilde{Y}^2
+49771008\, \tilde{X}\, \tilde{Y}^3 \, +95607040\, \tilde{X}^2\, \tilde{Y}
+95607040\, \tilde{X}\, \tilde{Y}^2
\nonumber \\
 \hspace{-0.95in}&&   \, \quad \,+19771392\, \tilde{X}\, \tilde{Y} \, \, \, -4096. 
\end{eqnarray}
The elimination of $\, z$ in 
\begin{eqnarray}
\label{elimin}
\hspace{-0.96in}&&  \quad   \quad   \quad   \quad   \quad   \quad  
  \, \, \tilde{X} \, = \, \, -\, {{4096 \, z } \over {(z+16)^2 }}, 
\quad \quad \quad \quad 
\tilde{Y} \, = \, \, z \cdot \, (z+16), 
\end{eqnarray}
or in
\begin{eqnarray}
\label{elimin2}
\hspace{-0.96in}&&  \quad   \quad   \quad  \quad    \quad   \quad  
  \, \, \tilde{X} \, = \, \, -\, {{4096 \cdot \, z } \over {(z+16)^2 }}, 
\quad \quad \quad \quad 
\tilde{Y} \, = \, \, {{4096  \cdot \,(z+16)} \over {z^2}} , 
\end{eqnarray}
or in
\begin{eqnarray}
\label{elimin2}
\hspace{-0.96in}&&  \quad   \quad   \quad  \quad    \quad   \quad  
  \, \, \tilde{X} \, = \, \, {{4096  \cdot \,(z+16)} \over {z^2}}, 
\quad \quad \quad \quad 
\tilde{Y} \, = \, \, z \cdot \, (z+16), 
\end{eqnarray}
corresponds to (\ref{pol1}), the  first polynomial
 $\,\, \,  {\hat p}_1(\tilde{X},\, \tilde{Y}) \, = \, \, 0$. 

The elimination of $\, z$ between any two Hauptmoduls among the three Hauptmoduls
 $\, {\cal P}_1 $, $\, {\cal P}_2$, or $\, {\cal P}_3$, yields the 
{\em same modular equation} (\ref{mod4}). 
In general for modular equations representing $\,\, \tau \, \rightarrow \, \, N \, \tau$,
one Hauptmodul is of the form $\, \alpha \cdot z \, + \, \, \cdots$ when the other one 
is of the form  $\,\, \alpha \cdot z^N \, + \, \, \cdots$ (see~\cite{Youssef}). Here with
$\, {\cal P}_2$ and $\, {\cal P}_3$, we have two Hauptmoduls algebraically related by
the  modular equation (\ref{mod4}) representing 
$\, \tau \, \rightarrow \, \, 4 \, \tau$, but each of them is of the form 
$\, \pm \, \alpha \cdot z \, + \, \, \cdots$ This result is reminiscent of
the {\em involutive} series solution of (\ref{mod4}), 
(given by equation (104) in~\cite{Youssef}):
\begin{eqnarray}
\label{othersolution}
\hspace{-0.95in}&&   \,\,
Y \, \,\,  = \, \, \,  \, \, 
-X \, \,   \, \, \,   -{\frac {31\,{X}^{2}}{36}}\,  \, \,  \, 
-{\frac {961 }{1296}} \cdot \, X^3 \,  \, \,  \, 
-{\frac {203713 }{314928}}\cdot \, X^4 \, \, \,  \,  
-{\frac {4318517 }{7558272}} \cdot \, X^5
 \nonumber \\
\hspace{-0.95in}&& \,   \,   \,   \, 
 -{\frac {832777775}{1632586752}}\cdot \, X^6\, \, 
-{\frac {729205556393 }{1586874322944}} \cdot \,{X}^{7}\, 
-{\frac {2978790628903 }{7140934453248}} \cdot \,{X}^{8}  
\, \, \,  + \,   \dots 
\end{eqnarray}
Replacing in (\ref{othersolution}), $\, (X, \, Y)$ by\footnote[1]{Or replacing 
$\, (X, \, Y)$ by $\, ({\cal P}_2^{(1)}, \, {\cal P}_2^{(1)})$.
}  $\, ({\cal P}_2, \, {\cal P}_3)$,
one verifies that the expansions in $\, z$, of LHS and RHS  
of (\ref{othersolution}) are equal. 

\vskip .1cm
\vskip .1cm

\section{$\, _2F_1([1/4,3/4],[1],{\cal P})$ hypergeometric as modular forms}
\label{identitiesmodular}

\vskip .1cm 

\subsection{$\, _2F_1([1/4,1/4],[1],{\cal P})$ and $\, _2F_1([1/2,1/2],[1],{\cal P})$  as modular forms}
\label{identitiesmodularsub1}

\vskip .1cm

In Table 15 of Maier~\cite{Maier1}, one sees that $\, _2F_1([1/4,1/4],[1],x)$ hypergeometric 
functions are related to  $\, \tau \, \rightarrow \, \, 2 \, \tau$: 
\begin{eqnarray}
\label{identity14}
\hspace{-0.96in}&& \,\,  \, \quad  \,  \, 
_2F_1\Bigl([{{1} \over {4}}, \, {{1} \over {4}} ], \, [1], \, - \,  {{x} \over {64}} \Bigr) 
 \, \, \, = \, \, \,\,
 \Bigl( {{ x \, +16} \over { 16 }} \Bigr)^{-1/4} 
\cdot \,  _2F_1\Bigl([{{1} \over {12}}, \, {{5} \over {12}} ], \, [1],
 \,   {{ 1728 \cdot \, x  } \over {(x \, +16)^3} } \Bigr)
\nonumber \\
\hspace{-0.96in}&& \,\,  \, \quad \quad \quad  \quad 
\, \, = \, \, \,
 \Bigl( {{ x \, +256} \over { 256 }} \Bigr)^{-1/4} 
\cdot \,  _2F_1\Bigl([{{1} \over {12}}, \, {{5} \over {12}} ], \, [1], 
\,   {{ 1728 \cdot \, x^2  } \over {(x \, +256)^3} } \Bigr).
\end{eqnarray}
One has the following identity\footnote[1]{It can be deduced from (\ref{identity14}) 
together with (\ref{4tauHauptthird2}) with (\ref{4tauHauptthird3last}), 
or  (\ref{4tauHauptthird3B}) with (\ref{4tauHauptthird3lastB}).
 }:
\begin{eqnarray}
\label{identitiesFIRSTBIS4}            
\hspace{-0.98in}&& \quad  \quad \quad \quad \quad  \, \,
  _2F_1\Bigl( [{{1} \over { 4}}, \,{{1} \over { 4}}], \, [1], 
\,  \, \,  - \, {{1 } \over {64}} \cdot \,  {{ x^2} \over { x \, +16 }}  \Bigr)
\nonumber \\
\hspace{-0.98in}&& \quad  \quad  \quad \quad \quad \quad \quad  \, \,
 \, \, = \, \, \,   \, 
 \Bigl({{ x+16} \over {16}} \Bigr)^{1/4}  \cdot  \,  
_2F_1\Bigl( [{{1} \over { 4}}, \,{{1} \over { 4}}], \, [1],
 \, \,  -\, {{ x \cdot \, (x \, +16)} \over { 64}}   \Bigr). 
\end{eqnarray}

\vskip .3cm 

One also sees in Table 15 of Maier~\cite{Maier1} 
that  $\,\,  _2F_1([1/2,1/2],[1],x)\, $ hypergeometric 
functions are related to a $\, \tau \, \rightarrow \, \, 4 \, \tau$ 
isogeny\cite{Youssef}: 
\begin{eqnarray}
\label{identity12}
\hspace{-0.95in}&&    \quad    \quad  \,  \,  \,
_2F_1\Bigl([ {{1} \over {2}}, \, {{1} \over {2}}],
 \, [1], \, - \, {{x} \over {16}} \Bigr)
\\
\hspace{-0.95in}&&    \quad    \quad        \quad  
  \,   \, \, \,  \,  \, = \,   \,
  \Bigl({{x^2 \, +16\,x\, +16} \over {16}} \Bigr)^{-1/4}  \cdot \, 
_2F_1\Bigl([{{1} \over {12}}, \, {{5} \over {12}}], \, [1], 
\,  {{ 1728 \cdot \, x  \cdot \, (x+16) } \over {(x^2 \, +16\,x\, +16)^3} } \Bigr)
 \nonumber
\\
\hspace{-0.95in}&&    \quad    \quad      \quad  
  \,   \, \, \,  \,  \, = \,   \,
 \Bigl({{x^2 \, +256\,x\, +4096} \over {4096 }} \Bigr)^{-1/4}  \cdot \, 
_2F_1\Bigl([{{1} \over {12}}, \, {{5} \over {12}}], \, [1], 
\,  {{ 1728 \cdot \, x^4  \cdot \, (x+16) } \over {(x^2 \, +256\,x\, +4096)^3} } \Bigr). 
\nonumber
\end{eqnarray}
One has the following identity:
\begin{eqnarray}
\label{identitiesFIRSTBIS}            
\hspace{-0.98in}&& \quad  \,  \quad  \, \, 
_2F_1\Bigl( [{{1} \over { 2}}, \,{{1} \over { 2}}], \, [1], 
\, {{ 8 \, x \cdot \, (1\, +x^2)} \over {(1+x)^4 }}   \Bigr) 
 \, \,\, = \, \, \, \, 
 (1 \, +x)^2 \cdot  \,  _2F_1\Bigl( [{{1} \over { 2}}, \,{{1} \over { 2}}], \, [1], \, x^4 \Bigr). 
\end{eqnarray}

\vskip .1cm 

\subsection{$\, _2F_1([1/4,3/4],[1],{\cal P})$ hypergeometric as modular forms}
\label{identitiesmodularsub}

Let us now focus on the $\, _2F_1([1/4,3/4],[1],{\cal P})$ hypergeometric function:
\begin{eqnarray}
\label{identity1bis}
\hspace{-0.95in}&&    \,   \, \,  \,  \, 
 _2F_1\Bigl([{{1} \over {4}}, \, {{3} \over {4}}], \, [1], \,x \Bigr) 
\,\,\, \,  = \, \, \, \,
 (1 \, +3\, x)^{-1/4}  \cdot \,
  _2F_1\Bigl([{{1} \over {12}}, \, {{5} \over {12}}],
 \, {{27 \, x \cdot \, (1\, -x)^2} \over { (1 \, +3 \, x)^3 }}\Bigr).
\end{eqnarray}
The emergence of $\, _2F_1([1/4,3/4],[1],{\cal P})$ hypergeometric functions 
in physics, walk problems in the quarter of a plane~\cite{Rook,76,77} 
in enumerative combinatorics, or in   
interesting subcases of diagonals (see  section (\ref{1434})),  
raises the question  if $\, _2F_1([1/4,3/4],[1],{\cal P})$
should be seen as associated to  the isogenies~\cite{Youssef}
 $\, \tau \, \rightarrow \, \, 2 \, \tau$
or  $\, \tau \, \rightarrow \, \, 4 \, \tau$.
The identity 
\begin{eqnarray}
\label{identity1bis}
\hspace{-0.95in}&&  \quad \quad   \, 
 _2F_1\Bigl([{{1} \over {4}}, \, {{3} \over {4}}], \, [1], \,64 \, x^2 \Bigr) 
\,\,\,  \, = \, \, \, \,
(1 \, + 8 \, x)^{-1/2} 
\cdot \,
 _2F_1\Bigl([{{1} \over {2}}, \, {{1} \over {2}}], \, [1], 
\,\, {{16  \, x} \over { 1 \, + 8 \, x }}  \Bigr), 
\end{eqnarray}
or equivalently
\begin{eqnarray}
\label{identityquat}
\hspace{-0.95in}&&    \, \quad \quad 
 _2F_1\Bigl([{{1} \over {4}}, \, {{3} \over {4}}], \, [1], 
\, \, \Bigl(   {{  x} \over { 2 \, -  \, x }} \Bigr)^2 \Bigr) 
\,\,\,  \, = \, \, \, \,
\Bigl({{2 \, -x} \over {2}} \Bigr)^{1/2} 
\cdot \,
 _2F_1\Bigl([{{1} \over {2}}, \, {{1} \over {2}}], \, [1], \, x \Bigr), 
\end{eqnarray}
seems to relate  $\, _2F_1([1/4,3/4],[1],{\cal P}) \, $ to $\,\,  _2F_1([1/2,1/2],[1],x)$, 
and thus seems to relate  $\, _2F_1([1/4,3/4],[1],{\cal P})\, $ rather 
$\,\,  \tau \, \rightarrow \, \, 4 \, \tau$.  Yet things are more subtle.

Let us see how  $\, _2F_1([1/4,3/4],[1],{\cal P})$ can be described as a modular form
corresponding to pullbacked $\, \, _2F_1([1/4,3/4],[1],{\cal P})\,  $ hypergeometric functions with
two different rational pullbacks. For instance, 
one deduces from (\ref{identitiesFIRSTBIS}) combined with (\ref{identityquat}),
 several identities on the hypergeometric function $\, _2F_1([1/4,3/4],[1],{\cal P})\, $ like
\begin{eqnarray}
\label{newident2bis}
\hspace{-0.98in}&&     \quad \quad \quad 
_2F_1\Bigl([{{1} \over {4}}, \, {{3} \over {4}}], \, [1], \,   {{ x^2} \over {(2\, -x)^2 }} \Bigr)
\nonumber \\ 
\hspace{-0.98in}&&    \quad \quad \quad \quad \quad \quad 
\, \,   = \,  \,  \,  
 \Bigl( {{2 \, -x} \over {2 \cdot \, (1 \, -2\,x)}} \Bigr)^{1/2}      \cdot \, 
_2F_1\Bigl([{{1} \over {4}}, \, {{3} \over {4}}], \, [1], 
\,   - \, 4 \cdot \,{\frac {{x} \cdot \, (1 \, - \, {x}) }{ (1 \, - \, 2\,{x})^{2}}}  \Bigr),
\end{eqnarray}
or
\begin{eqnarray}
\label{newident}
\hspace{-0.98in}&&  \quad \quad \quad    \quad    \, 
_2F_1\Bigl([{{1} \over {4}}, \, {{3} \over {4}}], \, [1], \,   {{ x^2} \over {(2\, -x)^2 }} \Bigr)
\nonumber \\ 
\hspace{-0.98in}&&    \quad \quad \quad \quad \quad    \quad \quad 
\,  \, = \,  \, \, 
 \Bigl( {{2\, -x} \over { 2 \cdot \, (1 \,  + \, x)}} \Bigr)^{1/2}      \cdot \, 
_2F_1\Bigl([{{1} \over {4}}, \, {{3} \over {4}}], \, [1], 
\,    {\frac { 4\, {x}}{ (1 \, + \, {x})^{2}}} \Bigr).
\end{eqnarray}
and thus:
\begin{eqnarray}
\label{newident3bis}
\hspace{-0.98in}&&   \quad \quad \quad   \quad  \quad       
_2F_1\Bigl([{{1} \over {4}}, \, {{3} \over {4}}], \, [1], \,    {\frac { 4\, {x}}{ (1 \, + \, {x})^{2}}} \Bigr)
\nonumber \\ 
\hspace{-0.98in}&&   \quad \quad  \quad  \quad \quad \quad \quad \, \, \, \, 
\,  = \,  \,  
 \Bigl( {{1\, +x} \over {1\, -2 \, x }} \Bigr)^{1/2}      \cdot \, 
_2F_1\Bigl([{{1} \over {4}}, \, {{3} \over {4}}], \, [1], 
\,   - \, 4 \cdot \,{\frac {{x} \cdot \, (1 \, - \, {x}) }{ (1 \, - \, 2\,{x})^{2}}}  \Bigr).
\end{eqnarray}
One also has the identity:
\begin{eqnarray}
  \label{Funquarttroisquart}  
\hspace{-0.98in}&& \,\, \quad \quad \quad 
_2F_1\Bigl([{{1} \over {4}}, \, {{3} \over {4}}], \, [1], \,  {{ x^4} \over {(2\, -x^2)^2 }} \Bigr)
\\
\hspace{-0.98in}&& \,\, \quad \quad  \quad \quad \quad  \quad \quad       
\, \, = \,  \,\, 
 \Bigl({{2\, -x^2 } \over {2\cdot \, (1 \, +6\,x \, \,+x^2) }}  \Bigr)^{1/2}  \cdot \, 
_2F_1\Bigl([{{1} \over {4}}, \, {{3} \over {4}}], \, [1], 
\,  \,  {{ 16 \cdot  \, x  \cdot  \,(1+x)^2} \over { (1 \, +6\,x \, \,+x^2)^2}} \Bigr). 
\nonumber 
\end{eqnarray}
Recalling the viewpoint developed in our previous paper~\cite{Youssef}
these identities can be seen to be of the form
\begin{eqnarray}
\label{FunquarttroisquartIDENTITY}
\hspace{-0.98in}&& \,\, \quad \quad \quad \quad  \quad 
_2F_1\Bigl([{{1} \over {4}}, \, {{3} \over {4}}], \, [1], \,  B \Bigr)
\, \, \, = \,  \, \,\,  \, G  \cdot \, 
_2F_1\Bigl([{{1} \over {4}}, \, {{3} \over {4}}], \, [1], 
\,  \,  A \Bigr),
\nonumber 
\end{eqnarray}
where $\, G$ is some algebraic factor. For instance in the case of the last identity 
(\ref{Funquarttroisquart})
 \begin{eqnarray}
\label{whereAB}
\hspace{-0.98in}&& \quad   \quad \, \,   \quad  \quad \quad\quad
A \,\, = \, \,  \, 
 {{ 16 \cdot  \, x  \cdot  \,(1+x)^2} \over { (1 \, +6\,x \, \,+x^2)^2}},
  \quad \quad  \quad \,  \, \,   
 B \,\, = \, \,  \,  {{ x^4} \over {(2\, -x^2)^2 }},
\end{eqnarray}
we have 
\begin{eqnarray}
\label{whereB}
\hspace{-0.98in}&&   
B \, \, = \, \, 
 {\frac {{A}^{4}}{262144}}  \,  \, 
+{\frac {5\,{A}^{5}}{524288}} \,  \, 
+{\frac {1069\,{A}^{6}}{67108864}}  \,  \, 
+{\frac {6003\,{A}^{7}}{268435456}} \, 
+{\frac {1961123\,{A}^{8}}{68719476736}}  \, 
 \, + \, \, \cdots 
\end{eqnarray}
and  $\, G$ is an algebraic factor 
\begin{eqnarray}
\label{Galgfactor}
\hspace{-0.98in}&& \, \, 
G \, \,  \, = \, \, \, \, 
1 \, \,\, -\, {{3} \over {16}} \,A \,
\, \, -{\frac {69\,{A}^{2}}{1024}} \, \, \,
-{\frac {633\,{A}^{3}}{16384}} \,  \,
-{\frac {55209\,{A}^{4}}{2097152}} \,
 \, -{\frac {659109\,{A}^{5}}{33554432}} \, 
\, \, \, + \, \, \, \cdots 
\end{eqnarray}
solution of:
\begin{eqnarray}
\label{algfactor}
\hspace{-0.98in}&& \,\,  \quad  \quad 
 65536 \cdot \, G^8 \,  \,  -16384\cdot \, G^6  \, \, +1536 \cdot \, (27\,A\,-26)  \cdot \, G^4 
 \, \,  + 64 \cdot \, (135\,A\,-136) \cdot \, G^2
\nonumber \\
\hspace{-0.98in}&& \,\, \quad  \quad  \quad   \quad \quad 
  \, +3969 \, A^2 \, -3456 \, A \,  \, \, -512 
 \,  \, \,  \, = \,  \, \, \,  \, 0.
\end{eqnarray}
The important result of~\cite{Youssef} is that after elimination of the algebraic 
factor $\, G$ one finds that the  two pullbacks $\, A$ and $\, B$
verify the following {\em Schwarzian equation}:
\begin{eqnarray}
\label{Schwa}
\hspace{-0.98in}&& \,\, \, \,
-{{1} \over {8}} \,{\frac {3\,{A}^{2}-3\,A+4}{{A}^{2} \left( A-1 \right)^{2}}} \, \, \,\, 
+ {{1} \over {8}}\,{\frac {3\,{B}^{2}-3\,B+4}{{B}^{2} \left( B-1 \right)^{2}}}  \,
\cdot \, \Bigl({{d B } \over {dA}}\Bigr)^2 \,
\, + \, \, \,  \{B, \, A\} 
\,\, \, \,= \, \, \,\,\, 0,
\end{eqnarray} 
where $\,  \{B, \, A\}$ denotes the Schwarzian derivative.

Do note that  $\, _2F_1([1/4,3/4],[1],{\cal P})$ is a selected 
hypergeometric function since the rational function in the Schwarzian derivative (\ref{Schwa})
\begin{eqnarray}
\label{WSchwa}
\hspace{-0.98in}&& \,\, \, \,  \, \, \quad  \quad  \quad  \quad  \quad  \quad  \quad  
W(A)\, \, = \, \, \, 
-{{1} \over {8}} \,{\frac {3\,{A}^{2} \, -3\,A \, +4}{{A}^{2} \cdot \, (A-1)^{2}}}, 
\end{eqnarray}
is invariant under the $\,\, A \, \rightarrow \, \, 1 \, -A \, \, $ 
transformation:  $\, \, \, W(A)\, \, = \, \, \,  W(1 \, -A)$.

This Schwarzian equation can be written in a more symmetric way between 
$\, A$ and $\, B$, namely:
\begin{eqnarray}
\label{SchwaSymmeq}
\hspace{-0.98in}&& \,\, \, \,     \quad \quad   \quad \quad   \quad \quad  
 {{1} \over {8}}\,{\frac {3\,{B}^{2}-3\,B+4}{{B}^{2} \left( B-1 \right)^{2}}}  \,
\cdot \, \Bigl({{d B } \over {dx}}\Bigr)^2 \,
\, + \, \, \,  \{B, \, x\} 
 \nonumber \\
\hspace{-0.98in}&& \,\, \quad \quad   \quad \quad   \quad \quad   \quad \quad \quad \quad
\,\, \, \,= \, \, \,\,\, 
 {{1} \over {8}}\,{\frac {3\,{A}^{2}-3\,A+4}{{A}^{2} \left( A-1 \right)^{2}}}  \,
\cdot \, \Bigl({{d A } \over {dx}}\Bigr)^2 \,
\, + \, \, \,  \{A, \, x\}.
\end{eqnarray}
Let us denote $\, \rho(x)$ the rational function of the LHS or the RHS of 
equality (\ref{SchwaSymmeq}). For the three identities (\ref{newident2bis}), 
(\ref{newident}), (\ref{newident3bis}) this rational function 
is (of course\footnote[1]{Since these identities share one pullback.}) 
the same rational function, namely 
\begin{eqnarray}
\label{SchwaSymmrho}
\hspace{-0.98in}&& \,\, \, \,   \quad \quad \quad \quad   \quad \quad  \quad \quad   
 \rho(x)
\,\, \, \,= \, \, \,\,\, 
{{1} \over {2}} \cdot \,{\frac { x^2 \, - x \, +1}{ x \cdot \, (x-1)^{2} }}, 
\end{eqnarray}
when the last identity (\ref{Funquarttroisquart}) corresponds to: 
\begin{eqnarray}
\label{SchwaSymmrho}
\hspace{-0.98in}&& \,\, \, \,   \quad \quad \quad \quad  \quad \quad   \quad \quad  
  \rho(x)
\,\, \, \,= \, \, \,\,\, 
{{1} \over {2}} \cdot \,{\frac { (x^2 +1)^{2}}{ x^2 \cdot \, (x^2-1)^{2}}}. 
\end{eqnarray}
Let us consider the first two identities (\ref{newident2bis}) and (\ref{newident}),
denoting by $\, A$ and $\, B$ the corresponding pullbacks: 
\begin{eqnarray}
\label{SchwaSymmrhoABapp}
\hspace{-0.98in}&& \,\, \, \, \, \, 
A \, \, = \, \, \,
 - \, 4 \cdot \,{\frac {{x} \cdot \, (1 \, - \, {x}) }{ (1 \, - \, 2\,{x})^{2}}},    
  \quad \quad  \hbox{or:}  \quad  \quad 
 {\frac { 4\, {x}}{ (1 \, + \, {x})^{2}}},  \quad  \quad \quad 
B \, \, = \, \, \,   {{ x^2} \over {(2\, -x)^2 }} 
\end{eqnarray}
These two pullbacks are related by the {\em asymmetric} modular equation:
 \begin{eqnarray}
\label{Mod2app}
\hspace{-0.98in}&& \, \, \, 
81\cdot \, A^2\, B^2 -18 \, A\, B \cdot \, (8\,B\,+A) \, 
+(A^2 \, +80 \cdot \,A\,B\, +64\, B^2)
\, \, -64 \, B\,\, = \,\,\,0. 
\end{eqnarray}
giving the following expansion for$\, A$ seen as an 
{\em algebraic series}\footnote[2]{We discard the other root expansion 
$\, \, \, B \, = \, \,  1 \, +A \, +{\frac{5}{4}}{A}^{2} \, +{\frac{25}{16}}{A}^{3}
\, +{\frac{31}{16}}{A}^{4} \,  \, + \cdots$  } in $\, B$:
 \begin{eqnarray}
\label{Mod2serapp}
\hspace{-0.98in}&& \quad  \,\, \, \,   \, 
B \, \, = \, \, \, \, {\frac{1}{64}}{A}^{2} \, \, 
+{\frac{5}{256}}{A}^{3} \, \, 
+{\frac{83}{4096}}{A}^{4}\, \,  +{\frac{163}{8192}}{A}^{5} \, \, 
+{\frac{5013}{262144}}{A}^{6} \,\, 
\,\,  + \, \,\,  \cdots
\end{eqnarray}
We will denote $\, {\cal M}_2(A, \, B)$ the LHS of the modular equation (\ref{Mod2app}):
such an algebraic series is clearly\footnote[9]{From (\ref{Mod2serapp}) see~\cite{Youssef}.} a 
$\, \tau \, \rightarrow \, 2 \, \tau$ 
 (or  $\, q \, \rightarrow \, q^2$ in the nome $\, q$) 
isogeny~\cite{Youssef}. Composing this algebraic transformation with itself 
in order to have  a  $\, \tau \, \rightarrow \, 4 \, \tau$
(or $\, q \, \rightarrow \, q^4$) 
representation, amounts to eliminating\footnote[5]{Performing the resultant: 
 $\, resultant({\cal M}_2(A, \, X), \,{\cal M}_2(X, \, B), \, X)$. } $\, X$ between 
$\, {\cal M}_2(A, \, X) \, \, = \, \, 0$ and $\, {\cal M}_2(X, \, B) \, \, = \, \, 0$ 
(i.e. two times the modular equation (\ref{Mod2})). This elimination gives 
the following {\em asymmetric} modular curve corresponding 
to identity (\ref{Funquarttroisquart}):
\begin{eqnarray}
  \label{ModeCurveunquarttroisquart}
\hspace{-0.98in}&& \quad  \quad  \quad  \quad  
15752961\,{A}^{4}{B}^{4} \, \,  \, 
-428652\,{A}^{3}{B}^{3} \cdot \, (64\,B+83\,A) \,
\nonumber \\
\hspace{-0.98in}&& \,\,  \quad  \quad \quad  \quad   \quad  
 \,+162\,{A}^{2}{B}^{2} \cdot \, (48640\,{B}^{2}+494208\,AB+124051\,{A}^{2}) 
\nonumber \\
\hspace{-0.98in}&& \,\,  \, \quad  \quad \quad  \quad   \quad  
 +108 \,\, A \, B \cdot \, (32768\,{B}^{3} -500480\,{B}^{2}A -491200\,B{A}^{2} \, -83\,{A}^{3})
\nonumber \\
\hspace{-0.98in}&& \,\,  \, \quad  \quad \quad  \quad   \quad  
\,   +262144\,{B}^{4} +10354688\,{B}^{3}A +46715904\,{B}^{2}{A}^{2} +159488\,B{A}^{3} +{A}^{4} 
\nonumber \\
\hspace{-0.98in}&& \, \, \quad  \quad \quad  \quad   \quad  
\, \, -3072\,B  \cdot \, (256\,{B}^{2}+4736\,AB+177\,{A}^{2})
\nonumber \\
\hspace{-0.98in}&& \,\,\,\, \quad \quad \quad \quad  \quad \quad  \quad   \quad  
\, \,  +131072 \, B \cdot  \, (6\,B+5\,A)
\, \,  \,  -262144 \,B
\, \,\, = \, \,\, \, 0,
\end{eqnarray}
parametrised by
\begin{eqnarray}
\label{whereAB}
\hspace{-0.98in}&& \quad   \quad \, \,     \quad  \quad \quad\quad
A \,\, = \, \,  \, 
 {{ 16 \cdot  \, x  \cdot  \,(1+x)^2} \over { (1 \, +6\,x \, \,+x^2)^2}},
  \quad \quad  \quad    
 B \,\, = \, \,  \,  {{ x^4} \over {(2\, -x^2)^2 }},
\end{eqnarray}
 where: 
\begin{eqnarray}
\label{whereBB}
\hspace{-0.98in}&&   
B \, \, = \, \, 
 {\frac {{A}^{4}}{262144}}  \,  \, 
+{\frac {5\,{A}^{5}}{524288}} \,  \, 
+{\frac {1069\,{A}^{6}}{67108864}}  \,  \, 
+{\frac {6003\,{A}^{7}}{268435456}} \, 
+{\frac {1961123\,{A}^{8}}{68719476736}}  \, 
 \, \, + \, \, \cdots 
\end{eqnarray}
Note that $\, B \, $ in (\ref{whereAB}) is nothing but the composition of  
$\, B$ in ( \ref{SchwaSymmrhoAB}) by $\, x \, \rightarrow \, x^2$ and  
 that $\, A$ in (\ref{whereAB}) in nothing but the composition of   
$\, A$  in (\ref{whereAB}) with itself:
\begin{eqnarray}
\label{whereAA}
\hspace{-0.98in}&&   \quad   \quad   \quad  \quad  \quad   \quad 
 {{ 16 \cdot  \, x  \cdot  \,(1+x)^2} \over { (1 \, +6\,x \, \,+x^2)^2}}
 \, \, \, \, = \, \,    \,   \, 
 {\frac { 4\, {x}}{ (1 \, + \, {x})^{2}}} \, \circ  \, \,  {\frac { 4\, {x}}{ (1 \, + \, {x})^{2}}}.
\end{eqnarray}

The modular curve (\ref{Mod2}) is unpleasantly asymmetric: the two pullbacks are not on the same 
footing. Note that, using the $\, A \, \leftrightarrow \, 1 \, -\, A \, $ symmetry (see (\ref{WSchwa})) 
on the Schwarzian equations (\ref{SchwaSymmeq}), and changing $\, A \, \rightarrow \, 1 \, -\, A \, $ in  
the asymmetric modular curve (\ref{Mod2}), 
one gets the {\em symmetric} modular curve: 
\begin{eqnarray}
\label{symMod2}
\hspace{-0.98in}&& \quad \quad \quad \quad \quad 
81 \cdot \, A^2\, B^2 \,\,  -18 \cdot  \, (A^2\,B  +A\, B^2)\,  \, +A^2 \, -44\, A\, B \, +B^2 
\nonumber \\ 
\hspace{-0.98in}&& 
\quad \quad \quad \quad \quad \quad \quad \quad 
\,\,  -2 \cdot  \, (A\, +B)  \,\,  +1 \, \, \,  = \,   \, \, 0.
\end{eqnarray}
Changing $\, B \, \rightarrow \, 1 \, -\, B \, $  in  
the asymmetric modular curve (\ref{Mod2}), 
one also gets another symmetric modular curve:
\begin{eqnarray}
\label{symMod2bis}
\hspace{-0.98in}&& \quad \quad \quad \quad \quad 
81 \cdot \, A^2 \, B^2 \,\,  -144 \cdot  \, (A^2\,B  +A\, B^2)
\nonumber \\ 
\hspace{-0.98in}&& 
\quad \quad \quad \quad \quad \quad \quad \quad 
 \,\,  +208\, A\, B \,\, \,+ 64 \cdot \, (A^2 \, +B^2 \, -A \, -B) 
\,   \,\,  = \,  \,  \, 0.
\end{eqnarray}
The two pullbacks for (\ref{symMod2bis}) read:
\begin{eqnarray}
\label{symMod2bispullback}
\hspace{-0.98in}&&   \quad \quad \quad \quad \quad \quad \quad \quad
A \, \, = \, \, \,  {{ 4 \, x } \over { (1\, + \, x)^2 }},
   \quad  \quad    \quad
B \, \, = \, \, \, {{4 \cdot \, (1\, - \, x)  } \over { (2 \, - \, x)^2 }} .
\end{eqnarray}
Similarly the {\em asymmetric} modular curve (\ref{ModeCurveunquarttroisquart}) 
 can be turned back into a {\em symmetric} modular curve 
by changing  $\, A \, \leftrightarrow \, 1 \, -\, A$,
 or $\, B \, \leftrightarrow \, 1 \, -\, B$.
The price to pay to restore the symmetry between the two pullbacks (\ref{symMod2bispullback})
is that the corresponding pullbacks do not yield hypergeometric identities 
{\em expandable for $\, x$ small}.

\vskip .1cm 

Finally, the identity (\ref{newident3bis}) corresponds to a symmetric 
relation between these two-pullbacks which reads:
\begin{eqnarray}
\label{newident4}
\hspace{-0.98in}&&  
\quad \quad \quad \quad 
   81 \cdot \,{C}^{2}{D}^{2} \,  \,  \,
 -144 \cdot \, ({C}^{2} \, D  \, +\,C{D}^{2})   \,  \,  \, 
+16 \cdot \,(4\, {C}^{2} \, +13\,C \,  D \, + 4\,{D}^{2}) 
\nonumber  \\ 
\hspace{-0.98in}&&  
\quad \quad \quad \quad \quad \quad \quad \quad \quad  \quad \quad
  \, -64 \cdot \, (C \, +\,D)
 \, \, \,  \, = \, \, \, \,  \, 0.
\end{eqnarray}
The corresponding series expansion 
\begin{eqnarray}
\label{newident4expans}
\hspace{-0.98in}&&   \quad \, 
D \,  \, = \, \, \,\,  -C \, \,\, - {{5} \over {4}} \,{C}^{2} 
\,\,\, -{\frac {25\,{C}^{3}}{16}} \,\, \,-{\frac {31\,{C}^{4}}{16}} \, \,\,
-{\frac {305\,{C}^{5}}{128}} \,\, \, -{\frac {2979\,{C}^{6}}{1024}} \,\,
\, -{\frac {14457\,{C}^{7}}{4096}} 
\nonumber \\
\hspace{-0.98in}&&  \quad \quad \, \, \,  \, 
\,  -{\frac {17445\,{C}^{8}}{4096}} \, \,\,
-{\frac {167615\,{C}^{9}}{32768}} \,\,\, -{\frac {801941\,{C}^{10}}{131072}} \,\, \,
-{\frac {3822989\,{C}^{11}}{524288}}
 \,  \, \, \,\,  \,  + \, \,  \,  \, \cdots 
\end{eqnarray}
is {\em an involutive series}.

\vskip .2cm 

\section{Modular forms: recalls on Maier's paper~\cite{Maier1} and the associated Schwarzian equations }
\label{modularAPP}

\vskip .1cm

In fact, the previous pullbacks in the pullbacked $\, _2F_1$
hypergeometric functions can be seen (and should be seen)
as {\em Hauptmoduls}~\cite{Maier1}.

In~\cite{Maier1}, Maier underlined the representation of a selected set of 
modular forms as pullbacked hypergeometric functions 
with two possible {\em rational pullbacks} 
(related by a genus zero modular equation).  In~\cite{Youssef}, we 
revisited that viewpoint: an identity on a hypergeometric function with 
a pullback and the same hypergeometric function with another pullback,
the (algebraic) 
map\footnote[2]{Called by Veselov~\cite{Veselov} in a mapping framework, ``correspondence''.},
changing one pullback into the other one, being a symmetry of 
{\em infinite order}\footnote[1]{Of course 
hypergeometric functions have finite order symmetries like $\, x\, \rightarrow \, 1\, -x$, 
that we discard. 
With infinite order symmetries one can associate some discrete dynamical map: 
in these particular cases algebraic function maps~\cite{Youssef,Youssef2,Veselov}.}, 
is such a strong constraint that it is almost characteristic of 
modular forms~\cite{Youssef}: the hypergeometric functions can be seen
as {\em automorphic functions} with respect to these {\em infinite order symmetries}.

\vskip .1cm

The two different modular equations (\ref{modcurve3}),  (\ref{mod4}) 
corresponding respectively to $\, \tau \, \rightarrow \, 3\, \tau$ and 
 $\, \tau \, \rightarrow \, 4\, \tau$, suggest
that a genus zero modular equation, corresponding to $\, \tau \, \rightarrow \, N\, \tau$,
could encapsulate these two subcases.  In such a scenario, $\, N$ must be a multiple of 
$\, 3$, $\, 4$, $5$, ... In fact, the set of values of $\, N$ corresponding 
to {\em modular equations} with a (genus zero) rational parametrization is 
obtained for a {\em finite set}~\cite{Maier1,GenusZero,GenusZero2} of integer values:
$\, 2$, $\, 3$, $\, 4$,  $\, 5$,  $\, 6$,  $\, 7$, $\, 8$, $\, 9$, 
$\, 10$, $\, 12$,  $\, 13$,  $\, 16$,  $\, 18$ and $\, 25$. 
Some canonical rational parametrizations of these selected 
genus zero modular equations are given in~\cite{Maier1}. The two Hauptmoduls 
read respectively for these selected values $\, 2$, 
$\, 3$, $\, 4$,  $\, 5$, ..., $\, 25$:
\begin{eqnarray}
\label{Weberapp}
\hspace{-0.95in}&& \quad 
N\, = \, 2: \quad \quad  \quad \quad \quad \quad 
 {{ 1728 \cdot \, z^2} \over { (z+256)^3}}, 
\quad  \quad \quad \quad \quad \, \quad 
 {{ 1728 \cdot \, z} \over { (z+16)^3}},
\\
\hspace{-0.95in}&& \quad 
N \, = \, \, 3: \quad \quad \quad 
{{ 1728 \cdot \, z^3} \over { (z \, +27)\cdot \, (z \, +243)^3}},
\quad \quad  \quad \quad 
 {{ 1728 \cdot \, z} \over { (z \, +27)\cdot \, (z \, +3)^3}},
\\
\label{60app}
\hspace{-0.95in}&& \quad  
N \, = \, \, 4: \quad  \quad \quad 
 {{1728 \cdot z^4 \cdot \, (z\, +16) } \over { (z^2\, +256 \, z \, + 4096)^3}},
  \quad \quad  \quad \quad 
  {{1728 \cdot z \cdot \, (z\, +16) } \over { (z^2\, +16 \, z \, + 16)^3}}, 
\\
\hspace{-0.95in}&& \quad 
N\, = \, \, 5: \quad \quad \quad 
{{ 1728 \cdot \, z^5} \over {(z^2 \, +250\,z\,+3125)^3 }},
 \quad \quad  \quad \quad 
  {{ 1728 \cdot \, z} \over {(z^2 \, +10\,z\,+5)^3 }}, 
\end{eqnarray}
\begin{eqnarray}
\label{z6app}
\hspace{-0.95in}&& \quad 
N \, = \, \, 6: \quad \quad  \quad \,  \,  
{{ 1728 \cdot \, z^6 \cdot \, (z+8)^2 \cdot \, (z+9)^3} \over {
(z+12)^3 \cdot \, (z^3 +252 \, z^2 +3888 \, z +15552)^3 }}, 
\nonumber \\
\hspace{-0.95in}&& \quad \quad  \quad  \quad  \quad 
\quad  \quad  \quad \quad  \quad  \quad \quad  \quad 
 {{ 1728 \cdot \, z \cdot \, (z+8)^3 \, (z+9)^2 } \over {
(z+6)^3 \cdot \, (z^3 +18 \, z^2 +84 \, z +24)^3 }}, 
\end{eqnarray}
\begin{eqnarray}
\label{z7app}
\hspace{-0.95in}&& \quad 
N\, = \, \, 7: \quad \quad  \quad \,  \, 
 {{ 1728 \cdot \, z^7 } \over {
(z^2  \, +13  \, z \, +49)\cdot \, (z^2 +245 \, z + 2401)^3 }}, 
\nonumber \\
\hspace{-0.95in}&& \quad \quad  \quad  \quad  \quad 
\quad  \quad  \quad   \quad   \quad \quad  \quad  \quad \quad 
 {{ 1728 \cdot \, z  } \over {
(z^2  \, +13  \, z \, +49) \cdot \, (z^2 +5 \, z + 1)^3 }}, 
\quad   \quad \,  \cdots 
\end{eqnarray}

The seven-parameter pullbacked $\, _2F_1$ hypergeometric function 
(\ref{2F15Hyp}) cannot correspond {\em generically} to 
rationally parametrized (genus zero) modular equations.
One cannot imagine to identify the Hauptmodul pullback in  (\ref{2F15Hyp}) 
with expressions~\cite{Maier1} like (\ref{j3}) or (\ref{4tauHaupt}) 
for  $\, N$ a multiple of $\, 3$, $\, 4$, $5$, ... In the generic 
seven-parameters case one gets the expression (\ref{2F15Hyp}) for 
the diagonal of the seven-parameters rational function (\ref{Ratfonc}) as a pullbacked
$\, _2F_1$ hypergeometric function with a {\rm rational pullback} (\ref{Q512}),
{\em the other pullback being algebraic} and deduced from the various 
modular equations~\cite{Youssef} (see section (\ref{modulargeneric})).

\vskip .1cm

\subsection{Landen transformation: $\, \tau \, \rightarrow \, 2 \, \tau$. }
\label{Landen}

\vskip .1cm

To describe this situation let us recall the result detailed in~\cite{Youssef}
for $\, \tau \, \rightarrow \, 2\, \tau$. The emergence 
of a {\em modular form}~\cite{Short,IsingCalabi}
corresponds to the identity on the {\em same}
hypergeometric function but where the pullback $\, x \, $ is changed 
$\, x \, \rightarrow  \, \, y(x) \, = \, y \, $ according to 
{\em modular equations}~\cite{Andrews,Atkin,Hermite,Hanna,Morain,Weisstein}. Let us
consider the  modular equation (\ref{modularcurveapp}) below corresponding to the 
Landen transformation~\cite{Abra,dlmf}, or inverse Landen transformation, 
and consider the corresponding 
$\, _2F_1$ hypergeometric identity
\begin{eqnarray}
\label{modularform2explicitapp}
\hspace{-0.95in}&& \quad \quad  \quad  \quad  \quad 
 _2F_1\Bigl([{{1} \over {12}}, \, {{5} \over {12}}], \, [1], \, y  \Bigr)
\, = \, \, \, \,\, 
 {\cal A}(x) \cdot \,
 _2F_1\Bigl([{{1} \over {12}}, \, {{5} \over {12}}], \, [1], \,  x  \Bigr), 
\end{eqnarray}
where $\, {\cal A}(x)$ is an algebraic function given by:
\begin{eqnarray}
\label{wherecalAapp}
\hspace{-0.95in}&& \quad \quad  \quad 
1024\,\,{\cal A}(x)^{12} \, \,  \, 
-1152\,\,{\cal A}(x)^{8}  \,\,  \, +132\,\,{\cal A}(x)^{4}
 \, \,  +125\,x \, \, \,  -4 \, \,  \,  \,= \, \, \, \, \,  0. 
\end{eqnarray}
The relation between $\, x$ and $\, y$ in (\ref{modularform2explicitapp}) is given by 
the {\em modular equation}~\cite{Andrews,Atkin,Hermite,Hanna,Morain,Weisstein}:
\begin{eqnarray}
\label{modularcurveapp}
\hspace{-0.95in}&& \quad 
1953125 \cdot \,{x}^{3}{y}^{3} \, \, \, -187500  \cdot \,{x}^{2}{y}^{2} \cdot \, (x+y)
\,  \, \,  +375 \cdot \, xy \cdot \, (16\,{x}^{2} \, -4027  \,xy +16\,{y}^{2})
\nonumber \\ 
\hspace{-0.95in}&& \quad \quad  \quad 
\quad  
 \,  -64  \cdot \, (x+y)  \cdot \, ({x}^{2}+1487\,xy+{y}^{2}) 
\,\, \,  +110592 \cdot \,xy
 \, \,  \,= \, \,\,  \, \, 0. 
\end{eqnarray}
Using this algebraic relation between $\, x$ and $\, y$ one can rewrite 
(\ref{modularform2explicitapp}) with (\ref{wherecalAapp}) as
\begin{eqnarray}
\label{modularform2explicit2app}
\hspace{-0.95in}&& \quad \quad  \quad  \quad  \quad   \quad 
 _2F_1\Bigl([{{1} \over {12}}, \, {{5} \over {12}}], \, [1], \, x  \Bigr)
\, = \, \, \, \,\, 
 {\tilde A}(y) \cdot \,
 _2F_1\Bigl([{{1} \over {12}}, \, {{5} \over {12}}], \, [1], \,  y  \Bigr), 
\end{eqnarray}
where $\, {\tilde A}(y)$ is an algebraic function 
given\footnote[5]{This result breaking the symmetry between 
the two variables $\, x$ and $\, y$ may look paradoxical. In fact assuming 
(\ref{wherecalAapp}) and (\ref{wherecalA2app}) and the modular equation 
(\ref{modularcurveapp}) one has 
$\, {\tilde A}(y) \cdot \, {\cal A}(x) \, \, = \, \, \, 1$.} by:
\begin{eqnarray}
\label{wherecalA2app}
\hspace{-0.95in}&& \quad \quad  \, \quad \quad 
\,\,{\tilde A}(y)^{12} \,  \,  \, 
-18 \,\,{\tilde A}(y)^{8}  \, \,  \, +33 \,\,{\tilde A}(y)^{4}
 \, \,  + 500\, y \, \,  \,  - 16 \, \,  \,  \,= \, \, \, \, \,  0. 
\end{eqnarray}
Using the fact that $\, {\tilde A}(y)$ is the reciprocal of $\,{\cal A}(x)$,
one can rewrite (\ref{wherecalA2app}) as:
\begin{eqnarray}
\label{wherecalA3app}
\hspace{-0.95in}&& \quad \,   \quad \quad \quad 
4 \,\,{\cal A}(x)^{12} \,\cdot (125 \, y \, -4)   \, 
 \, \, + 33 \,\,{\cal A}(x)^{8} \,\,  -18 \,{\cal A}(x)^{4}
 \, \, \, \,   +1 \, \,  \,  \,= \, \, \, \, \,  0. 
\end{eqnarray}
In other words the elimination of $\, {\cal A}(x)$ 
between (\ref{wherecalAapp}) and (\ref{wherecalA3app})
gives the modular curve (\ref{modularcurveapp}).  Thus, introducing  
$\, A \, = \, \, {\cal A}(x)^{4}$,
(\ref{wherecalAapp}) and (\ref{wherecalA3app}) can be seen as  alternative
rational parametrizations of the modular equation (\ref{modularcurveapp}):
\begin{eqnarray}
\label{otherparamapp}
\hspace{-0.95in}&& \, \, \, \,
x \, = \, \,
 \, {{4} \over {125 }} \cdot \, (1\, -A)\cdot \, (1\, -16\, A)^2,
 \quad  \, \, \,
y \, = \, \, 
 -\, {{1 } \over { 500 }} \cdot \, {{ (1\, -A)^2 \cdot \, (1\, -16\, A) } \over {A^3}}.
\end{eqnarray}
Note that $\, y$, in terms of $\, A$ is nothing 
but $\, x$, in terms of $\, A$, 
 $\, A \, $ taken to be $\, 1/16/A$. The two variables $x$ and $\, y$ 
{\em are thus on the same footing}: permuting $\, x$ and $\, y$ 
corresponds to the involutive transformation 
$\, A \, \, \leftrightarrow \, \, 1/16/A$. 
Finally changing $\, A$ into $\, A \, = \, \, (z\, +256)/(z+16)/16$,  
the previous parametrization 
(\ref{otherparamapp}) becomes the known parametrization~\cite{Youssef,Maier1} 
of the fundamental modular equation (\ref{modularcurveapp}), namely 
$\, \,x \, = \, 1728 \, z/(z+16)^3\,$ and $\,\, y \, = \,  1728 \, z^2/(z+256)^3$.

\vskip .1cm

\subsection{Schwarzian equations }
\label{modularAPPSch}

\vskip .1cm

In general, one can rewrite a remarkable hypergeometric identity 
like (\ref{Many}), (\ref{identi}), in the form
\begin{eqnarray}
\label{modularform2app}
\hspace{-0.95in}&& \quad \quad  \quad \quad \quad \quad 
  {\cal A}(x) \cdot \, 
_2F_1\Bigl([\alpha, \, \beta], \, [\gamma], \,  x  \Bigr)
\,\, = \, \, \, \,  
 _2F_1\Bigl([\alpha, \, \beta], \, [\gamma], \, y(x)  \Bigr), 
\end{eqnarray}
where $\, {\cal A}(x)$ is an algebraic function
and where $\, y(x)$ is an algebraic function (more precisely an 
{\em algebraic series}) corresponding to the previous modular equation 
$\,\, M(x, \, y(x)) \, = \,\, \, 0$.

The Gauss hypergeometric function 
$\,\, _2F_1([\alpha, \, \beta], \, [\gamma], \, x) \, $
is solution of the second order linear 
differential operator\footnote[1]{Note that $\, A(x)$ is 
the log-derivative of 
$\, \, u(x) \, \,= \,\, \, x^{\gamma} \cdot \, (1 \, -x)^{\alpha+\beta+1-\gamma}$.}: 
\begin{eqnarray}
\hspace{-0.95in}&& \quad \quad  \quad \quad 
 \Omega \,\,  = \, \, \, \,  
D_x^2    \,  \,\,   + \, A(x) \cdot \,  D_x  \, \, + \, B(x), 
\quad \quad \quad  \quad \quad  \hbox{where:} 
\nonumber \\
\label{Gaussdiffapp}
\hspace{-0.95in}&& \quad    
A(x) \,\,  = \, \, \, 
{{ (\alpha +\beta+1) \cdot  \, x \,  \, -\gamma} \over { x \cdot \, (x\, -1)}}
  \, \, = \, \,  \,  {{u'(x)} \over { u(x)}}, 
\quad  \quad \quad 
B(x) \,\,  = \, \, \,  {{\alpha  \, \beta } \over {x \cdot \, (x\, -1) }}.
\end{eqnarray}
A  straightforward calculation 
enables us to find the algebraic function $\, {\cal A}(x)$
in terms of the algebraic function pullback $\, y(x)$
in (\ref{modularform2app}):
\begin{eqnarray}
\label{modularform3app}
\hspace{-0.95in}&& \quad   \quad \quad \quad \quad  \quad \quad  \quad \quad
{\cal A}(x)  \, \, = \, \,  \,  
 \Bigl( {{u(x)} \over { u(y(x)) }}
 \cdot \,  y'(x) \Bigr)^{-1/2}. 
\end{eqnarray}

The identification of the two operators, 
$\, 1/v(x) \cdot \, \Omega \cdot \, v(x)$ and $\, \Omega^{pull}$  
(the pullback of operator $\, \Omega$ for a pullback $\, y(x)$),  thus
corresponds (beyond (\ref{modularform3app})) to just one condition that can 
be rewritten (after some algebra ...) in the following 
Schwarzian form~\cite{Youssef,Youssef2}:
\begin{eqnarray}
\label{condition1app}
\hspace{-0.95in}&& \quad   \quad \quad  \quad \quad  \quad 
 W(x) 
\, \, \,  \,-W(y(x)) \cdot  \, y'(x)^2
\, \, \,  \,+ \,  \{ y(x), \, x\} 
\, \,\, \, = \,\, \, \,  \, 0, 
\end{eqnarray}
or: 
\begin{eqnarray}
\label{conditionapp}
\hspace{-0.95in}&& \quad   \quad \quad  \quad \quad  \quad 
 {{W(x)} \over {y'(x)}}  
\, \, \,  \,-W(y(x)) \cdot  \, y'(x)
\, \, \,  \,+ \, {{ \{ y(x), \, x\} } \over {y'(x)}}
\, \,\, \, = \,\, \, \,  \, 0, 
\end{eqnarray}
where  
\begin{eqnarray}
\label{wherecond}
\hspace{-0.95in}&& \quad   \quad \quad  \quad \quad  \quad 
W(x)  \, \, = \, \,  \, \, \,  A'(x) \, \, \,  + \, \, 
   {{A(x)^2} \over {2 }} \, \,  \, \,  -2 \cdot \, B(x),
\end{eqnarray}
and where $\,\{ y(x), \, x\} \, $ denotes the 
{\em Schwarzian derivative}~\cite{What}:
\begin{eqnarray}
\label{Schwaapp}
\hspace{-0.95in}&&  \, \,  \,    \, 
\{ y(x), \, x\}    \, \, = \,    \,  \,
{{y'''(x) } \over{ y'(x)}} 
 \,  \,  - \, \, {{3} \over {2}}
 \cdot \, \Bigl({{y''(x)} \over{y'(x)}}\Bigr)^2
 \, \, = \, \,     \,
{{ d } \over { dx  }} \Bigl( {{y''(x) } \over{ y'(x)}}   \Bigr) 
\, \, - {{1} \over {2}} \cdot \, \Bigl( {{y''(x) } \over{ y'(x)}}   \Bigr)^2. 
\end{eqnarray}
For (\ref{Gaussdiffapp}) the function $\, W(x)$ reads:
\begin{eqnarray}
\label{SchwaWapp}
\hspace{-0.98in}&&
 {{ 
 (\alpha-\beta+1)\cdot \, (\alpha-\beta-1)\cdot \, x^2  \,
+2\cdot \, (2\, \alpha \, \, \beta
 -\alpha \, \gamma \, -\beta \, \gamma \, +\gamma)\cdot \, x 
\,  + \gamma \cdot \, (\gamma \, -2)
} \over { 2 \cdot \, x^2 \cdot \, (x\, -1)^2}}. \, \, 
\end{eqnarray}
The hypergeometric functions such that $\, W(x) \, = \, \, W(1\, -x) \, $ 
correspond to the two conditions: 
\begin{eqnarray}
\label{SchwaCondapp}
\hspace{-0.98in}&& \quad \quad \quad \quad  \quad 
 \alpha \, + \, \beta \, \, = \, \, \, 1 \, \, 
 \,\quad \quad  \quad \hbox{or:} \, \,  \quad \quad  \quad  \, 
\alpha \, + \, \beta \, \, = \, \, \, \, 2 \, \gamma \, -1.
\end{eqnarray}
This is the case, for instance~\cite{Garvan}, with the hypergeometric functions
 $\, _2F_1([1/2,1/2][1], \, x)$,  $\, _2F_1([1/3,2/3][1], \,x)$, 
$\, _2F_1([1/4,3/4][1], \,x)$, $\, _2F_1([1/6,5/6][1], \,x)$. 

\vskip .1cm
\vskip .2cm

{\bf Remark:} Denoting  $\, W_H(x)$ the rational function $\, W(x)$ given by (\ref{wherecond}) 
for the second order linear differential operator annihilating (\ref{2F15HypformA}), 
i.e. $\, _2F_1([1/12,5/12],[1],{\cal H}(x))$ where $\, {\cal H}(x)$
is the Hauptmodul $\,  1 \, -P_4(x)^2/P_2(x)^3$ in  (\ref{2F15HypformA}), 
with $\, P_2$ and $\, P_4$ given by (\ref{P2}) and (\ref{P4}).
$\, W_H(x)$ can be deduced from the $\, W(x)$ for the order-two linear differential
 operator annihilating $\, _2F_1([1/12,5/12],[1],\, x)$,  
from the relation~\cite{Youssef2}
\begin{eqnarray}
\label{deduced}
\hspace{-0.98in}&&  \, \,  \, \, \, \, 
W_H(x) \, \, = \, \, \, 
  W\Bigl({\cal H}(x) \Bigr) \, - \, \{{\cal H}(x), \, x\},  \, \,  \quad  \, \, 
 W(x) \, = \, \, -{{32 \, x^2 \, -41\, x \, +36} \over {72 \cdot \, (x \, -1)^2 \cdot\, x^2}}, 
\end{eqnarray}
where $\, \{{\cal H}(x), \, x\}$ denotes the Schwarzian derivative of $\,{\cal H}(x)$. $\, W_H(x)$ is of the form
\begin{eqnarray}
\label{deducedoftheform}
\hspace{-0.98in}&&  \quad \quad \quad \quad \quad
 W_H(x) \, \, = \, \, \, \, {{p_{16}(x) } \over { x^2 \cdot \, p_{3}(x)^2 \cdot \, p_{5}(x)^2  }}
 \, \, \, = \, \, \,  \, -{{1} \over {2 \, x^2}} \,\,\,   + \, \, \cdots 
\end{eqnarray}
where $\, p_{16}(x)$,  $\, p_{3}(x)$ and $\, p_{5}(x)$ are polynomials of degree 
respectively sixteen, three and five in $\, x$. These polynomials are 
{\em homomogeneous polynomials} 
in the seven parameters $\, a$,  $\, b_i$,  $\, c_i \,\, $ of (\ref{Ratfonc}).
For instance  $\, p_{16}(x)$ is a homogeneous polynomial of homogeneous degree $\, 44$,
 $\, p_{3}(x)$ is a homogeneous polynomial of homogeneous degree $\, 10$
and  $\, p_{5}(x)$ is a homogeneous polynomial of homogeneous degree $\, 12$.

\vskip .1cm

\section{Exact expression of  polynomial $\, P_6 \, $ for the ten-parameter rational function (\ref{Ratfoncplusplusplus})}
\label{DeltaP6}

The diagonal of the  ten-parameters rational function (\ref{Ratfoncplusplusplus}) 
is the pullbacked hypergeometric function
\begin{eqnarray}
\label{2F15HypformAplusplusplusapp}
\hspace{-0.7in}&&\quad \quad \quad \quad \quad 
{{1} \over { P_3(x)^{1/4}}} \cdot \, 
 _2F_1\Bigl([{{1} \over {12}}, \, {{5} \over {12}}], \, [1],
 \, \, 1 \, - \, {{P_6(x)^2 } \over {P_3(x)^3}}\Bigr),
\end{eqnarray}
where $\, P_3(x)$ is given by (\ref{oftheformP3}) and 
$\, P_6(x)$ is a polynomial of degree six in $\, x$
of the form 
\begin{eqnarray}
\label{oftheformP6}
\hspace{-0.99in}&&  \quad \quad  \quad \quad \quad \quad  \quad  \,  \, 
P_6(x) \, \,   \,  \,  = \, \,  \,  \,  \,   p_4 \, \, \, \,   + \, \, \Delta_6(x), 
\end{eqnarray}
where $\, p_4$ is the  polynomial  $\, P_4(x)$ given by (\ref{P4})  
in section (\ref{R}), and where $\, \Delta_6(x)$
is the following polynomial of degree six in $\, x$:
\begin{eqnarray}
\label{DeltaP6exact}
\hspace{-0.98in}&&\quad \quad \quad 
\Delta_6(x)  \,\,    = \, \, \, -5832 \cdot \, d_1^2\, d_2^2\, d_3^2 \cdot \, x^6
\nonumber \\
\hspace{-0.98in}&&\quad \quad  \quad \quad  \quad 
+3888 \cdot \, d_1\, d_2\, d_3 \cdot \, 
(b_1\, c_2 \, d_2\, +\, b_2\, c_3 \, d_3 +\, b_3\, c_1\, d_1) \cdot \, x^5
\nonumber \\
\hspace{-0.98in}&&\quad \quad  \quad \quad  \quad 
-864 \cdot \, (c_1^3\, d_1^2\, d_3 +\, c_2^3\, d_1\, d_2^2 \, +\, c_3^3\, d_2\, d_3^2) \cdot \, x^5 
\nonumber \\
\hspace{-0.98in}&&\quad \quad  \quad \quad  \quad 
 -1296 \cdot \,  c_1\, c_2\, c_3\, d_1\, d_2\, d_3 \cdot \, x^5
\nonumber 
\end{eqnarray}
\begin{eqnarray}
\hspace{-0.98in}&&\quad \quad  \quad \quad  \quad 
 -1296 \cdot \, b_1\, b_2\, b_3\, d_1\, d_2\, d_3 \cdot \, x^4 \, \, 
\nonumber \\
\hspace{-0.98in}&&\quad \quad  \quad \quad  \quad 
-1296 \cdot \, a \cdot \, d_1\, d_2\, d_3\, (b_1\, c_1+b_2\, c_2+b_3\, c_3) \cdot \, x^4 
\nonumber \\
\hspace{-0.98in}&&\quad \quad  \quad \quad  \quad 
-1296 \cdot \, 
(b_1\, b_2\, c_2\, c_3\, d_2\, d_3+b_1\, b_3\, c_1\, c_2\, d_1\, d_2+b_2\, b_3\, c_1\, c_3\, d_1\, d_3) \cdot \, x^4
\nonumber \\
\hspace{-0.98in}&&\quad \quad  \quad \quad  \quad 
+864 \cdot \, (c_1^2\, c_3\, d_1\, d_3+c_1\, c_2^2\, d_1\, d_2+c_2\, c_3^2\, d_2\, d_3)
 \cdot \, a\,\cdot \, x^4
\nonumber \\
\hspace{-0.98in}&&\quad \quad  \quad \quad  \quad 
-864 \cdot \, (b_1^3\, d_2^2\, d_3+b_2^3\, d_1\, d_3^2  +b_3^3\, d_1^2\, d_2) \cdot \, x^4
\nonumber \\
\hspace{-0.98in}&&\quad \quad  \quad \quad  \quad 
+864 \cdot \, \Bigl(b_1^2\, c_1\, c_3\, d_2\, d_3+b_1\, b_2\, c_1^2\, d_1\, d_3+b_1\, b_3\, c_3^2\, d_2\, d_3
\nonumber \\
\hspace{-0.98in}&&\quad \quad \quad  \quad \quad \quad \quad  \quad 
+b_2^2\, c_1\, c_2\, d_1\, d_3+b_2\, b_3\, c_2^2\, d_1\, d_2+b_3^2\, c_2\, c_3\, d_1\, d_2\Bigr) \cdot \, x^4
\nonumber \\
\hspace{-0.98in}&&\quad \quad  \quad \quad  \quad 
+216 \cdot \, (b_1^2\, c_2^2\, d_2^2+b_2^2\, c_3^2\, d_3^2+b_3^2\, c_1^2\, d_1^2) \cdot \, x^4
\nonumber \\
\hspace{-0.98in}&&\quad \quad  \quad \quad  \quad 
+288 \cdot \, (b_1\, c_1^3\, c_2\, d_1+b_2\, c_2^3\, c_3\, d_2+b_3\, c_1\, c_3^3\, d_3) \cdot \, x^4
\nonumber \\
\hspace{-0.98in}&&\quad \quad  \quad \quad  \quad 
-576 \cdot \, (b_1\, c_1^2\, c_3^2\, d_3+b_2\, c_1^2\, c_2^2\, d_1+b_3\, c_2^2\, c_3^2\, d_2) \cdot \, x^4
\nonumber \\
\hspace{-0.98in}&&\quad \quad  \quad \quad  \quad 
-144 \cdot \, c_1\, c_2\, c_3 \cdot \, (b_1\, c_2\, d_2+b_2\, c_3\, d_3+b_3\, c_1\, d_1)  \cdot \, x^4
\nonumber
\end{eqnarray}
\begin{eqnarray}
\hspace{-0.98in}&&\quad \quad  \quad \quad  \quad  \quad  \quad 
+540 \cdot \, d_1\, d_2\, d_3\, a^3 \cdot \, x^3
\nonumber \\
\hspace{-0.98in}&&\quad \quad  \quad \quad  \quad  \quad  \quad 
-648 \cdot \, (b_1\, c_3\, d_2\, d_3+b_2\, c_1\, d_1\, d_3+b_3\, c_2\, d_1\, d_2)
 \cdot \, a^2 \cdot \, x^3
\nonumber \\
\hspace{-0.98in}&&\quad \quad \quad \quad  \quad \quad  \quad  \quad  \quad 
-72 \cdot \, (c_1^2\, c_2\, d_1+c_1\, c_3^2\, d_3+c_2^2\, c_3\, d_2) \cdot \, a^2 \cdot \, x^3
\nonumber \\
\hspace{-0.98in}&&\quad \quad \quad \quad  \quad \quad  \quad 
+288 \cdot \, (b_1^3\, b_3\, c_1\, d_2+b_1\, b_2^3\, c_2\, d_3+b_2\, b_3^3\, c_3\, d_1) \cdot \, x^3
\nonumber \\
\hspace{-0.98in}&&\quad \quad \quad \quad \quad  \quad  \quad  \quad 
-576 \cdot \, (b_1^2\, b_2^2\, c_1\, d_3+b_1^2\, b_3^2\, c_3\, d_2+b_2^2\, b_3^2\, c_2\, d_1) \cdot \, x^3
\nonumber \\
\hspace{-0.98in}&&\quad \quad \quad \quad \quad  \quad  \quad  \quad 
-144 \cdot \, b_1\, b_2\, b_3\, (b_1\, c_2\, d_2+b_2\, c_3\, d_3+b_3\, c_1\, d_1) \cdot \, x^3
\nonumber \\
\hspace{-0.98in}&&\quad \quad \quad \quad \quad  \quad  \quad  \quad 
+864 \cdot \, (b_1^2\, b_2\, d_2\, d_3+b_1\, b_3^2\, d_1\, d_2+b_2^2\, b_3\, d_1\, d_3) \cdot \, a \cdot \, x^3
\nonumber \\
\hspace{-0.98in}&&\quad \quad \quad \quad \quad  \quad  \quad  \quad 
-144 \cdot \, \Bigl(b_1^2\, c_1\, c_2\, d_2+b_1\, b_2\, c_2^2\, d_2+b_1\, b_3\, c_1^2\, d_1
\nonumber \\
\hspace{-0.98in}&&\quad \quad \quad \quad \quad \quad \quad \quad  \quad \quad  \quad 
  +b_2^2\, c_2\, c_3\, d_3+b_2\, b_3\, c_3^2\, d_3+b_3^2\, c_1\, c_3\, d_1 \Bigr) \cdot \, a \cdot \, x^3
\nonumber \\
\hspace{-0.98in}&&\quad \quad \quad \quad  \quad  \quad \quad  \quad 
+720 \cdot \, (b_1\, b_2\, c_1\, c_3\, d_3+b_1\, b_3\, c_2\, c_3\, d_2+b_2\, b_3\, c_1\, c_2\, d_1)
 \cdot \, a \cdot \, x^3
\nonumber 
\end{eqnarray}
\begin{eqnarray}
\hspace{-0.98in}&&\quad \quad \quad \quad \quad \quad  \quad 
 \, +36 \cdot \, a^3 \cdot \, (b_1\, c_2\, d_2 +\, b_2\, c_3\, d_3 +\, b_3\, c_1\, d_1)  \cdot \, x^2
 \nonumber \\
\hspace{-0.98in}&&\quad \quad \quad \quad \quad \quad \quad  \quad   \quad  \quad   
 \,   -72  \cdot \, a^2 \cdot \, (b_1^2\, b_3\, d_2 +\, b_1\, b_2^2\, d_3 +\, b_2\, b_3^2\, d_1)  \cdot \, x^2.
\end{eqnarray}

\vskip .1cm 

\section{Polynomials $\, P_3(x)$ and  $\, P_5(x)$ for the nine-parameter rational function (\ref{Ratfoncplusplusplus})}
\label{polynomials}

The two polynomials  $\, P_3(x)$ and  $\, P_5(x)$ encoding the
pullback of the pullbacked hypergeometric function (\ref{2F15HypformAplusplusplusbis}) 
for the nine-parameter rational function (\ref{Ratfoncplusplusplus})
in section (\ref{justtwosteps}), read 
\begin{eqnarray}
\label{P3screw}
\hspace{-0.7in}&&\quad  \quad  \, \,\,
P_3(x) \, \,  = \, \, \,  \,  \,  p_2 \, \, \, 
+ \, \, 48 \cdot \, c_2 \cdot \, (3\, b_3\, d_1\, d_2\,-c_1^2\, d_1\, -c_2\, c_3\, d_2) \cdot \, x^3
\nonumber \\
\hspace{-0.7in}&&\quad \quad \quad  \quad  
\, \, \, \,\, +24 \cdot \,
 (a\, b_1\, c_2\, d_2\,+a\, b_3\, c_1\, d_1\,-2\, b_1^2\, b_3\, d_2\,-2\, b_2\, b_3^2\, d_1) \cdot \, x^2, 
\end{eqnarray}
and 
\begin{eqnarray}
\label{P5screw}
\hspace{-0.98in}&&\quad  \quad  
P_5(x) \, \,  = \, \,  \,  \, 
 p_4 \,  \, \,  \,  -864 \cdot \, c_2^3 \,  \, d_1\, d_2^2 \cdot \, x^5
\nonumber \\
\hspace{-0.7in}&&\quad \quad  \quad   \quad  
+ 864 \cdot \, (a\, c_1\, c_2^2\, d_1\, d_2 +\, b_2\, b_3\, c_2^2\, d_1\, d_2 \,
 +\, b_3^2\, c_2\, c_3\, d_1\, d_2 - \, b_3^3\, d_1^2\, d_2) \cdot \, x^4
\nonumber \\
\hspace{-0.98in}&&\quad\quad  \quad  \quad   
-576 \cdot \, (b_2\, c_1^2\, c_2^2\, d_1 \, + \, b_3\, c_2^2\, c_3^2\, d_2 ) \cdot \, x^4
\nonumber \\
\hspace{-0.98in}&&\quad \quad  \quad   \quad  
+ 288  \cdot \, (b_1\, c_1^3\, c_2\, d_1 \, +\, b_2\, c_2^3\, c_3\, d_2  ) \cdot \, x^4
 \nonumber \\
\hspace{-0.98in}&&\quad \quad  \quad   \quad  
 -144 \cdot \, (b_1\, c_1\, c_2^2\, c_3\, d_2  \, +\, b_3\, c_1^2\, c_2\, c_3\, d_1 ) \cdot \, x^4
 \nonumber \\
\hspace{-0.98in}&&\quad \quad  \quad   \quad  
+ 216  \cdot \, ( b_1^2\, c_2^2\, d_2^2  \,  +\, b_3^2\, c_1^2\, d_1^2 \, 
- 6 \, b_1\, b_3\, c_1\, c_2\, d_1\, d_2 ) \cdot \, x^4
\nonumber \\
\hspace{-0.98in}&&\quad \quad  \quad   \quad  
 -72 \cdot \, (9\, a^2\, b_3\, c_2\, d_1\, d_2 
+a^2\, c_1^2\, c_2\, d_1+a^2\, c_2^2\, c_3\, d_2) \cdot \, x^3
\nonumber \\
\hspace{-0.98in}&&\quad \quad  \quad   \quad  
 -144 \cdot \, a \cdot \,
 (b_1^2\, c_1\, c_2\, d_2 \,  +\, b_1\, b_2\, c_2^2\, d_2 +\, b_1\, b_3\, c_1^2\, d_1 
 \, + \, b_3^2\, c_1\, c_3\, d_1) 
\cdot \, x^3 \nonumber \\
\hspace{-0.98in}&&\quad\quad  \quad   \quad  \quad    \quad  \quad   
 \,  -144 \cdot \, (b_1^2\, b_2\, b_3\, c_2\, d_2 +\, b_1\, b_2\, b_3^2\, c_1\, d_1) 
\cdot \, x^3
\nonumber \\
\hspace{-0.98in}&&\quad \quad  \quad   \quad  
 + 720  \cdot \, (a\, b_1\, b_3\, c_2\, c_3\, d_2 \, +\, a\, b_2\, b_3\, c_1\, c_2\, d_1 ) 
\cdot \, x^3
\nonumber \\
\hspace{-0.98in}&&\quad\quad  \quad   \quad  
 - 576 \cdot \, (b_1^2\, b_3^2\, c_3\, d_2 \, +\, b_2^2\, b_3^2\, c_2\, d_1  ) 
\cdot \, x^3
 \nonumber \\
\hspace{-0.98in}&&\quad \quad  \quad   \quad  
 +288 \,  \cdot \, (b_1^3\, b_3\, c_1\, d_2 
\, +\, b_2\, b_3^3\, c_3\, d_1 \, +3 \, a\, b_1\, b_3^2\, d_1\, d_2) \cdot \, x^3
\nonumber \\
\hspace{-0.98in}&& \quad \quad \quad  \quad  \, \, 
+36 \cdot \, a^2 \cdot \, 
(a\, b_1\, c_2\, d_2+a\, b_3\, c_1\, d_1 -2\, b_1^2\, b_3\, d_2 -2\, b_2\, b_3^2\, d_1) \cdot \, x^2, 
\end{eqnarray}
where the polynomials  $\, p_2$ and $\, p_4$  are the  polynomials  
$\, P_2(x)$ and  $\, P_4(x)$ of degree two and four in $\, x$
 given by (\ref{P2}) and  (\ref{P4}) in section (\ref{R}): 
 $\, p_2$ and $\, p_4$   correspond to the $\, d_1 \, = \, d_2 \, = \, 0$ limit. 

\vskip .1cm 

\section{Monomial symmetries on diagonals}
\label{monomialdiag}

Let us sketch the demonstration of the monomial symmetry results 
of section (\ref{monomial}), with the condition that the determinant of (\ref{3x3})
is not zero and 
the conditions (\ref{defdiagBIScond}) are verified. We will 
denote by $\, n$ the integer in the three equal  sums (\ref{defdiagBIScond}): 
$\, n \, = \, A_i\, +B_i\, +C_i$.
The diagonal of the rational function of three variables ${\cal R}$ is defined 
through its multi-Taylor expansion (for small $\, x$, $\, y$ and $\, z$): 
\begin{eqnarray}
\label{defdiagBIS}
\hspace{-0.90in}&&\quad \quad \, \, \,
{\cal R}\Bigl(x, \, y, \, z\Bigr)
\, \, \,\, = \, \,  \,  
\sum_{m_1 \, = \, 0}^{\infty} \, \sum_{m_2 \, = \, 0}^{\infty} \, \sum_{m_3\, = \, 0}^{\infty} 
 \,R_{m_1,  \, \ldots, \, m_n}
\cdot  \, x^{m_1} \cdot \, y^{m_2} \cdot \, z^{m_3}, 
\end{eqnarray}
as the series in one variable $\, x$:
\begin{eqnarray}
\label{defdiag2BIS}
\hspace{-0.7in}&&\quad \quad \quad  \, 
\Phi(x) \, \,\, = \, \, \, \, Diag\Bigl({\cal R}\Bigl(x, \, y, \, z \Bigr)\Bigr)
\, \, \, = \,  \, \,  \sum_{m \, = \, 0}^{\infty}
 \,R_{m, \, m, \, m} \cdot \, x^{m}.
\end{eqnarray}
The monomial transformation (\ref{monomial}) changes the multi-Taylor 
expansion (\ref{defdiagBIS}) into
\begin{eqnarray}
\label{defdiagBISmonom}
\hspace{-0.90in}&& 
\tilde{{\cal R}}\Bigl(x, \, y, \, z\Bigr)
\, \, = \, \, \, \sum_{M_1 \, = \, 0}^{\infty} \sum_{M_2 \, = \, 0}^{\infty} \sum_{M_3\, = \, 0}^{\infty} 
 \, {\tilde R}_{M_1,  \, M_2 \, M_3}  \cdot \,  x^{M_1} \cdot \, y^{M_2} \cdot \, z^{M_3} \,\, \, \, = 
\nonumber \\
\hspace{-0.90in}&& \,\sum_{m_1 \, = \, 0}^{\infty} \sum_{m_2 \, = \, 0}^{\infty} \sum_{m_3\, = \, 0}^{\infty} 
 \,R_{m_1,  \, m_2 \, m_3}
\cdot  \, \Bigl( x^{A_1} \, y^{A_2} \, z^{A_3}  \Bigr)^{m_1}  \,
 \Bigl(x^{B_1} \, y^{B_2} \, z^{B_3}   \Bigr)^{m_2}  \, \Bigl(  x^{C_1}  \, y^{C_2}  \, z^{C_3} \Bigr)^{m_3}
\nonumber \\
 \hspace{-0.90in}&& \quad  \quad \quad \quad \quad
\, \, = \, \, \,
 \sum_{m_1 \, = \, 0}^{\infty} \sum_{m_2 \, = \, 0}^{\infty} \sum_{m_3\, = \, 0}^{\infty} 
 \,R_{m_1,  \, m_2 \, m_3}  \cdot \,  x^{M_1} \cdot \, y^{M_2} \cdot \, z^{M_3}
\nonumber
\end{eqnarray}
where:
\begin{eqnarray}
\label{M1}
 \hspace{-0.90in}&& \quad \quad \quad \quad \quad \quad \quad
M_1 \, \, = \, \,  \,  \,  A_1 \cdot \, m_1 \, + B_1 \cdot \, m_2  \, + \,C_1 \cdot \,  m_3,  \\
\label{M2}
 \hspace{-0.90in}&& \quad  \quad\quad \quad \quad \quad \quad
M_2 \, \, = \, \,  \, \, A_2 \cdot \,  m_1  \, + B_2 \cdot \,   m_2 \, + \,C_2 \cdot \, m_3,  \\
\label{M3}
 \hspace{-0.90in}&& \quad  \quad\quad \quad \quad \quad \quad
 M_3 \, \, = \, \,  \, \, A_3 \cdot \,  m_1  \, + B_3  \cdot \,  m_2  \, + \, C_3 \cdot \,  m_3.
\end{eqnarray}
Taking the diagonal amounts to forcing the exponents $\, m_1$, $\, m_2$ and $\, m_3$ to be equal. 
It is easy to see that when condition (\ref{defdiagBIScond}) is verified, 
$\,  m_1 =\, m_2 =\, m_3$ yields $\, M_1 \, = M_2 \,= \,  M_3$. Conversely 
if the determinant of (\ref{3x3}) is not zero it is straightforward to see
that the conditions $\,\, M_1 \, = M_2 \,= \,  M_3\,$ yield $\,\,  m_1 =\, m_2 =\, m_3$.

Then if one knows an exact expression for the diagonal of a rational function, 
the diagonal of this rational function changed by the monomial 
transformation (\ref{monomial}) reads
\begin{eqnarray}
\label{defdiag2BIS}
\hspace{-0.98in}&&
 Diag\Bigl(\tilde{{\cal R}}\Bigl(x, \, y, \, z \Bigr)\Bigr)
\,  = \,   \sum_{M \, = \, 0}^{\infty}
 \,\tilde{R}_{M, \, M, \, M} \cdot \, x^{M}
\,  = \,   \sum_{m \, = \, 0}^{\infty}
 \, R_{m, \, m, \, m} \cdot \, x^{n \cdot \, m} \, = \, \,  \Phi(x^n),
\end{eqnarray}
and thus equal to the previous   exact expression  $\,  \Phi(x)$, 
where we have changed $\, x \, \rightarrow \, \, x^n$, where $\, n$ is the integer  
$\,n \, = \, \,  A_1\, +B_1\, +C_1$
$ \, = \, \, A_2\, +B_2\, +C_2 \, = \, \, A_3\, +B_3\, +C_3$.
These monomial symmetries for diagonal of rational functions are not specific of 
rational functions of three variables: they can be straightforwardly generalized 
to an arbitrary number of variables. 

\vskip .1cm 

\section{Rescaling symmetries on diagonals}
\label{rescalingdiag}

\vskip .1cm 

We sketch the demonstration of the result in section (\ref{breath}).  One recalls
that the diagonal of the rational function of three variables ${\cal R}$ is defined 
through its multi-Taylor expansion (for small $\, x$, $\, y$ and $\, z$) 
\begin{eqnarray}
\label{defdiagBISter}
\hspace{-0.90in}&&\quad \quad \, \, \,
{\cal R}\Bigl(x, \, y, \, z\Bigr)
\, \, \,\, = \, \,\, \,
\sum_{m_1 \, = \, 0}^{\infty} \, \sum_{m_2 \, = \, 0}^{\infty} \, \sum_{m_3\, = \, 0}^{\infty} 
 \,R_{m_1,  \, \ldots, \, m_n}
\cdot  \, x^{m_1} \cdot \, y^{m_2} \cdot \, z^{m_3}, 
\end{eqnarray}
as the series in one variable $\, x$:
\begin{eqnarray}
\label{defdiag2BISter}
\hspace{-0.7in}&&\quad \quad 
\Phi(x) \, \,\, = \, \, \, \, Diag\Bigl({\cal R}\Bigl(x, \, y, \, z \Bigr)\Bigr)
\, \, \, = \,  \, \,  \sum_{m \, = \, 0}^{\infty}
 \,R_{m, \, m, \, m} \cdot \, x^{m}.
\end{eqnarray}
The (function rescaling) transformation (\ref{Scaling}) transforms the 
multi-Taylor expansion  (\ref{defdiagBISter}) into:
\begin{eqnarray}
\label{defdiagBISternew}
\hspace{-0.98in}&&\quad \quad \, \, \,
{\cal R}\Bigl(x, \, y, \, z\Bigr)
\, \, \,\, = \, \,\, \,
\\
\hspace{-0.98in}&&\quad \quad \quad  \quad \, \, \,
\sum_{m_1 \, = \, 0}^{\infty} \, \sum_{m_2 \, = \, 0}^{\infty} \, \sum_{m_3\, = \, 0}^{\infty} 
 \,R_{m_1,  \, \ldots, \, m_n}
\cdot  \, x^{m_1} \cdot \, y^{m_2} \cdot \, z^{m_3} \cdot \, F(x\, y\, z)^{m_1\, +m_2\, +m_3}. 
\nonumber 
\end{eqnarray}
We assume that the function $\, F(x)$ has some simple Taylor series expansion.
Each time taking the diagonal of (\ref{defdiagBISternew}) forces the exponents 
$\, m_1$, $\, m_2$ and $\, m_3$ to be equal in the term  $ \, x^{m_1} \cdot \, y^{m_2} \cdot \, z^{m_3}$ 
of the multi-Taylor expansion (\ref{defdiagBISternew}), one  gets a factor 
$\, F(x\, y\, z)^{m_1\, +m_2\, +m_3} \, = \, F(x\, y\, z)^{3 \, m}$. 
Consequently, the diagonal of (\ref{defdiagBISternew}) becomes: 
\begin{eqnarray}
\label{defdiag2BISter}
\hspace{-0.98in}&& \quad  \quad  \quad   \quad \quad   \quad 
 Diag\Bigl(\tilde{{\cal R}}\Bigl(x, \, y, \, z \Bigr)\Bigr)
\, \,\,  = \, \,  \,  \sum_{m \, = \, 0}^{\infty}
 \, R_{m, \, m, \, m} \cdot \, x^{n} \cdot \, F(x)^{3 \, n} 
\nonumber \\
\hspace{-0.98in}&& \quad  \quad   \quad   \quad   \quad \quad   \quad   \quad  \quad   \quad 
 \,  = \, \,
 Diag\Bigl({\cal R}\Bigl(x, \, y, \, z \Bigr)\Bigr)\Bigl( x \cdot \, F(x)^{3}\Bigr).
\end{eqnarray}
Clearly, these function-dependent rescaling symmetries for diagonals of rational 
functions are not specific of 
rational functions of three variables: they can be straightforwardly generalized 
to an {\em arbitrary number of variables}.

\vskip .1cm 

\vskip .1cm

\end{document}